\documentclass[fleqn,usenatbib]{mnras}

\usepackage{newtxtext,newtxmath}

\usepackage[T1]{fontenc}

\DeclareRobustCommand{\VAN}[3]{#2}
\let\VANthebibliography\thebibliography
\def\thebibliography{\DeclareRobustCommand{\VAN}[3]{##3}\VANthebibliography}

\usepackage{graphicx}	
\usepackage{amsmath}	

\usepackage[dvipsnames]{xcolor} 
\usepackage{color,soul} 
\usepackage{rotating}
\usepackage{longtable}

\usepackage{lscape}
\usepackage{caption}        
\usepackage{float}
\usepackage[none]{hyphenat}
\usepackage{booktabs}
\usepackage{caption}
\usepackage{subcaption}
\usepackage{textcomp}
\usepackage{marvosym}

\title[Extraplanar emission in isolated edge-on galaxies] 
{Extraplanar emission in isolated edge-on late-type galaxies. \\ I. 
The H$\alpha$ distribution versus to the old and young 
stellar discs.\thanks{Based on observations obtained at the Observatoire de 
Haute Provence (OHP, France), operated by the French CNRS.} 
}

\author[M. M. Sardaneta et al.]{
Minerva M. Sardaneta,$^{1}$\thanks{E-mail: mminerva@astro.unam.mx}
Philippe Amram,$^{2}$
Roberto Rampazzo,$^{3}$
Margarita Rosado,$^{1}$
M\'onica S\'anchez-Cruces,$^{1}$ 
\newauthor
Isaura Fuentes-Carrera$^{4}$
and Soumavo Ghosh$^{5}$
\\
$^{1}$Universidad Nacional Aut\'onoma de M\'exico. Instituto de Astronom\'ia.
A.P. 70-264, 04510. Ciudad de M\'exico, M\'exico\\
$^{2}$Aix Marseille Univ, CNRS, CNES, LAM, Marseille, France\\
$^{3}$INAF-Osservatorio Astrofisico di Asiago,Via dell’Osservatorio 8, 36012 Asiago, Italy\\
$^{4}$Escuela Superior de F\'isica y Matem\'aticas, Instituto Polit\'ecnico Nacional, U.P. Adolfo L\'opez Mateos, C.P.07738, Ciudad de M\'exico, M\'exico\\
$^{5}$Max-Planck-Institut f\"{u}r Astronomie, K\"{o}nigstuhl 17, D-69117 Heidelberg, Germany
}

\date{Accepted XXX. Received YYY; in original form ZZZ}

\pubyear{2023}

\begin{document}
\label{firstpage}
\pagerange{\pageref{firstpage}--\pageref{lastpage}}
\maketitle

\begin{abstract}
Isolated galaxies are the ideal reference sample to study the galaxy structure minimising potential environmental effects. We selected a complete sample of 14 nearby, late-type, highly inclined ($i\geq80^{\circ}$), isolated galaxies from the Catalogue of Isolated Galaxies (CIG) which offers a vertical view of their disc structure. 
We aim to study extraplanar Diffuse Ionized Gas (eDIG) by comparing the old and young disc components traced by near-infrared (\textit{NIR}) and Ultraviolet (\textit{UV})  imaging with the H$\alpha$ emission structure. 
We obtained H$\alpha$ monochromatic maps from the Fabry-Perot (FP) interferometry, while the old and young discs structures are obtained from the photometric analysis of the 2MASS \textit{K$_{s}$}-band, and GALEX \textit{NUV} and \textit{FUV} images, thereby identifying the stellar disc and whether the eDIG is present. 
The H$\alpha$ morphology is peculiar in CIG~71, CIG~183, CIG~593 showing clear asymmetries. 
In general, geometric parameters (isophotal position angle, peak light distribution, inclination) measured from H$\alpha$, \textit{UV} and \textit{NIR} show minimal differences (e.g. $\Delta i\leq\pm$10$^{\circ}$), suggesting that interaction does not play a significant role in shaping the morphology, as expected in isolated galaxies. 
From H$\alpha$ maps, the eDIG was detected vertically in 11 out of 14 galaxies. 
Although the fraction of eDIG is high, the comparison between our sample and a generic sample of inclined spirals suggests that the phenomenon is uncorrelated to the galaxy environment. As suggested by the extraplanar \textit{UV} emission found in 13 out of 14  galaxies the star formation extends well beyond the disc defined by the H$\alpha$ map. 
\end{abstract}

\begin{keywords}
galaxies: haloes -- galaxies: evolution -- galaxies: ISM -- 
galaxies: photometry -- galaxies: stellar content -- galaxies: fundamental parameters
\end{keywords}



\section{Introduction}\label{Sec:introduction}

Galaxy interactions dramatically impact several galaxy properties, such as their morphology and star-formation history \citep[see e.g.][and references therein]{rampazzo-2016}. The idea of gathering catalogues of isolated galaxies has been developed at least since the 1970s \citep[e.g. the \textit{Catalogue of Isolated Galaxies} (CIG),][]{Karachentseva-1973} as a baseline for comparison with galaxies located in denser environments subject to interaction-induced phenomena. In the last decades, catalogues of isolated galaxies, selected from very low galaxy density regions, have been either implemented or revised using redshift measurements \citep{verdesm-2005,Karachentseva-2009,Argudo-F-2013,Argudo-F-2015}. A recent H\,\textsc{i} study of isolated galaxy samples suggests a picture of "nurture free" galaxies  \citep{Jones2018}. Other works \citep[see e.g.][]{Rampazzo2020} have shown that isolated elliptical galaxies suffered past interactions from which they still show the `scars', although they have probably not experienced gravitational influences from their close neighbours over the past billion years.

This paper concentrates on the stellar versus H$\alpha$ components in
isolated spiral galaxies seen nearly edge-on ($i\geq80^{\circ}$). This galactic configuration relative to the sky plane provides more information on the vertical structure of the disc 
\citep[e.g.][]{burstein-1979, fraternalli-2006, kamphuis-2007,
peters-2017} revealing the extraplanar Diffuse Ionized Gas (eDIG),
detected in galaxies with relatively high star formation rate (SFR)
surface densities \citep[][]{rossa-2003-i, rossa-2003-ii}. 
The eDIG is usually traced at optical wavelengths using the H$\alpha$ recombination line \citep[e.g][]{Rosado-2013}. In addition, as the eDIG might be spatially correlated with the star formation (SF)-gas relations, it has been observed to trace regions of \ion{H}{i} dominance within galactic discs, especially those located outside active star-forming regions, making the 21-cm emission line a helpful tool for the eDIG detection \citep[e.g.][]{Zschaechner-2015-a}.  
Ultraviolet (\textit{UV}) imaging can also detect the low-intensity
outer star formation disc by detecting OB stars located in very low SFR
density regions \citep[e.g.][]{hoopes-2001,
thilker-2002,gil-de-paz-2005, thilker-2005,  thilker-2007-i, 
gil-de-paz-2007-ii}. From the comparison between the H$\alpha$ and the
\textit{UV} images of spiral galaxies with low inclination
($i\leq80^{\circ}$), \cite{thilker-2007-i} showed that the Near
Ultraviolet (\textit{NUV}) emitting component is radially more extended
than the emission of the old stellar population in at least 30\% of
galaxies from the Local Universe (z$\approx$0.02). In high inclined
galaxies, in cases where an eDIG layer does exist, the \textit{UV}
haloes tend to form a thick disc having a similar morphology to the
\textit{UV} halo and occurring in about the same place
\citep{hodges-k-2016}. Moreover, it has been observed that the
H$\alpha$-\textit{UV} flux ratio is lower in the eDIG than in \ion{H}{ii}
regions, indicating that the field OB stars, which are situated outside the star-forming regions, are important contributors to the eDIG ionization in most of the galaxies, and even in some galaxies they may be the dominant ionization source \citep{hoopes-2001, hodges-k-2016, jo-2018}.

The eDIG morphology shows a wide variety of local morphological features \cite[][]{rand-1998, rossa-2003-ii, Rosado-2013} such as prominent layers of diffuse gas, filamentary structures, or just one or a few patches of extraplanar emission \cite[][]{rand-1998, rossa-2003-ii, Rosado-2013}. 
Some eDIG originating phenomena have been proposed, such as 
Inter Stellar Medium (ISM) activity in the galaxy disc \citep[e.g.][]{heald-2006, heald-2006-b, heald-2007, Rosado-2013, ho-2016, jones-manga-2017, bizayaev-2017},  
intergalactic medium accretion \citep[e.g.][]{putman-2012, putman-2017,  levy-2019, bizyaev-2022},  
satellite galaxies interaction \citep[e.g.][]{walker-1996, Zschaechner-2015-a} 
and accretion or intra-cluster interaction \citep[e.g.][]{tomicic-2021, sardaneta-2022, boselli-2022}. 
In parallel, several studies have demonstrated that the \textit{UV} halo emission is consistent with a reflection nebula produced by dust in the halo \citep[e.g.][]{hodges-k-2014, shin-2015, hodges-k-2016, jo-2018, shin-2019}. Various mechanisms have been suggested as potential sources for diffuse-and-global \textit{UV}  haloes, including galactic radiation, magneto-hydrodynamic phenomena, and dust accretion from the circumgalactic or intergalactic medium \citep[see][for a comprehensive examination]{shin-2015}. The contribution of each process over cosmic time is still unclear as they both produce hot and cold components that occupy roughly the same space \citep[e.g.][]{hodges-k-2014}.

Several studies from large surveys such as SAMI\footnote{Sydney-AAO Multi-object Integral field spectrograph (SAMI)} \citep{ho-2016}, MaNGA\footnote{Mapping Nearby Galaxies at Apache Point (MaNGA)} \citep[][]{jones-manga-2017, bizayaev-2017, bizyaev-2022} and CALIFA\footnote{Calar-Alto Legacy Integral Field Area (CALIFA)} \citep{levy-2019}, and from specific surveys as CHANG-ES\footnote{Continuum haloes in Nearby Galaxies—an EVLA Survey (CHANG-ES)} 
\citep{lu-changes-2023} have employed various environmental definitions in the selection of their samples of edge-on galaxies, contributing to varying interpretations of the relationship between extraplanar gas properties and their host galaxies. 
For instance, they have either discarded major mergers with clear tidal features \citep[][]{ho-2016, bizyaev-2022}, 
rejected galaxies that have another object in the field, including stars \citep{bizayaev-2017} or 
excluded galaxies with distorted discs in their H$\alpha$ images \citep{lu-changes-2023}.  
Recently, two studies agreed that the eDIG could be a consequence of the accretion of the CircumGalactic Medium (CGM) \citep[][]{levy-2019, bizyaev-2022} which has also been proposed as probable \textit{UV} halo origin \citep[e.g.][]{hodges-k-2014, shin-2015}. 
However, most of the literature still lacks a general agreement on the sources of ionization for eDIG.  Since the eDIG may be related to the CGM, 
\cite{bizyaev-2022} visually examined the outer regions of each galaxy of their sample 
aiming to locate environmental structures that might be connected to the eDIG. Their findings ranged from small satellites lacking noticeable structure to large satellites even larger than the primary object. Therefore, the consensus or discrepancy in explaining the eDIG origin is likely a consequence of the different definitions of environment adopted. As a reference study case, isolated star-forming galaxies with the possibility of edge-on observations is needed.

In order to study the environmental effects on the eDIG distribution, we selected a sample of isolated nearby late-type, high-inclined galaxies from the CIG catalogue \citep[][]{Karachentseva-1973}.  Indeed, nearby galaxies are better resolved, late-type galaxies are those which contain the most H$\alpha$ emission, and almost edge-on galaxies are those for which it is easier to detect the gas emission at high galactic latitude and for which the column density is higher, even in the disk outskirts.  
The extraplanar gas is commonly defined as the detectable H$\alpha$ emission where the stellar continuum is not detected.  This could be done from the continuum-subtracted emission-line image which delimits the stellar disc \citep[e.g.][]{miller-2003-i, Rosado-2013, levy-2019, tomicic-2021}. This is typically what is done when observations with narrow- and broad-band filters around H$\alpha$ emission are combined. 
Alternatively, the old stellar disk in spiral galaxies (Population~II stars) could be better defined from redder broad-band in the near-infrared \citep[\textit{NIR}, e.g.][]{kamphuis-2007, ho-2016, bizayaev-2017, bizyaev-2022} and young stellar population from bluer broad-band, e.g. in the \textit{UV}, which traces the young stellar populations of ages up to tenths of Myrs \citep[][]{hoopes-2001, thilker-2007-i, bianchi-2011, Kennicutt2012}. 
We used this method in this work.  We present the H$\alpha$ emission maps of a sample of galaxies. The net H$\alpha$ maps (i.e. continuum free) were obtained using the scanning Fabry-Perot spectroimager (FP) GHASP, providing complete two-dimensional coverage of very extended line emission regions, ideal for studying the faint diffuse gas emission.  We compared H$\alpha$ FP monochromatic maps with \textit{NIR}, \textit{NUV} and Far \textit{UV} (\textit{FUV}) images available in the literature.  We defined a certain threshold in the \textit{NIR} image under which the stellar density is low and consider that the eDIG is the H$\alpha$ emission below this limit. 
The geometric parameters of the galaxies, such as the isophotal position angle, peak light distribution and inclination, are obtained from the photometric analysis of the \textit{UV} and \textit{NIR} images, and will serve as reference when studying the H$\alpha$ kinematics in a forthcoming paper.

The paper is structured as follows. 
In Section \ref{Sec:Sample}, we provide details about the sample selection. 
In Section \ref{Sec:Obs}, we describe the data acquisition and reduction processes. 
In Section \ref{Sec:stellarDisc}, we explain how the stellar disc and the extraplanar components are identified. 
In Section \ref{Sec:results}, we present the individual results of the photometric analysis of each galaxy in our sample. 
In Section \ref{Sec:discussion}, we discuss our main results. 
Finally, in Section \ref{Sec:Conc}, we provide a summary and the
conclusions of our results. 
In order to compare our results with previous surveys, throughout this work we assumed a Hubble constant of $H_{0}=70\,\,\,\mathrm{km\, s^{-1}\, Mpc^{-1}}$ \citep[e.g.][]{thilker-2007-i, ho-2016, levy-2019}.

\section{The sample selection}\label{Sec:Sample}

\begin{table*}
\caption{General parameters of the highly inclined isolated galaxies sample}

\begin{center}
\begin{tabular}{ccccccccccccc}
\hline
CIG & Other & RA (J2000) & Dec (J2000) & $V_{\mathrm{sys}}$ & Distance & $i$ &  $K_{s}$ & $D_{25}(B)$  & Type  & $S_{60}$ & $S_{100}$ & $L_{\mathrm{FIR}}$\\ 
Name & Name & (hh mm ss) & ($^{\circ}$ ‘ “) &  ($\mathrm{km\, s^{-1}}$) & (Mpc) & (deg) &   (mag) & (kpc) & & (Jy) & (Jy) &  ($10^{43}\,\mathrm{erg\,s^{-1}}$) \\  
(1) & (2) & (3) & (4) &  (5) & (6) & (7) &  (8) & (9) & (10) & (11) & (12) & (13) \\  
\hline 
71 & UGC 01391 & 01 55 15.8  & +10 00 49.2  & 5901 & 84.3 & 83.8 & 10.8 & 35.4 & Sc & 0.66 & 1.38 & 6.79 \\ 
95 & UGC 01733 & 02 15 20.6  & +22 00 22.0  & 4418 & 63.1 & 86.5 & 11.6 & 33.4 & Sc-w & 0.22 & 0.64 & 1.49 \\ 
159 & UGC 03326 & 05 39 37.1  & +77 18 44.9  & 4121 & 58.9 & 85.4 & 9.6 & 60.8 & *Scd: & 1.10 & 3.57 & 3.26 \\ 
171 & UGC 03474 & 06 32 37.6  & +71 33 39.5  & 3634 & 51.9 & 84.0 & 10.1 & 33.8 & *Scd: & 0.57 & 1.43 & 4.40 \\ 
183 & UGC 03791 & 07 18 31.8  & +27 09 28.7  & 5090 & 72.7 & 80.4 & 11.4 & 26.0 & *Sd : & 0.49 & 1.38 & 2.78 \\ 
201 & UGC 03979 & 07 44 31.0  & +67 16 24.9  & 4061 & 58.0 & 80.9 & 10.6 & 31.4 & SA(rs)c & 1.18 & 2.86 & 3.39 \\ 
329 & UGC 05010 & 09 24 55.1  & +26 46 28.8  & 4096 & 58.5 & 81.3 & 9.3 & 42.8 & SA(rs)b & 0.30 & 1.34 & 1.61 \\ 
416 & UGC 05642 & 10 25 41.8  & +11 44 20.8  & 2322 & 33.2 & 81.1 & 11.6 & 18.0 & Sd -pec & 0.52 & 1.16 & 2.69 \\ 
593 & UGC 08598 & 13 36 40.7  & +20 12 00.5  & 4909 & 70.1 & 83.2 & 10.8 & 35.5 & SBx(s)b: & 0.09 & 0.40 & 0.53 \\ 
847 & UGC 11132 & 18 09 26.2  & +38 47 39.9  & 2837 & 40.5 & 81.2 & 10.7 & 24.6 & *Sb & 0.59 & 1.98 & 5.74 \\ 
906 & UGC 11723 & 21 20 17.5  & -01 41 03.6  & 4899 & 70.0 & 80.9 & 10.0 & 37.9 & Sbc & 1.97 & 5.95 & 11.51 \\ 
922 & UGC 11785 & 21 39 26.8  & +02 49 37.6  & 4074 & 58.2 & 84.2 & 11.0 & 29.4 & Scd-w & 0.31 & 1.62 & 2.57 \\ 
936 & UGC 11859 & 21 58 07.4  & +01 00 32.3  & 3011 & 43.0 & 85.7 & 11.4 & 38.7 & Sc: & 0.71 & 1.49 & 1.13 \\ 
1003 & UGC 12304 & 23 01 08.3  & +05 39 15.7  & 3470 & 49.6 & 82.5 & 10.3 & 22.9 & Scd & 2.06 & 4.57 & 15.07 \\ 
\hline
\end{tabular}
\end{center}
Columns: (1) CIG galaxy name; (2) UGC galaxy name; (3) and (4) \textsc{wcs} coordinates (J2000); (5) $V_{\mathrm{sys}}$: systemic velocity from NED; (6) heliocentric distance to the galaxy; (7) $i$: inclination computed using the relation~\ref{eq:i1} (see the text); (8) $K_{s}$: apparent \textit{K$_{s}$}-band magnitude from NED;  (9) $D_{25}(B)$: optical diameter in the $B$-band from NED; (10) Hubble classification from \cite{buta-2019} and, if not available, from NED~(*); (11)\,and\,(12)\,flux densities at 25$\,\mu$m,  60$\,\mu$m and 100$\,\mu$m in Jy from \cite{lisenfeld-2007}; (13)\,\textit{FIR} luminosity computed with the equation\,\ref{ec:LFIR-rossa00Eq8} (see Section~\ref{Sec:ddd}).
\label{table:general}
\end{table*}

Isolated galaxies have been the subject of selection and study for many decades. Evidences are accumulating that over at least a few billion years the evolution of isolated galaxies has not been driven by interactions with physically associated companions \citep[see e.g.][and references therein]{verdesm-2005, Karachentseva-2009, Karachentsev-2011, Rampazzo2020}. In this sense, isolated galaxies are the ideal reference sample to study internal galactic forces and galaxy morphological, dynamical and photometric evolution minimizing  possible environmental effects.  
\medskip

The CIG catalogue \citep{Karachentseva-1973}  and its more recent revisions \citep{verdesm-2005,verley-2007, Argudo-F-2013}  are examples  of a successful attempt to compile a sample of isolated galaxies. 
Historically, \cite{Karachentseva-1973} obtained her catalogue from the visual inspection of the \textit{Digitized Sky Survey} (DSS) images of the 27\,837 galaxies --\,and of their surroundings\,-- (at Galactic latitudes $|b|\geq 20^{\circ}$)  contained in the CGCG\footnote{\textit{Catalogue of galaxies and of clusters of galaxies} \citep[CGCG,][]{Zwicky-1968} \url{https://cdsarc.cds.unistra.fr/viz-bin/cat/VII/190} .}. A list of 1050 galaxies  were found to meet the  isolation criterion she adopted \citep[see also][]{Karachentseva-2009}.

Since we are looking for the incidence of eDIG among isolated galaxies we have selected  a sample from the CIG catalogue \citep{Karachentseva-1973} of late-type galaxies (Spirals with morphologycal type\,$>5$, i.e. Sb to be checked) \citep[see][]{buta-2019} dominated by Population\,I stars that satisfy the following additional characteristics we derived from the  NED data-base:

\begin{enumerate}

\item galaxies with high inclination ($i\geq 80^{\circ}$) computed with the expression
\begin{equation}
i=\cos^{-1}(b/a),
\label{eq:i1}
\end{equation}
where $b$ and $a$ are the apparent optical major and minor axes of the galaxy respectively;

\item galaxies with redshift $z\leq 0.02$ to ensure that the emission of the redshifted H$\alpha$ line fits in the optical wavelength;

\item galaxies with  apparent \textit{K$_{s}$}-band magnitude $K_{s}\leq 12$. The near-infrared is sensitive to nuclear rings and large-scale bars, which might fuel active nuclei \citep[e.g.][]{kormendy-1982, Eskridge-2002, jarrett-2003}.

\end{enumerate}

A complete sample of 14 nearby isolated late-type edge-on galaxies was obtained (see Figure~\ref{fig:rgb}).
General parameters of our galaxy sample are listed in Table\,\ref{table:general}.

\begin{figure*}
  \centering 
  \subfloat{\includegraphics[width=0.31\hsize]{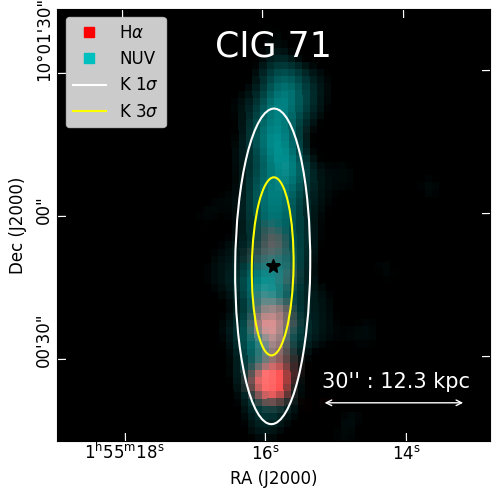}}%
   \qquad
  \subfloat{\includegraphics[width=0.31\hsize]{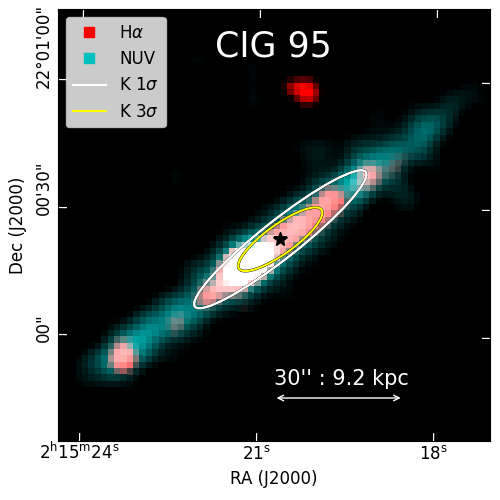}}
     \qquad
 \subfloat{\includegraphics[width=0.31\hsize]{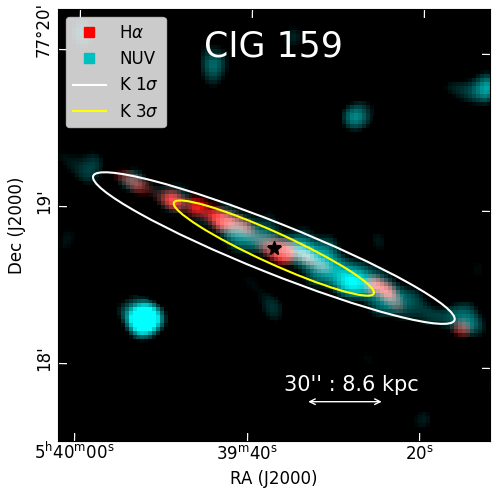}}%
 
  \subfloat{\includegraphics[width=0.31\hsize]{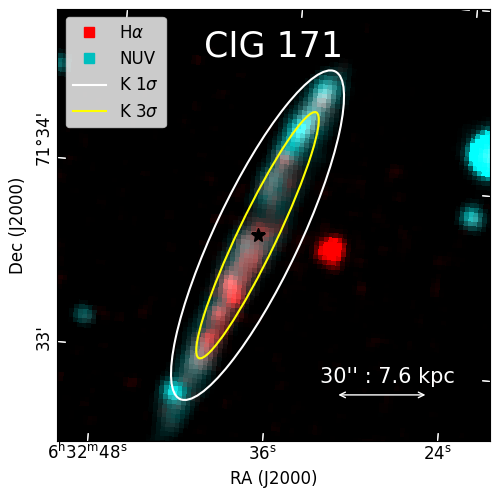}}
     \qquad
    \subfloat{\includegraphics[width=0.31\hsize]{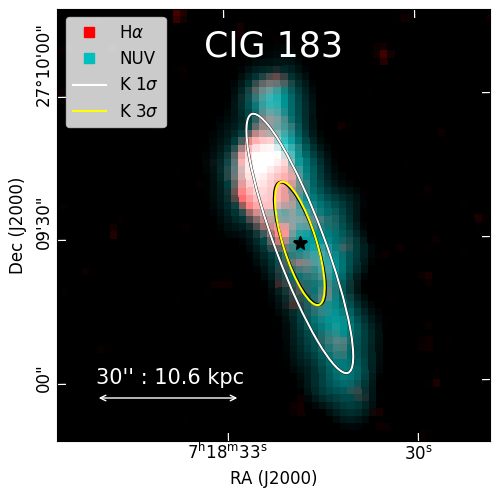}}%
       \qquad
  \subfloat{\includegraphics[width=0.31\hsize]{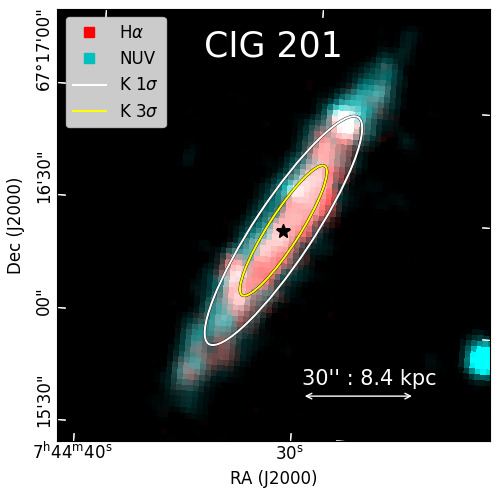}}

\subfloat{\includegraphics[width=0.31\hsize]{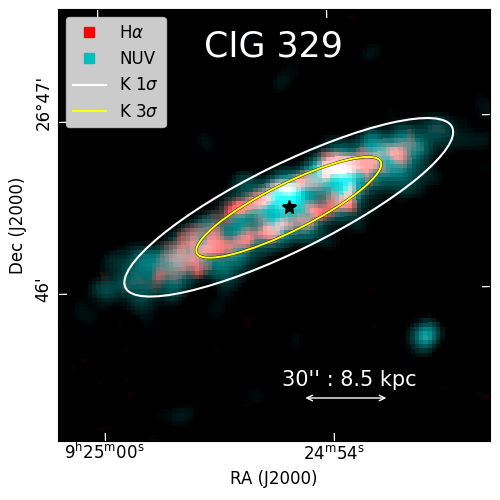}}%
  \qquad
  \subfloat{\includegraphics[width=0.31\hsize]{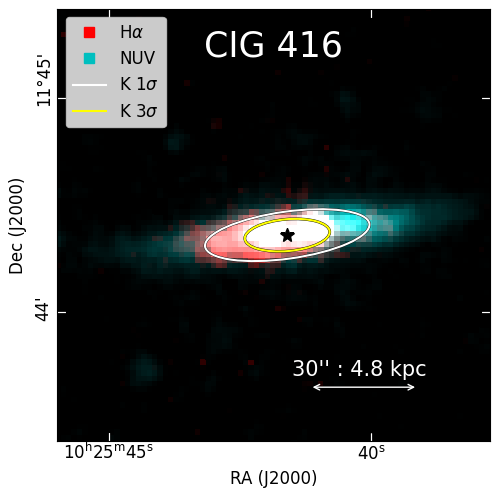}}
    \qquad
 \subfloat{\includegraphics[width=0.31\hsize]{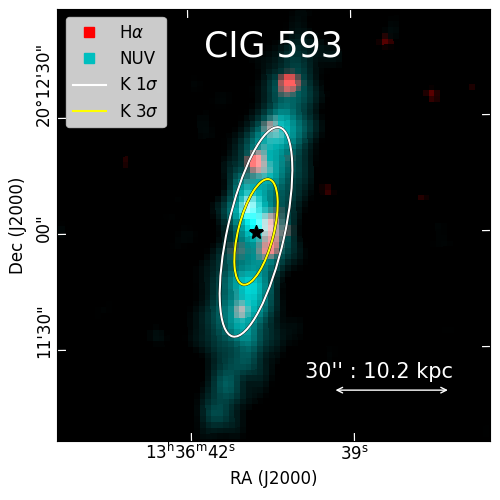}}%

    \caption{
    Superposition of the respective FP H$\alpha$ monochromatic map and GALEX~\textit{NUV} image of the galaxies in our sample. The ellipses fitted to the surface brightness level at 1$\sigma$ (white) and 3$\sigma$ (yellow) of the typical background noise \citep[see e.g.][]{jarrett-2000} of the respective 2MASS \textit{K$_{s}$}-band image are overlain on the respective map. 
    To distinguish the extraplanar material of each galaxy, the old stellar disc plane is traced by the ellipse fitted to the 3$\sigma$ surface brightness level. We have masked the nearest and brightest field stars to each galaxy in the \textit{UV}-band images. At the distance of each galaxy (see Table~\ref{table:general}) the kpc measure equivalent to 30~arcsec is indicated with an arrow. 
  }
  \label{fig:rgb}
\end{figure*}

\begin{figure*}
  \ContinuedFloat 
  \centering 
  
     \subfloat{\includegraphics[width=0.31\hsize]{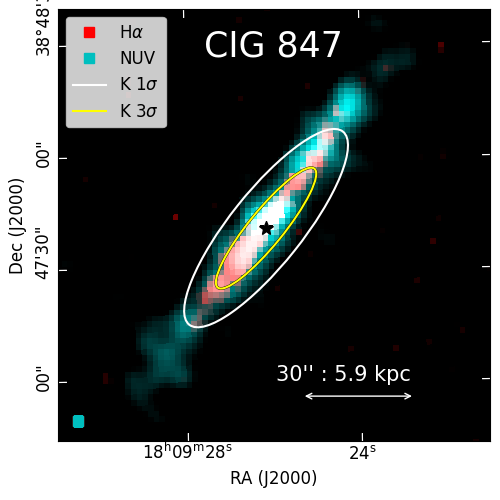}}
     \qquad
    \subfloat{\includegraphics[width=0.31\hsize]{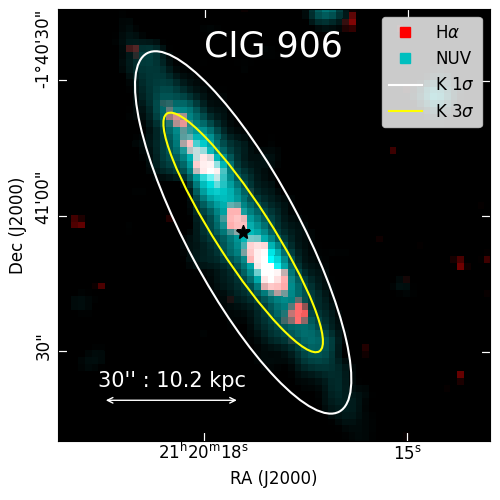}}%
    \qquad
\subfloat{\includegraphics[width=0.31\hsize]{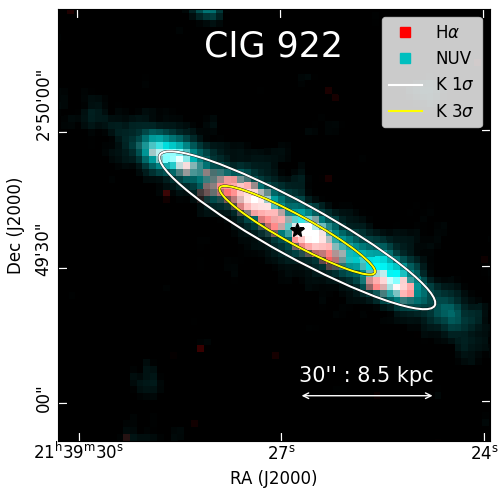}}%

\subfloat{\includegraphics[width=0.31\hsize]{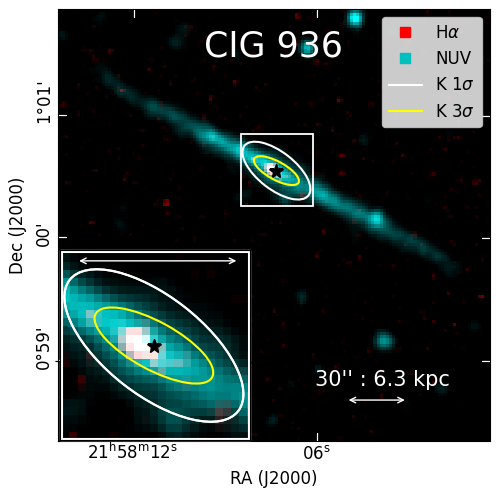}}
    \qquad
  \subfloat{\includegraphics[width=0.31\hsize]{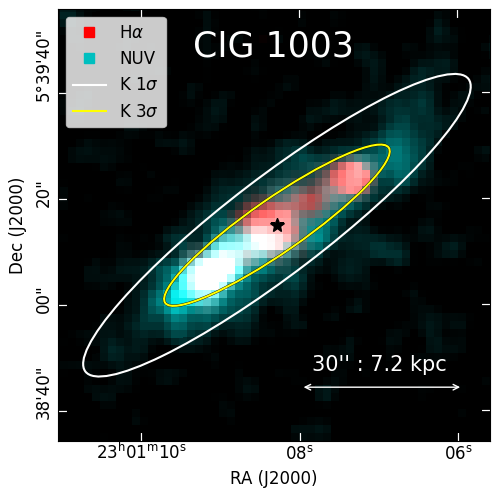}}
  
  \caption{Continue.}
  \label{fig:rgb}
\end{figure*}


\subsection*{Isolation degree}\label{sec:isolation}

To be an isolated galaxy, the CIG catalogue required that no similar size neighbours were found close to the galaxy. 
This requirement can be described as follows: a galaxy with a standard angular diameter denoted as $a_{1}$ is classified as isolated when the angular separation $X_{1i}$ 
between the galaxy and its $i$-th neighbor with angular diameter $a_{i}\,\epsilon\,[0.25,\,4]\cdot a_{1}$, meets or exceeds a threshold of $X_{1i}\geq20\,a_{i}$ \citep[see equations 1 and 2 in][]{Karachentseva-2009}. 
Since galaxies from the CIG catalogue were collected by visual inspection of DSS images, the companion galaxies are expected to be faint or mostly dwarf galaxies. 
Recently, the CIG catalogue has been refined using new quantification methods \citep[e.g.][]{verley-2007} and modern tools like images with higher resolution than the plates of Palomar \citep[e.g.][]{Argudo-F-2013} and new images from \ion{H}{i} emission data \citep[e.g.][]{Jones2018}. In this Section, we discuss how effectively our galaxy sample selected from the CIG catalogue meets these new isolation checks. However, even if a subset of our sample does not meet the strictest isolation criteria, all galaxies in our sample still represent a significant part of the overall isolated galaxy sample.

In the frame of the long-term AMIGA project (Analysis of the interstellar Medium of Isolated GAlaxies\footnote{Analysis of the interstellar Medium of Isolated GAlaxies (AMIGA) \url{(http://www.iaa.es/AMIGA.html/)}}), \cite{verley-2007-i, verley-2007} revised all the positions of the CIG galaxies based upon the digitized photographic plates from POSS-1 and POSS-2 providing a quantification of the degree of isolation of CIG galaxies with the local number density of neighbour galaxies ($\eta_{k}$), and the tidal strength ($Q_{k}$) affecting the central galaxy by its neighbourhood. These parameters depend on the position, diameter and mass of the primary galaxy and its $k$-th neighbour. 
Thus, these parameters provide a clear picture of the environment around the isolated galaxies: the presence of one single similar size neighbour at a small distance to the CIG galaxy would result in a high value of the tidal strength ($Q_{k}$) estimation, while the local number density ($\eta_{k}$) remains quite low as it is averaged over more nearest neighbours \citep{verley-2007}. 
In Table\,\ref{tab:neighbours} we list the number of neighbours and the nearest-neighbour distance to each galaxy in our sample according to \cite{verley-2007-i}. 
Then, \citet{Argudo-F-2013} re-evaluated the isolation criteria of AMIGA galaxies within a field radius of 1\,Mpc using both photometric and spectroscopic data available from the \textit{Sloan Digital Sky Survey} (SDSS) to refine the $\eta_{k}$ and $Q_{k}$ parameters.

\begin{figure}
\centering
\includegraphics[width=0.9\hsize]{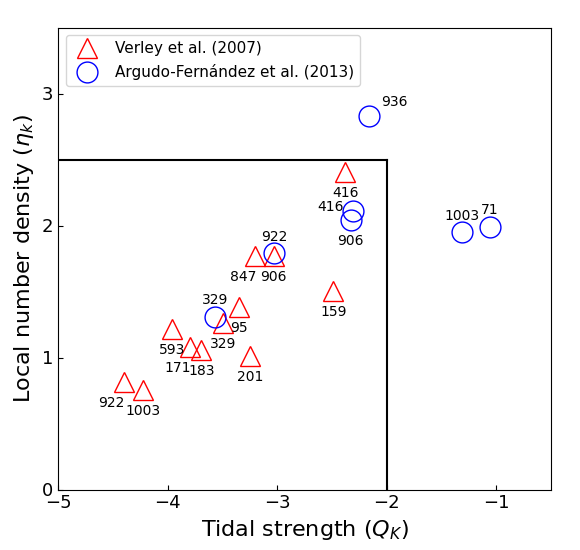}
\caption{Comparison between the local number density ($\eta_{k}$)  and tidal strength ($Q_{k}$) parameters for the CIG galaxies in our sample.  These parameters were obtained first, based on digitized photographic plates from POSS-1 and POSS-2 \citep[][]{verley-2007, verley-2007-i} (red triangles) and later, on available  photometric and spectroscopic  data from SDSS-DR9 \citep{Argudo-F-2013} (blue circles). The CIG catalogue numbers are indicated on top the circles and below the triangles. 
According to \citet{verley-2007}, the horizontal ($\eta_{k}=2.4$)  and vertical ($Q_{k}=-2$) lines enclose the {\it region of bona fide} isolated galaxies in the AMIGA sample. 
}
\label{Fig:isolation}
\end{figure}

Figure~\ref{Fig:isolation} shows the comparison between the local number density~($\eta_{k}$)  and tidal strength~($Q_{k}$) of the galaxies in our sample making the distinction between the results of \citet{verley-2007} and \citet{Argudo-F-2013}. 
From Figure~\ref{Fig:isolation} and Table\,\ref{tab:neighbours}, we observe that some galaxies classified as isolated in the AMIGA sample have its nearest neighbour at an angular distance of twice or three times their optical diameters, for example CIG~171 and 201. According with \cite{verley-2007-i}, most of the neighbouring galaxies of CIG galaxies have a diameter $\leq0.25 a_{1}$, implying they are dwarf companions which were not took into account by \cite{Karachentseva-1973} when collecting the CIG catalogue but perhaps they can influence the evolution of the main galaxy. 
Therefore, if CIG~171 and 201 are in region of bona fide isolated galaxies in the AMIGA sample, we may infer that the nearest neighbours are small enough to not exert significant gravitational perturbations on the main galaxy. On the other hand,   
\citet{verley-2007} had already determined that the galaxies of our sample, CIG\,71 and CIG\,936, were outside this region, excluding both galaxies from the AMIGA sample. As \citet{Argudo-F-2013} improved the quantification of the isolation degree, another galaxy of our sample, CIG\,1003, is also out of the area of isolated galaxies in the $Q_{k}-\eta_{k}$ plane. In this way, 3 out of 14 galaxies in our sample failed the isolation criteria of the AMIGA sample, probably because they have some faint  companions  previously undetected by \cite{Karachentseva-1973}.

Later, \cite{Jones2018} provided integrated  \ion{H}{i} fluxes of AMIGA galaxies by presenting a catalogue of \ion{H}{i} single dish observations.  
With these new data, \cite{Jones2018} included a review of integral features of AMIGA galaxies allowing to ensure isolation level of the AMIGA catalogue. Since the \cite{Argudo-F-2013} work is limited to the SDSS data, \cite{Jones2018} adopted the criteria of \cite{verley-2007}, and excluded all sources with heliocentric velocities below 1500\,$\mathrm{km\,s^{-1}}$, removed all dwarf galaxies that were in the CIG, and discarded all sources with potentially spurious spectral parameters, leaving a final sample of 544~CIG galaxies. In these terms, 8 out of 14 galaxies in our sample would not be considered well isolated: CIG~71, 159, 171, 201, 416, 922, 936 and 1003.

While some members of our sample, selected from the CIG catalogue \citep{Karachentseva-1973}, may not fully meet the strictest current isolation criteria, it is important to stress that all galaxies in our sample are still classified as isolated galaxies. 
The different arguments stated above make us aware of the different degrees of isolation of the galaxies in our sample. 
However, previous studies indicate that late-type galaxies are usually located in very low-density environments. 
In this context, we decided not to exclude any galaxy a priori but rather discuss each of them knowing their isolation properties.

\begin{table}
\caption{Number of neighbours of galaxies in our sample and the distance to the nearest one measured by \citet{verley-2007-i}.}
\begin{center}
\begin{tabular}{cccc}
\hline
CIG & Neighbours & \multicolumn{2}{c}{Distance}  \\
\cmidrule(lr){3-4} 
Name &  & (arcsec) & (kpc) \\ 
(1) & (2) & (3) & (4) \\ 
\hline
71 & 4 & 941 & 384.6 \\ 
95 & 71 & 362 & 110.7 \\ 
159 & 34 & 456 & 130.2 \\ 
171 & 91 & 152 & 38.2 \\ 
183 & 25 & 784 & 276.3 \\ 
201 & 57 & 109 & 30.6 \\ 
329 & 101 & 316 & 89.6 \\ 
416 & 168 & 362 & 58.3 \\ 
593 & 26 & 739 & 251.2 \\ 
847 & 79 & 644 & 126.4 \\ 
906 & 28 & 391 & 132.7 \\ 
922 & 97 & 197 & 55.6 \\ 
936 & 88 & 234 & 48.8 \\ 
1003 & 77 & 118 & 28.4 \\ 
\hline
\end{tabular}
\end{center}
Columns: (1)\,CIG galaxy name; (2)\,number of neighbours in a physical radius of 0.5\,Mpc; (3) and (4)~distance to the nearest neighbour in arcseconds and kpc respectively  given the heliocentric galactic distance listed in Table~\ref{table:general}.  
\label{tab:neighbours}
\end{table}

\section{The data}\label{Sec:Obs}

\subsection{H$\alpha$ observations and data reduction}\label{Sec:ha-data}

Observations of the H$\alpha$ emission line were performed using the scanning Fabry-Perot (FP) interferometer, GHASP, attached at the Cassegrain focus of the 1.93m telescope at the \textit{Observatoire de Haute-Provence} (OHP, see Table \ref{table:instrument}). A GaAs Image Photon Counting System (IPCS) is used to reach low detection levels, offering a total FoV of about $5.8\times 5.8$\,arcmin$^{2}$ with a pixel scale of $\sim$ 0.68~arcsec\,pix$^{-1}$  \citep{gach-2002}. Since the IPCS has no readout noise, exposures of 10~seconds were chosen for each of the 32 scanning steps required to cover the GHASP's free spectral range (FSR). In order to detect the eDIG emission, the scanning sequence was typically repeated  $\sim$35~times, resulting in a total exposure time per galaxy of $\sim$180\,min. We used narrow-band interference filters (mainly FWHM$\simeq$15\,\AA\, and in a few cases slightly larger up to 24\,\AA) enabling to select the redshifted H$\alpha$ line of ionized hydrogen (6562.78\,\AA). 
The scanning Fabry-Perot interferometer we used, with an interference order of $p\simeq$798 and a finesse $\mathcal{F}\simeq$13 at H$\alpha$, allowed GHASP to reach a typical spectral resolution of $R\sim 10\,000$.

Two calibration cubes were obtained, one at the beginning and another at the end of the observation, using a neon lamp with selected narrow line at $6598.95$\,\AA, close to the redshifted nebular wavelength in order to minimize phase shift effect \citep[e.g.][]{gomezl-2019}. The observation date, filter's central wavelength and FWHM, the total exposure time and the astronomical seeing of each galaxy are listed in Table \ref{table:obs}.

For the data reduction, we used the packages based on homemade \textsc{IDL}\footnote{Interactive Data Language, ITT Visual  Information Solutions: \url{https://www.l3harrisgeospatial.com/Software-Technology/IDL}} routines, \textsc{reducWizard}  
and \textsc{ComputeEverything} \citep[see e.g.][]{epinat-2008-ii}. 
The data reduction technique has been widely reported in
\cite{daigle-2006-fp}. In summary, it consists in the following steps: 
(1)\,integration with guiding correction of the total data cubes obtained during the observation; 
(2)\,application of a Hanning spectral smoothing on the data cube; 
(3)\,calibration in wavelength of the integrated data cubes through the computation of a parabolic phase map which is computed from the calibration cube in order to obtain the reference wavelength for the line profile observed inside each pixel, creating a wavelength-sorted data cube by applying the phase map correction to the interferogram data cube; 
(4)\,subtraction of the OH sky-lines emission; 
(5)\,a Gaussian spatial smoothing with a FWHM selected on the wavelength data cubes; and 
(6)\,computation of H$\alpha$ monochromatic, continuum, radial velocity and velocity dispersion maps for each wavelength calibrated cube. Finally, astrometric information was attached to the processed files by using the \textsc{IDL} task \textsc{koords} from the \textsc{KARMA} package \citep{gooch-1996}. In Table~\ref{table:obs}, we also listed the seeing and the  final spatial resolution in kpc~arcsec$^{-1}$ after the spatial Gaussian smoothing of the H$\alpha$ maps.

We performed two separate processes on the wavelength data cube. 
First, we applied Gaussian spatial smoothing, and then, independently, we performed adaptive spatial binning using a Voronoi tessellation with a target SNR$\simeq$5. 
To ensure a signal-to-noise ratio SNR$\geq5$ in the Gaussian smoothed data cube, we compared the monochromatic H$\alpha$ maps derived from both processes and set a threshold for the emission detected in the Gaussian smoothed maps. 
For a SNR$\simeq$5, the GHASP surface brightness detection limit is $F=2.5\times10^{-17}\,\mathrm{erg\,sec^{-1}\,cm^{-2}\,arcsec^{-2}}$ \citep[see][]{epinat-2008-ii, gomezl-2019, sardaneta-2022}. 
As no H$\alpha$ calibrated image of any galaxy in our sample has been previously published, we followed the calibration procedure outlined by \cite{epinat-2008-ii} to determine the total H$\alpha$ flux for the GHASP data (see Appendix~\ref{sec:flux}).

The data analysis was made with the program
\textsc{ADHOCw}\footnote{`Analyse et Depouillement Homogene des
Observations Cigale for Windows'
\url{http://cesam.lam.fr/fabryperot/index/softwares} developed by J.
Boulesteix at Marseille Observatory in 2005.},
\textsc{IRAF}\footnote{`Image Reduction and Analysis Facility'
\url{http://iraf.noao.edu/}} tasks, 
the SAO Image DS9 software\footnote{SAO Image DS9. An image display and visualization tool for astronomical data \citep{ds9}  \url{https://sites.google.com/cfa.harvard.edu/saoimageds9}} 
and our own \textsc{Python} scripts.

\begin{table}
\caption{Instrumental and observational parameters}
\footnotesize{
\begin{center}
\begin{tabular}{lc}
 \hline
\multicolumn{1}{c}{Parameter} &   {Value} \\ 
 \hline
Telescope & 1.93 m OHP \\ 
Aperture ratio of the focal reducer & $f/3.9$  \\ 
Instrument &   GHASP  \\ 
Detector type & IPCS GaAs system   \\  
Detector size (pix$^{2}$) & 512$\times$512 \\ 
Image scale (arcsec pix$^{-1}$) & 0.68  \\ 
Field of view (arcmin$^{2}$) & 5.9$\times$5.9    \\ 
Interference order at H$\alpha$ & 798 \\ 
FSR at H$\alpha$ (\AA \, / $\mathrm{km\,s^{-1}}$) & 8.23 / 376 \\ 
Finesse observed & $\sim$13 \\
Resolution & $\sim$ 10000 \\
Spectral sampling at H$\alpha$ (\AA \, / $\mathrm{km\,s^{-1}}$)  & $\sim$0.26 / $\sim$11.5  \\ 
 \hline
\end{tabular}
\end{center}
}

\label{table:instrument}
\end{table}

\begin{table*}
\caption{Journal of observations}
\footnotesize{
\begin{center}
\begin{tabular}{ccccccc}
 \hline
CIG  & Date & $\lambda_{c}$  & FWHM & $t_{\mathrm{exp}}$ & Seeing & Resolution\\ 
Name  & (aaaa/mm/dd) & (\AA) & (\AA) & (min) & (arcsec) & (kpc arcsec$^{-1}$) \\ 
(1) & (2) & (3) & (4) & (5) & (6) & (7) \\ 
\hline
71 & 2019/10/26 & 6700 & 24 & 176 & 4.4  & 2.0 \\ 
95 & 2019/10/27 & 6665 & 15 & 192 & 2.4  & 1.0 \\ 
159 & 2021/01/13 & 6655 & 15 & 181 & 2.6 & 1.0 \\ 
171 & 2021/01/14 & 6645 & 15 & 187 & 2.8 & 0.9 \\ 
183 & 2021/01/21 & 6675 & 15 & 213 & 3.0 & 1.3 \\ 
201 & 2021/01/16 & 6655 & 15 & 187 & 3.3 & 1.1 \\ 
329 & 2021/01/13 & 6655 & 15 & 181 & 3.8 & 1.2 \\ 
416 & 2021/01/21 & 6615 & 15 & 187 & 4.1 & 0.7 \\ 
593 & 2021/03/09 & 6675 & 15 & 219 & 2.9 & 1.2 \\ 
847 & 2019/10/29 & 6630 & 20 & 187 & 2.3 & 0.6 \\ 
906 & 2019/10/26 & 6675 & 15 & 187 & 1.9 & 0.9  \\ 
922 & 2019/10/25 & 6655 & 15 & 192 & 2.2 & 0.8 \\ 
936 & 2019/10/27 & 6630 & 20 & 187 & 3.0 & 0.8  \\ 
1003 & 2019/10/26 & 6645 & 15 & 187 & 2.5 & 0.8  \\ 
 \hline
\end{tabular}
\end{center}
Columns: (1)~CIG galaxy name; (2)~date of observation; (3)~$\lambda_{c}$: non-tilted filter central wavelength; (4)~FWHM: non-tilted full width half-maximum; (5)~$t_{\mathrm{exp}}$:~total exposure time; (6)~Seeing in arcs; (7)~Final spatial resolution in kpc per arcseconds after having applied a spatial Gaussian smoothing on the H$\alpha$ wavelength data cubes (see Section~\ref{Sec:ha-data}). 
}
\label{table:obs}
\end{table*}

\subsection{Archival data}

The DSS is available at the site of the \textit{Space Telescope Science Institute} (STScI)\footnote{The Digitized Sky Surveys were produced at the Space Telescope Science Institute under U.S. Government grant NAG\,W-2166. The images of these surveys are based on photographic data obtained using the Oschin Schmidt Telescope on Palomar Mountain and the UK Schmidt Telescope. The plates were processed into the present compressed digital form with the permission of these institutions. \url{https://archive.stsci.edu/cgi-bin/dss_form}}. The DSS $6.5\times6.5\,\mathrm{deg^{2}}$ plates have been scanned using a modified PDS microdensitometer with a binned pixel scale of about $\sim$1.0\,arcsec\,pix$^{-1}$. In this work, we use the DSS R-band images (5900-7150\,\AA) of our galaxy sample as a reference to provide a large-band optical view of the target. However, we do not calibrate these images in flux since their zero-point magnitude is currently unknown. The average seeing obtained from the image headers was $\sim$1.5~arcsec.

The \textit{Two Micron All-Sky Survey} \citep[2MASS,][]{jarrett-2000} project uses two highly-automated 1.3\,m telescopes, one at Mt. Hopkins, Arizona, and the other at Cerro Tololo, Chile. Each telescope is equipped with a three-channel camera, each channel consisting of a $256\times 256$\,pix$^{2}$ array of HgCdTe detectors, capable of observing the sky simultaneously at \textit{J}\,(1.24\,$\mu$m), \textit{H}\,(1.66\,$\mu$m) and \textit{K$_{s}$}\,(2.16\,$\mu$m) near-infrared bands with an angular resolution of 2.0\,arcsec and a pixel size of 1.0\,arcsec\,pix$^{-1}$. Owing to the transparency of interstellar dust within the galaxies and to the dominance of late-type stellar populations in producing galactic near-infrared flux, the 2MASS galaxy images  trace the overall stellar mass distribution in these galaxies\footnote{2MASS Data Release Documentation \url{https://irsa.ipac.caltech.edu/data/2MASS/docs/releases/docs.html}}. 
To calibrate in flux, the zero-point conversion values are published in the \textit{Explanatory Supplement to the 2MASS All Sky Data Release and Extended Mission Products}\footnote{Section VI.4.a of the \textit{Explanatory Supplement to the 2MASS All Sky Data Release and Extended Mission Products}: \url{https://www.ipac.caltech.edu/2mass/releases/allsky/doc/sec6_4a.html}}. In this work we use the \textit{K$_{s}$}-band images of the nearby isolated late-type edge-on galaxies selected to sketch the stellar Population\,II.

\textit{Galaxy Evolution Explorer} (GALEX) was the first ultraviolet (UV) all-sky survey covering a field of view (FoV) of $\sim 1.25$\,deg using microchannel plate detectors to obtain direct images in the  \textit{NUV} ($\lambda_{\mathrm{eff}}=2271$\,\AA) and \textit{FUV} ($\lambda_{\mathrm{eff}}=1528$\,\AA) with a resolution of 4.2/5.3\,arcsec (\textit{FUV/NUV}) and a pixel scale of 1.5\,arcsec\,pixel$^{-1}$ \citep{bianchi-2011}\footnote{\textit{Galaxy
Evolution Explorer} (GALEX) \url{http://www.galex.caltech.edu/index.html}}. Conversion values between GALEX count rates, fluxes, and AB magnitudes are available in the \textit{GALEX Guest Investigator Web Site}\footnote{Section Instrument and Calibration of the \textit{GALEX Guest Investigator Web Site}: \url{https://asd.gsfc.nasa.gov/archive/galex/FAQ/counts_background.html}}. We use  GALEX-\textit{NUV} and \textit{FUV} images as tracers of stellar Population\,I in galaxies in our sample.

Thus, in this work we study H$\alpha$, \textit{NIR}, and \textit{UV} images, each one with different resolutions. In the case of the H$\alpha$ emission data, the seeing conditions varied across the sample of galaxies, with different galaxies experiencing different levels of atmospheric turbulence due to the diversity of observational conditions during data gathering (see Table~\ref{table:obs}). For example, the closest galaxy, CIG~71, experienced the worst seeing conditions with a resolution of $\sim$4.4~arcsec~(1.8~kpc), as did the most distant galaxy, CIG~416, with a resolution of $\sim$4.1~arcsec~(0.7~kpc). However, CIG~922, a galaxy located at an average distance, had relatively average seeing conditions with a resolution of $\sim$2.2~arcsec~(0.6~kpc). The seeing for the \textit{NIR}~(2~arcsec) and \textit{NUV/FUV}~(4.2/5.3\,arcsec) images is considered constant for the whole sample. However, it is important to be aware that the different seeing disc sizes combined with the different distances of the galaxies blur the data and impact the spatial physical resolution making it more difficult to distinguish fine details such as eDIG filaments or patches.

Finally, since the \textit{Infrared Astronomical Satellite} (IRAS) satellite did not cover the whole sample of CIG galaxies, \citet{lisenfeld-2007} reprocessed the IRAS\,\textit{MIR/FIR} survey data using the ADDSCAN/SCANPI utility for 1\,030 out of 1\,050 CIG galaxies as part of the AMIGA project. In this work, we use the IR flux densities at 60 and 100$\,\mu m$ from AMIGA survey (see Table~\ref{table:general}). 

\section{Identifying the stellar disc}\label{Sec:stellarDisc}

To disentangle the galactic disc from the gaseous extraplanar component, 
we performed a photometric analysis of the 2MASS~2.2$\mu$m image, tracer of the old stellar population, and of the GALEX \textit{NUV} and \textit{FUV} images, connected to the young stellar population in the galaxy. Because of the particular characteristics of the images in each band, to derive the surface brightness profile, we used two different software packages, 
the \textsc{ellipse} task from the \textsc{IRAF} \textsc{STSDAS} package  \citep{Jedrzejewski-1987} and 
the  \textsc{isophote} package from the \textsc{Python} \textsc{Photutils} package\footnote{\textsc{Photutils} is an open source \textsc{Python} package affiliated of {\tt Astropy} that primarily provides tools for detecting and performing photometry of astronomical  sources. \url{https://photutils.readthedocs.io/en/stable/index.html} \url{https://photutils.readthedocs.io/en/stable/api/photutils.isophote.Ellipse.html}} \citep{photoutils_larry_bradley_2020_4044744}. 
Both programs fit an ellipse via an iterative method \citep[devised by][]{Jedrzejewski-1987} as well as quantify the distortion from a perfect ellipse by means of higher-order Fourier harmonics: 
\begin{equation}
\mu(\theta)=\mu_{0}+\displaystyle\sum_{n=1}^{4}\,(A_{n}\sin \,n\theta+B_{n}\cos\,n\theta),\label{ec:ellipse-fourier}
\end{equation}
where $\mu_{0}$ is the surface brightness averaged over the ellipse as a function of the azimuthal angle $\theta$ and, $A_{n}$ and $B_{n}$ the higher order Fourier coefficients. The first- and second-order coefficients ($A_{1}$, $B_{1}$, $A_{2}$, $B_{2}$) indicate the errors in the fitting procedure, being zero for a best-fit ellipse, while the third-order coefficients ($A_{3}$, $B_{3}$) characterize the deviation 
of the fitted ellipse from the isophote shape. 
The coefficient $B_{4}$ measures symmetric distortions from pure ellipticity: when $B_{4}>0$, the isophotes have a ‘disc-like' (circular) shape and, when $B_{4}<0$, the isophotes have a ‘boxy' (rectangular) shape. 
The inclination angle ($i$) with respect to the sky-plane of the galactic disc can be measured through the ellipticity ($\varepsilon$) of the fitted ellipse. Using diverse tracers of the stellar population serves as a crucial means to differentiate between the particular cases presenting \textit{UV} extended discs or \textit{UV} haloes which represent at least one-third of the galaxies in the local universe  \citep{thilker-2007-i} and, as we will show for our sample in Section~\ref{Sec:results}, approximately two-thirds of the late-type isolated high-inclined galaxies.

The analysis of edge-on galaxies is more difficult than the analysis of galaxies with lower inclination because the high inclination affects the apparent shape of the galaxy along the line of sight (LoS). 
In photometric analysis of low-inclination galaxies, a smooth bulge-disc transition is typically observed in the resulting plots \citep[e.g.][]{marino-2010}. However, in highly inclined galaxies, the brightness of the inner regions is dominated by the bulge with a small contribution from the thin disc. As the radius increases, the contribution from the disc becomes more important, leading to an abrupt bulge-disc transition in the resulting plots. This provides a noisier behaviour at small radii in the photometric
analysis for highly inclined galaxies.

Considering that in late-type galaxies old and young stellar populations are mixed in the disc \citep[e.g.][]{Nersesian-2019}, we define the stellar homogeneous disc by using the ellipse delimiting  the locus of the stars belonging to the old stellar population since they trace the bulk of the mass and probe the global potential of the disc galaxy.

\begin{table*}
\caption{Photometric parameters resulting from the ellipse fitting to the 2MASS \textit{K$_{s}$}-band and GALEX \textit{NUV} and \textit{FUV} images.}
\begin{center}
\begin{tabular}{ccccccccccccc}
\hline
& \multicolumn{3}{c}{\textit{K$_{s}$}-band, 1$\sigma$} &  \multicolumn{3}{c}{\textit{K$_{s}$}-band, 3$\sigma$} & \multicolumn{3}{c}{\textit{NUV}} & \multicolumn{3}{c}{\textit{FUV}}   \\ 
\cmidrule(lr){2-4}  \cmidrule(lr){5-7} \cmidrule(lr){8-10} \cmidrule(lr){11-13}  
CIG & $r$ & PA & $\varepsilon$ & $r$ & PA & $\varepsilon$ & $r$ & PA & $\varepsilon$ & $r$ & PA & $\varepsilon$ \\
Name & (arcsec) & (deg) &  & (arcsec) & (deg) &  & (arcsec) & (deg) &  & (arcsec) & (deg) &  \\ 
(1) & (2) & (3) & (4) & (5) & (6) & (7) & (8) & (9) & (10) & (11) & (12) & (13)   \\ 
\hline
71 & 33.1 & 179.1 & 0.76 &  18.7 & 178.8 & 0.77 & 61.0 & 177.9 & 0.80 & 34.6 & 177.9 & 0.84 \\ 
95 & 25.6 & 128.8 & 0.83 &  12.0 & 126.2 & 0.71 & 73.3 & 127.3 & 0.85 & 60.2 & 126.4 & 0.90 \\ 
159 & 74.4 & 68.9 & 0.87 &  42.0 & 66.2 & 0.860 & 114.0 & 67.0 & 0.90 & 53.1 & 71.0 & 0.90 \\ 
171 & 59.1 & 159.9 & 0.77 & 44.4 & 159.7 & 0.858 & 110.9 & 156.9 & 0.83 & 64.9 & 160.0 & 0.90 \\ 
183 & 28.9 & 19.8 & 0.82 &  13.5 & 16.6 & 0.74 & 66.2 & 19.8 & 0.72 & 48.8 & 18.0 & 0.73 \\ 
201 & 36.4 & 150.9 & 0.78 & 20.5 & 151.7 & 0.79 & 85.8 & 151.2 & 0.69 & 81.6 & 151.6 & 0.86 \\ 
329 & 63.6 & 115.1 & 0.76 &  35.9 & 115.4 & 0.77 & 86.8 & 110.0 & 0.79 & 71.3 & 113.0 & 0.80 \\ 
416 & 23.2 & 97.9 & 0.75 &  11.9 & 95.0 & 0.63 & 59.7 & 97.8 & 0.43 & 48.3 & 97.1 & 0.78 \\ 
593 & 27.8 & 167.0 & 0.73 &  14.1 & 165.5 & 0.67 & 73.5 & 165.0 & 0.81 & 44.8 & 168.0 & 0.75 \\ 
847 & 33.1 & 140.8 & 0.71 &  20.5 & 140.2 & 0.77 & 87.5 & 142.0 & 0.73 & 76.7 & 142.0 & 0.87 \\ 
906 & 45.0 & 28.3 & 0.72 &  31.3 & 32.8 & 0.80 & 60.6 & 35.0 & 0.63 & 52.3 & 31.8 & 0.88 \\ 
922 & 34.7 & 61.1 & 0.83 &  19.6 & 60.0 & 0.84 & 72.1 & 61.5 & 0.82 & 55.4 & 61.5 & 0.88 \\ 
936 & 19.8 & 52.3 & 0.58 &  12.3 & 61.6 & 0.64 & 111.7 & 61.0 & 0.86 & 71.4 & 61.1 & 0.90 \\ 
1003 & 45.5 & 127.3 & 0.80 & 25.7 & 124.7 & 0.80 & 52.8 & 129.4 & 0.54 & 53.3 & 119.4 & 0.81 \\ 
\hline
\end{tabular}
\end{center}
CIG galaxy name listed in column (1). Major axis length ($r$), position angle (PA) measured from the North to the East and ellipticity ($\varepsilon$) of the ellipses fitted to the \textit{K$_{s}$}-band image at 1$\sigma$ level on the background are listed in columns (2), (3) and (4), respectively, and at 3$\sigma$ level in columns (5), (6) and (7). Same parameters determined for the \textit{NUV} image are listed in columns (8), (9) and (10), and for the \textit{FUV} image in columns (11), (12) and (13). 
\label{table:ellipseParams}
\end{table*}

\subsection{\textit{NIR} emission}\label{sec:nir-emission}

For the 2MASS 2.2\,$\mu$m image, we applied the method of elliptical isophote fitting using the  task \textsc{ellipse} \citep{Jedrzejewski-1987} from the \textsc{IRAF} \textsc{STSDAS} package. 
As initial parameters we set the galaxy coordinates as the centre, the \textit{B}-band optical diameter ($D_{25}$) as the maximum semi-major axis length, and the optical position angle (PA) and ellipticity ($\varepsilon$) published in the NED (see Table \ref{table:general}). 
We used logarithmic radial sampling with an initial step of $\sim$0.5\,pixel along the semimajor axis in order to derive the surface brightness profile of each galaxy in an iterative process. 
The centre, PA, and ellipticity of the ellipse were allowed to vary in each iteration.

\cite{jarrett-2000PASP} defined the isophotal aperture $r_{20}$, derived from the \textit{K$_{s}$}-band isophote at $\mu_{K}=20\, \mathrm{mag\,arcsec^{-2}}$, as the 2MASS standard aperture, corresponding to roughly 1$\sigma$ of the typical background noise in the $K$-band   images. However, because of the high levels of background noise in the 1-2~$\mu$m atmospheric windows, the 2MASS is not as sensitive to low surface brightness emission from galaxies \citep{jarrett-2000PASP}. In consequence, the \textit{K$_{s}$}-band benchmark of $\mu_{K}=20 \, \mathrm{mag\,arcsec^{-2}}$ elliptical isophote aperture might underestimate the total flux between 10\% and 20\%, depending on the radial distribution according to the Hubble type \citep{jarrett-2000PASP, jarrett-2003}. On the other hand, discrepancies in the measurements at the $\mu_{K}=20 \, \mathrm{mag\,arcsec^{-2}}$ level  between the 2MASS and other \textit{K$_{s}$}-band surveys \citep[see e.g.][]{fingerhut-2010} have suggested potential errors in the 2MASS data reduction process. In fact,  \cite{jarrett-2000PASP} explained that systematic components such as \textit{H}-band airglow variations were not well understood by then and might have induced large errors in the photometry\footnote{Explanatory Supplement to the 2MASS All Sky Data Release and Extended Mission Products~IV   \url{https://www.ipac.caltech.edu/2mass/releases/allsky/doc/explsup.html}. 2MASS Data Processing.~5.~Extended Source Identification and Photometry \url{https://www.ipac.caltech.edu/2mass/releases/allsky/doc/sec4_5e.html}}. 
Consequently, the choice of a brighter isophote to define the galactic disc boundaries, beyond the 1$\sigma$ level at $\mu_{K}=20 \, \mathrm{mag\,arcsec^{-2}}$, may help to mitigate potential data reduction errors in the 2MASS data set.

To be consistent with previous optical photometric results,  \cite{jarrett-2003} derived some galactic properties from the 3$\sigma$ isophote,  such as the ellipticity ($\varepsilon$) and position angle (PA), in the 2MASS\,\textit{Large Galaxy Atlas} (LGA) and the 2MASS~\textit{Extended Source Catalogue}. These photometric parameters are consistent with our \textit{K$_{s}$}-band results as well (see Table~\ref{table:ellipseParams}). 
Furthermore, the \textit{K$_{s}$} elliptical isophotal photometry should include the core, bulge, and disk components, capturing most of the flux from a galaxy \citep[e.g.][]{jarrett-2000}. Therefore, to ensure that a significant portion of the galaxy's light is captured and to avoid empty sky background noise, hereafter we use the ellipse at 3$\sigma$ on the background as reference for the disc traced by the stellar Population\,II. Although this value may not represent the true galactic radii, it is a practical starting point to define the old population stellar disc boundaries.

The older stellar population in spiral galaxies points out internal structures such as spiral arms, bulges, warps, rings, and bars \citep{jarrett-2003}. 
Bars play a major role in the secular evolution of galaxies and, through vertical resonances, drive stars above the plane to form peanut/box-shaped pseudo-bulges \citep{combes-1981}. The non-axisymmetry of the bars produces the gas to flow inwards fuelling SF in the central regions of their host galaxies \citep[see e.g.][and references therein]{kim-2021}. 
Bottom frame of panel\,(d) of Figure~\ref{Fig:maps-example}, and from Figures~\ref{Fig:maps-c95} to \ref{Fig:maps-c1003}, shows the variation of the $B_{4}$ parameter with respect to the galactic semi-major axis (radius). From these graphics, we found that the only one galaxy in our sample that presents a coefficient $B_{4}<0$ is CIG\,329 (see Figure~\ref{Fig:maps-c329}) implying it has a `boxy' bulge. %

\subsection{\textit{UV} emission}

The  \textsc{ellipse} task from \textsc{IRAF} works in an iterative manner, if there is  contamination such as stars and \ion{H}{ii} regions or regions of too low SNR, the program continues generating ellipses but it outputs parameters with undefined values. 
To overcome this difficulty, other fitting software was used to derive the surface brightness profiles of the \textit{NUV} and \textit{FUV} images. 
The  \textsc{isophote} package from \textsc{Photutils} 
\citep{photoutils_larry_bradley_2020_4044744}, 
provides tools to fit elliptical isophotes to a galaxy image  using an iterative method described by \cite{Jedrzejewski-1987} (see equation\,\ref{ec:ellipse-fourier}) for each ellipse as well. 
The \textsc{isophote} package can also work in an iterative manner by measuring a maximum acceptable relative error in the local radial intensity gradient (typically 0.5). However, when two consecutive isophotes exceed the value specified by the parameter the program prevents ellipses either from prematurely stopping due to the stellar contamination or from growing in low SNR regions. 
Nonetheless,  if the maximum semimajor axis is specified, the intensity gradient is set to `none' and the algorithm proceeds inwards to the galaxy centre. Although it is not the general case and may not be the best fit, this procedure allows to analyse data with too low SNR in the inner regions of the galaxy, such as in some GALEX \textit{NUV} and \textit{FUV} images of galaxies in our sample. This fit algorithm is quite sensitive to the initial guesses. 
The iteration begins by giving a fixed semimajor axis length selected initially to grow linearly with steps of 0.5\,pix and specifying the initial value of the galactic centre, PA and $\varepsilon$, then the algorithm proceeds to fit isophotes inwards to the galaxy centre.

We initialized the fitting parameters using the galaxy coordinates as the centre, the optical PA and ellipticity obtained from the NED database, and the \textit{B}-band optical diameter ($D_{25}$) as the maximum length of the semi-major axis  obtained from the NED database (see Table \ref{table:general}). To overcome any discrepancies between the optical and \textit{UV} radii, we made adjustments to the initial parameters by comparing the \textit{UV} radius measured using the DS9 image analysis software with the optical radius, and modifying the corresponding initial parameter values if the optical radius was smaller than the \textit{UV} radius. For most of our targets, we allowed the centre position, ellipticity and PA to vary during the fitting process. Some galaxies (CIG~71, 95, 847, 906 and 1003) present the \textit{NUV/FUV} brightest knots beyond $\sim10$~arcsec from the peak light distribution of the \textit{NIR} emission, this last position is assumed the center of the galactic disc and it is probable that the \textit{NIR} emission is shadowing the actual peak of the \textit{UV} light.  Hence, for these galaxies, we fixed the centre position at the \textit{NIR} peak light distribution, assuming it indicates the galactic center, and we allowed the ellipticity and PA to vary during the fitting process.  Thus, we ensure that the fitted ellipses accurately reflect the morphology of the target galaxies, taking into account any deviations from the ideal elliptical shapes.

We computed the isophotal radii at surface brightness levels of $\mu_{\mathrm{\textit{NUV}, \textit{FUV}}}=28\,\mathrm{mag\,arcsec^{-2}}$ in  \textit{NUV} and \textit{FUV}, which roughly correspond to the average surface brightness at the optical diameter \citep[see e.g.][]{gil-de-paz-2007, marino-2010, cortese-2012}, using the limit AB magnitude in \textit{NUV} of 20.08\,mag and in \textit{FUV} of 18.82\,mag\footnote{GALEX Guest Investigator Web Site  \url{https://asd.gsfc.nasa.gov/archive/galex/FAQ/counts_background.html}}.

\subsection{Photometric and geometric parameters at different wavelengths}

Photometric parameters, such as the PA and galaxy inclination, are useful observational constraints 
to reveal galaxy peculiarities when the galaxy configuration is examined, particularly when  different wavelengths are considered. 
In Table \ref{table:ellipseParams} we list the photometric parameters resulting from the  ellipse fitting process. 
Panel~(g) of Figure~\ref{Fig:maps-example} and Figures~\ref{Fig:maps-c95} to \ref{Fig:maps-c1003} shows the surface brightness profiles of GALEX \textit{NUV} and \textit{FUV} images, while the resulting ellipses were overlaid on the \textit{NUV} (panel e) and \textit{FUV} (panel~f) images, respectively  (see for example Figure~\ref{Fig:maps-example}).   The 1$\sigma$ \textit{NUV} and \textit{FUV} isophotes computed after applying a Gaussian spatial smoothing of $\sigma\approx$4.5\,arcsec\,(3\,pix) on the \textit{NUV} and \textit{FUV} images were overlaid on the H$\alpha$ monochromatic map (panel~a) in order to allow a better comparison of the morphology traced at different wavelengths.

Inclination plays a critical role in the analysis of galaxy structure as it is used to correct its main parameters as the brightness \citep[e.g.][]{stone-2021} and the rotation velocity \citep[e.g.][]{epinat-2010}. 
Assuming that the image of a spiral galaxy is the projection of a disc with the shape of an oblate spheroid, the inclinations of spiral galaxies can be roughly derived from the ellipticity of apertures used for photometry.  
If the thickness of a disc is neglected the relation between the inclination ($i$) to the celestial plane and the apparent diameter ratio can be described using the relation~\ref{eq:i1}. 
If the thickness is taken in consideration,  the  inclination can be estimated using the expression: 
\begin{equation}
\cos^{2} i = \frac{(1-\varepsilon)^{2}-q_{0}^{2}}{1^{2}-q_{0}^{2}},
\label{ec:3 barbosa2015}
\end{equation}
where $\varepsilon=1-b/a$ is the ellipticity at the isophotal radius and $q_{0}=c/a$ is the flattening, in this case $a$, $b$ and $c$ are the three spheroid's axes \citep[see e.g.][]{barbosa-2015}. 
Many works have shown that 
the flattening is type dependent \citep[see e.g.][]{haynes-1984}.  In this sense, 
\cite{hall-2012} determined an intrinsic flattening for spiral galaxies, $q_{0}=0.13$, by studying the relationship between the axial ratios in the $i$-band versus the $g-i$ galaxy colour computed at the effective radius for 871 edge-on galaxies from the SDSS. They showed that, although $q_{0}$ increases with redder, progressively bulge dominated galaxies, the $q_{0}$ distribution was rather flat for late-type galaxies only with an upturn in the transition between Sa to S0 galaxies.

Nonetheless, deriving the inclination from ellipticity may be problematic. For instance, prominent bulges can dominate the axial ratio measurement or  in other cases, galaxies may not be axially symmetric owing to tidal effects or high surface brightness bars embedded in low surface brightness
discs \citep{epinat-2008-ii, epinat-2010, hall-2012,  kourkchi-2019, stone-2021}. All these rather common configurations can lead to large errors. Some solutions have been proposed, for example, in order to avoid contamination due to clumps of star formation, \cite{epinat-2010} suggested that the inclination should be preferentially derived from broad-band images with high resolution, as in the \textit{NIR} rest frame of the galaxy tracer of the old stellar disc. %

In addition to photometric techniques, to compute the galaxy inclination, some studies use visual inspection to evaluate the inclinations. For example, 
to select the high inclined galaxies from SDSS and CALIFA surveys, \cite{bizayaev-2017} and \cite{levy-2019}, respectively, used the visibility of the dust lane in optical images of the galaxies to confirm the edge-on nature of the galaxies, excluding galaxies with
dust lanes that were not centred in the midplane or displaying any visible track of spiral arms, bars, or other features. \cite{kourkchi-2019} evaluated this technique by designing an online graphical tool, achieving uncertainties of $\pm4^{\circ}$ and suggesting to replace the human eye with a machine learning algorithm to compute more accurate inclinations.

In this work, the galactic inclination $i$ was one of the selection paramaters (see Section\,\ref{Sec:Sample}) for our sample. 
Firstly, the inclination of the CIG galaxies in our sample was computed 
by the relation~\ref{eq:i1} %
using the galactic apparent major and minor axes from the optical wavelength published in the NED. Next, we computed the inclinations of galaxies in our sample applying the relation~\ref{ec:3 barbosa2015}, for the the 2.22\,$\mu$m and \textit{NUV} emission images we employed the ellipticity obtained from the ellipse fitting performed to them, and for the optical wavelength image, we used the galactic apparent major and minor axes from the NED. In all cases, we used the constant flattening $q_{0}=0.13$ determined by \cite{hall-2012}. In the case of the \textit{NIR} image we have computed the inclination of both 1$\sigma$ and 3$\sigma$ isophotal levels.

\begin{table}
\caption{Inclinations computed with different methods and wavelengths.}
\begin{center}
\begin{tabular}{cccccc}
\hline
CIG &  \multicolumn{2}{c}{$i_{\mathrm{Optic}}$} & $i_{\textit{NIR},\,1\sigma}$ & $i_{\textit{NIR},\,3\sigma}$  & $i_{\textit{NUV}}$ \\ 
\cmidrule(lr){2-3} %
(1) & (2) & (3) & (4) & (5) & (6) \\ 
\hline
71 & 83.8 & … & 78.4 & 78.6 & 81.2 \\ 
95 & 86.5 & … & 83.3 & 75.0 & 85.4 \\ 
159 & 85.4 & … & 90.0 & 86.9 & 90.0 \\ 
171 & 84.0 & … & 78.7 & 86.6 & 83.7 \\ 
183 & 80.4 & 84.0 & 82.9 & 76.9 & 75.5 \\ 
201 & 80.9 & 84.8 & 79.9 & 80.6 & 73.6 \\ 
329 & 81.3 & 85.5 & 78.4 & 79.1 & 80.4 \\ 
416 & 81.1 & 85.2 & 77.4 & 69.3 & 56.0 \\ 
593 & 83.2 & … & 76.4 & 72.3 & 81.7 \\ 
847 & 81.2 & 85.4 & 74.6 & 79.2 & 76.2 \\ 
906 & 80.9 & 84.8 & 75.7 & 81.3 & 69.6 \\ 
922 & 84.2 & … & 83.5 & 84.6 & 82.6 \\ 
936 & 85.7 & … & 66.6 & 70.2 & 86.4 \\ 
1003 & 82.5 & 89.1 & 81.1 & 81.5 & 63.5 \\ 
\hline
\end{tabular}
\end{center}
\label{tab:inclination}
Column\,(1):\,CIG name.  
Column\,(2)~and~(3):\,inclination computed assuming a flat disc (see equation~\ref{eq:i1}) and taking into account the galactic  thickness (see  equation~\ref{ec:3 barbosa2015}) respectively, using the optical ellipse's semi-axes available in NED in both cases.  
Columns (4), (5), (6) and (7)\,inclination  computed taking into account the galactic  thickness (see the relation \ref{ec:3 barbosa2015}), using the ellipses fitted to the 1$\sigma$ and 3$\sigma$ level of the \textit{NIR} image and to the one fitted to the brightness level $\mu_{\mathrm{\textit{NUV}}}=28\,\mathrm{mag\,arcsec^{-2}}$ of the \textit{NUV} image. 
In column (3), by using the flattening constant $q_{0}=0.13$ determined by \cite{hall-2012}, the lower limit of ellipticity $1-\varepsilon=0.13$, for which equation~\ref{ec:3 barbosa2015} is valid, is exceeded by some of the galaxies in our sample in the optical band.  
\end{table}

In Table\,\ref{tab:inclination} we compare the inclinations obtained. 
Equation~\ref{ec:3 barbosa2015} has a lower threshold of effectiveness depending on the flattening value ($q_{0}$) considered. Since \cite{hall-2012} determined the flattening value $q_{0}$ in the $i$-band, in the optical band we obtained inconsistent results for  some galaxies in our sample having ellipticity $1-\varepsilon\geq q_{0}=0.13$.  
Therefore, in the following we will discuss the inclination calculated  considering the galaxy as infinitesimally thin only in the optical band (see relation~\ref{eq:i1}). 
Excluding particular cases, the inclination computed with the ellipse fitted to the \textit{NIR} 3$\sigma$ isophotal level has an averaged difference of $\sim$2$^{\circ}$ with respect to the optical inclination, while the inclination from the \textit{NIR} 1$\sigma$ isophotal level has an averaged difference of $\sim$4$^{\circ}$ and the mean difference with the inclination from the \textit{NUV} emission is $\sim$3$^{\circ}$.  
Hence, the inclinations computed with the parameters of the ellipse fitted to the \textit{NIR} image at the 3$\sigma$ isophotal level are the most congruent with the optical inclination, and following \cite{epinat-2010}, this may be the best estimation to measure the galaxy inclination.

When computing the inclination, particular cases showing $\Delta i\gtrsim$10$^{\circ}$ with respect to the optical inclination were detected. Inclinations computed with the \textit{NIR} 3$\sigma$ isophotal level presented four particular cases: CIG\,95, 416, 593 and 936, which have a bright central bulge dominating the \textit{NIR} emission. In the case of the \textit{NUV} emission, the particular cases are CIG\,201, 416, 906 and 1003, which have the most extended \textit{UV} haloes and the largest filaments above the disc. In contrast, CIG\,936 is the only one particular case obtained for the \textit{NIR} 1$\sigma$ isophotal level. 
The galaxy CIG\,936 (Figure\,\ref{Fig:maps-c936}) is an extreme case in our sample displaying its \textit{NIR} and H$\alpha$ emission concentrated in the inner central $\sim$30\,arcsec\,($\sim$6.3\,kpc) of a disc with optical diameter of $D_{25}(B)=185.4$\,arcsec\,($\sim$38.9\,kpc). This galaxy might be compared with the Sombrero galaxy (NGC\,4594), an S0 galaxy whose bulge governs the axial ratio \citep[e.g.][]{hall-2012}, though we cannot consider CIG\,936 an early-type system whilst its disc is dominated by the \textit{UV} emission, tracer of the young stellar population. The results previously described seem to indicate that the apparent discrepancies in the inclination values  determined using images in different wavelengths are due to actual asymmetries of the old stellar disc and the presence of an extended \textit{UV} disc and halo.

In conclusion, inclination measurement is essential for unraveling the complex structures of galactic components, in particular \textit{UV} extended discs and haloes. By applying different stellar population tracers, this study allowed us to better separate these different components.

\section{Results}\label{Sec:results}

\begin{figure*}
\centering
\includegraphics[width=0.95\hsize]{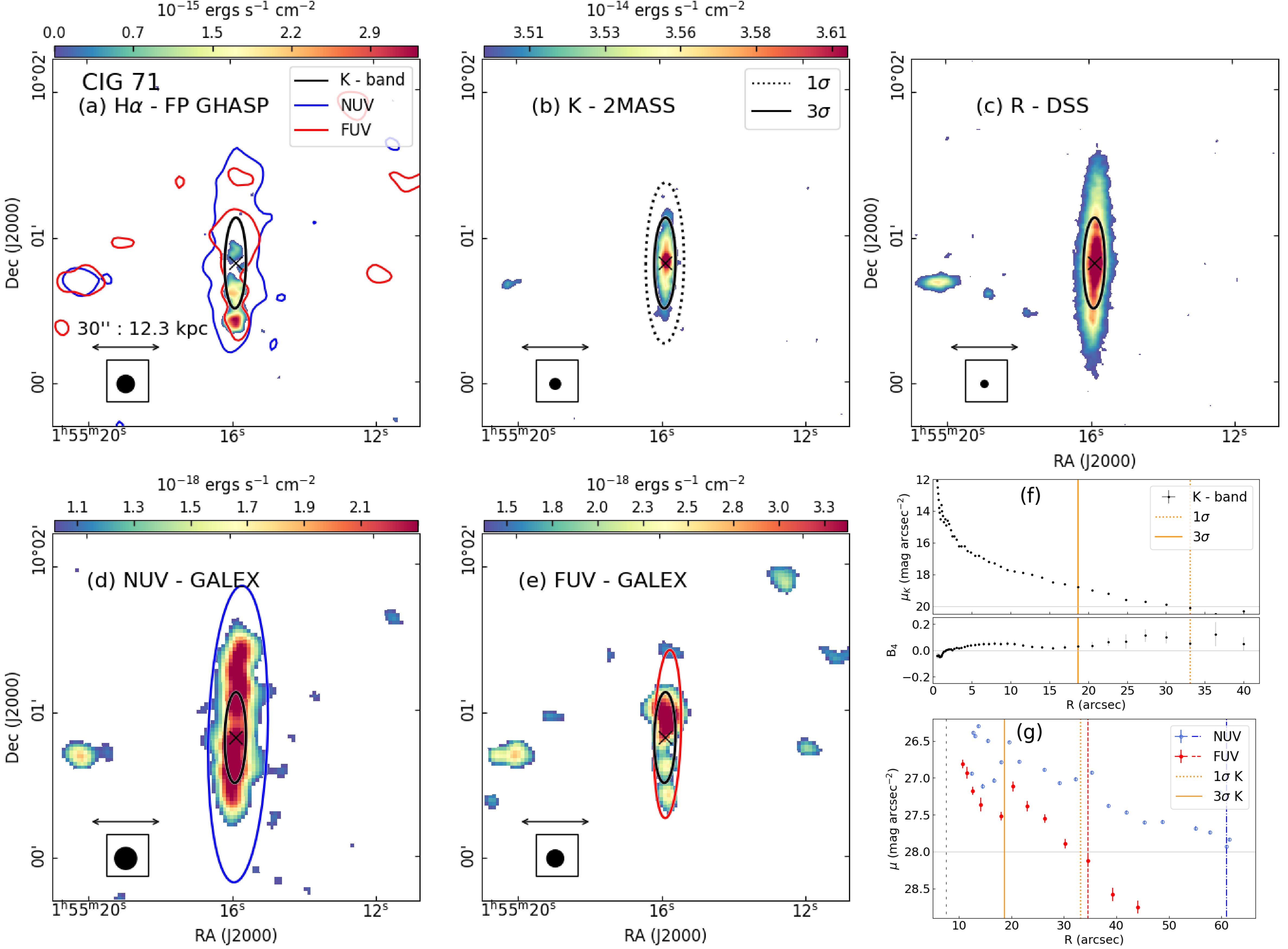}
\caption{
Example of the layout of the graphs and maps of each galaxy. CIG\,71 (UGC\,1301). 
Multiwavelength maps of the galaxy: (a)\,the H$\alpha$ 	monochromatic map from FP-GHASP data; (b)\,the 2MASS \textit{K$_{s}$}-band image; (c)\,the  DSS~\textit{R}-band image; (d) and (e)\,the GALEX \textit{NUV} and \textit{FUV} images, respectively. 
In\,panels\,(f)\,and\,(g)\,the surface brightness profile of the \textit{K$_{s}$}-band and \textit{NUV/FUV} emissions are plotted, respectively.  
In both panels, yellow vertical lines mark the surface brightness level at 1$\sigma$ (dotted~line) and 3$\sigma$ (solid~line) of the \textit{K$_{s}$}-band image. 
In panel\,(g), vertical lines indicate the radii corresponding to the isophotes at a surface brightness level at $\mu_{\textit{NUV},\textit{FUV}}=$\,28\,mag\,arcsec$^{-2}$ for the \textit{NUV} (blue~dot-dashed~line) and \textit{FUV} (red~dashed~line) maps, respectively. Additionally, the vertical black loosely-dashed line at $\sim$5\,arcsec shows the approximate FWHM of the GALEX point spread function \citep[see e.g.][]{marino-2010}. 
The ellipses fitted to these isophotes are superimposed on the corresponding map with their associated colour. 
The common elements in in panels (a), (b), (d) and (e) are: 
a colour-bar indicating the flux in units of $\mathrm{erg\,sec^{-1}\,cm^{-2}}$, 
the arrow tracing a scale of  30\,arcsec, 
a black circle embedded in a box indicating the spatial resolution of the image, 
a black ellipse representing the surface brightness level at 3$\sigma$ on the background of the \textit{K$_{s}$}-band image 
and, a black cross~($\times$) pointing the location of the \textit{NIR}~band peak light distribution. 
At the bottom of panel~(f) we show the relationship between the coefficient $B_{4}$ and the radius  using the \textit{K$_{s}$}-band map to determine if the isophotes tend to have `disc-like’~($B_4 < 0$) or `boxy’~($B_4>0$) shape (see Section~\ref{Sec:stellarDisc}). 
Finally, in panel\,(a), on the H$\alpha$ monochromatic map were overlaid the 1$\sigma$ isophotes on the background of the \textit{NUV} and \textit{FUV} images.
} 
\label{Fig:maps-example}
\end{figure*}

In this section we present  individual results obtained by 
comparing properties of the  DSS R-band, the 2MASS \textit{K$_{s}$}-band, the GALEX \textit{NUV} and \textit{FUV}  images versus the H$\alpha$ monochromatic map from  FP GHASP data. For each galaxy we  indicate the old stellar population disc using the ellipse fitting to the isophotal level at 3$\sigma$ on the background of the 2MASS \textit{K$_{s}$}-band image,  and we trace the young stellar population using both the  ellipse  corresponding to the $\mu_{\mathrm{\textit{NUV}, \textit{FUV}}}=28\,\mathrm{mag\,arcsec^{-2}}$ and the 1$\sigma$ isophote of the GALEX \textit{NUV} and \textit{FUV} images.  An example is shown in Figure~\ref{Fig:maps-example}.

\paragraph*{CIG\,71 (UGC\,1391):} 
The H$\alpha$ monochromatic map shows a discontinuous disc  mainly shifted Southwards from the photometric centre of the \textit{NIR}-band exceeding the radius of the old stellar disc. A prominent and luminous knot is observed at $\sim$30~arcsec Southwards from the NIR photometric centre. 
The \textit{K$_{s}$}-band image shows a symmetrical disc within the ellipse fitted to the 3$\sigma$~brightness level, but the disc becomes more extended and warped towards the South. 
The \textit{FUV} emission is concentrated near the galactic disc centre and then spreads  Southwards at lower brightness. 
This map also shows a \textit{FUV} emission cloud in the North. 
The \textit{NUV} emission displays a continuous disc extended along the major axis beyond the old stellar population disc. The photometric centre does not match with the H$\alpha$, \textit{NUV}, and \textit{FUV} emission maxima. 
The isolation criteria of \cite{verley-2007} placed CIG\,71 outside the isolation plane (see Figure~\ref{Fig:isolation}),  which was confirmed using \ion{H}{i} emission by \cite{Jones2018}.  
Moreover, we observed that CIG\,71 has common features observed in interacting galaxies \citep[e.g. NGC\,5258 from KPG\,389,][]{fuentesc-2019}. Likely, this galaxy experienced an interaction despite its nearest neighbor is 384\,kpc away (see Table \ref{tab:neighbours}).

\medskip
Remaining galaxies are discussed in Appendix~\ref{Sec:individualResults}.

\section{Discussion}\label{Sec:discussion}

In the following we discuss our findings at the light of the current literature focusing on three aspects: the stellar and warm gas emission in the disc, the detection of extraplanar warm gas emission and the extension of the \textit{UV} emission with respect to  \textit{NIR} and  H$\alpha$ emission.

\subsection{The emission in the disc}\label{Sec:eDIGdetection}

In this work, we have examined the location of both the old and young stellar population in order to determine the boundaries of the galactic disc in highly inclined galaxies to reveal the existence or lack of a gaseous extraplanar component.  Since in late-type galaxies the old and young stellar populations are mixed in the disc and arms, and to be consistent with previous photometric results \citep[see e.g.][]{jarrett-2003}, we normalised the stellar disc using the ellipse fitted to the 3$\sigma$ on the background level of the \textit{NIR} image, which delimits the region occupied by both old and young stellar populations.

Nonetheless, the definition of the galactic disc, which is used to find an extraplanar component, is not immediately obvious. 
In Figures~\ref{Fig:maps-example}, and~\ref{Fig:maps-c95} to~\ref{Fig:maps-c1003}, we plot the DSS~\textit{R}-band image of galaxies in our sample to compare images in different wavelengths with the optical one.  We observe that the radius of the \textit{R}-band image is  larger than the radius of the isophotal $20 \, \mathrm{mag\,arcsec^{-2}}$ radius in \textit{K$_{s}$}-band. In fact, \cite{jarrett-2000PASP} compared the \textit{K$_{s}$}-band to optical-band (\textit{B}-band) imaging properties (size, total flux, central surface brightness and elongation) in order to understand the stellar light coming from different components.  He found that the \textit{K}-band elliptical isophotal diameter corresponding to $\mu_{K}=20 \, \mathrm{mag\,arcsec^{-2}}$ surface brightness ($D_{20}$) represents about 70\% of the optical radius measured at $\mu_{B}=25 \, \mathrm{mag\,arcsec^{-2}}$, showing that the \textit{NIR} sensitivity  decreases for the later-type (i.e., `blue') galaxies, and concluding that as late-type galaxies evolve, their cores gradually shift towards reddening and brightening, resembling early-type disks or spheroids. Certainly, the \textit{K$_{s}$}-band predominantly traces the oldest stellar population \citep[stars older than $\sim$4~Gyr,][and references therein]{fingerhut-2010}, while optical bands like \textit{R} or \textit{B}-band are more sensitive to the younger, bluer stars. Therefore, it is expected that \textit{K$_{s}$}-band measurements correspond to a smaller region. Since our main goal is to study the extraplanar ionized gas component in edge-on galaxies,  we use 2MASS \textit{K$_{s}$}-band imagery looking specifically the location of the old stellar population, tracer of the galactic fundamental structure, with the aim of isolating the oldest stars from the effects of young, blue stars.

On the other hand, the \textit{Wide-field Infrared Survey Explorer} (WISE) have completed a mid-infrared survey of the entire sky in bands centered at wavelengths of 3.4, 4.6, 12, and 22$\mu$m. However, its angular resolution of 6.1~arcsec, 6.4~arcsec, 6.5~arcsec, and 12.0~arcsec at these bands respectively \citep[e.g.][]{wright-2010, jarrett-2012}, is notably inferior compared to the higher angular resolution of 2.0~arcsec achieved by the 2MASS for the \textit{K$_{s}$}-band images.  
Therefore, the selection of the narrow-band \textit{K$_{s}$}-band imagery for our analysis offers a clear advantage over the DSS~\textit{R}-band or WISE data, as it allows a more precise distinction of the galactic components we are interested in studying in the galaxies in our sample. This choice allows us to discriminate between galactic disc emission and extraplanar emission, providing us with a more solid basis for our analysis and interpretation of the eDIG in galaxies of our sample.

Considering that the mass of our galaxy sample ranges between $10^{9}$ and  $10^{10}\,M_{\odot}$ (see Table~\ref{table:edig}), one might infer that our galaxies are dwarf galaxies with fainter emission in older stellar populations. To explore the classification of dwarf galaxies, we compare our sample with the  dwarf irregular (dI)  galaxies from \cite{fingerhut-2010}’s  survey. Firstly, it is worth noting that most of these dI galaxies were too faint to be detected by the 2MASS, implying  magnitudes of $K\geq 13.5$~mag \citep{jarrett-2000, jarrett-2000PASP, jarrett-2003}. Secondly, the mass range of their dI sample falls between $10^{7}$ to $10^{9}\,M_{\odot}$ \citep{mccall-2012}. In contrast, our sample consists of galaxies with apparent \textit{K$_{s}$}-band magnitude $K\leq 13.5$~mag (see selection criteria in Section~\ref{Sec:Sample}), exhibiting masses one order magnitude higher and having been previously studied in the context of the 2MASS LGA and the 2MASS \textit{Extended Source Catalogue}. This prior analysis ensures that our galaxies are indeed extended objects with bright \textit{NIR} photometry. Therefore, in the context of \cite{fingerhut-2010}, our sample does not align with the category of dwarf galaxies. Since we are not studying dwarf galaxies, we could claim that the surface brightness isophote at 3$\sigma$ of the \textit{K$_{s}$}-band image, selected to trace the stellar disc, encompasses the light of both the bulge and the disc.

Inside the ellipses fitted to the 3$\sigma$ level of the 2.22\,$\mu$m images, we observed regular discs, while the isophote at 1$\sigma$ level shows that the old stars at the borders follow the shape of the halo. For instance, the faintest \textit{NIR} emission in CIG\,71 suggests that the old stars are  displaced to the South as well as the H$\alpha$ emission, and in CIG\,201 the faint old stars trace the same shape of the filamentary structure outlined by the H$\alpha$ and \textit{UV} emission. 
Regarding the behaviour of the $B_{4}$ parameter versus the major axis relation (see Section~\ref{Sec:stellarDisc}), in most galaxies the pattern of the curve changes increasing its dispersion from the radial position of the 3$\sigma$ level. This is more evident in CIG\,71, 171, 329, 847 and 936. While a similar break in the \textit{UV}  surface brightness profiles is traced at the radial position of the isophote at 1$\sigma$ level of the \textit{K$_{s}$}-band image, as observed in CIG\,171, 201 and 906. This break in the surface  brightness profiles could define the edge of the stellar disc, however, we still relay in our selection because of the better resolution of the 2MASS than the one of the GALEX images (see Section\,\ref{Sec:Obs}).  Moreover, for some galaxies as CIG\,329, 922 or 1003, if the stellar disc is defined by the  the isophote enclosing $\mu_{K}=20\,\mathrm{mag\,arcsec^{-2}}$ (1$\sigma$ level), they would appear to be dominated by an old stellar disc and no extraplanar gas would have been detected. In this way, besides to the environmental criteria adopted for selecting a sample, the paradox in selecting the edge of the stellar disc might be another parameter causing discordance in the published results regarding the detection of eDIG.

We have found a variety of morphologies in different wavelengths in galaxies of our sample. With respect to the old stellar population disc, most of the \textit{K$_{s}$}-band images (12 out of 14) showed regular discs with a central bulge and some small filaments along the disc that do not break the stellar disc symmetry in general. 
Two galaxies showed an asymmetric or perturbed stellar disc: CIG\,416 (Figure~\ref{Fig:maps-c416}) presenting a tail shape drawn in its \textit{K$_{s}$}-band image, and CIG\,593 (Figure~\ref{Fig:maps-c593}) with a truncated old stellar disc that follows the morphology of the disturbed and faint H$\alpha$ disc.

On the other hand, the only one galaxy in our sample reliably identified as LINER \citep{sabater-2012} and whose photometric analysis revealed a coefficient $B_4<0$ suggesting a "boxy" bulge is CIG\,329 (Figure~\ref{Fig:maps-c329}). 
Although it is difficult to detect a bar in edge-on galaxies, a plausible explanation could be that there is a bar in CIG\,329 which may have undergone a vertical bending instability to form this box-shaped bulge. 
It is well known that two-thirds of star-forming disc galaxies in the local universe are barred \cite[e.g.][]{Eskridge-2000,Whyteetal2002,Aguerrietal2009,Mastersetal2011}. Numerical simulations of isolated galaxies have shown that a stellar bar can be formed spontaneously \citep[see e.g.][]{Milleretal1970,Hohl1971,CombesandSanders1981,SellwoodandWilkinson1993,Athanassoula2003, Athanassoula2013,Sellwood2013,saha-2018, ghosh-2022}, but also, that a bar can be formed by a 1:1 prograde flyby interaction and remain several Gyr after the interaction when galaxies are already separated by several Mpc \cite[e.g.][]{lang-2014}. 
If the stellar bar of CIG\,329 was formed by a secular process, why has no bar formed in other galaxies in our sample? It might be that these galaxies are hostile to bar formation or that the bar has been destroyed in the course of their evolutionary pathways. Indeed, the presence of a massive central bulge can suppress the bar instability \citep[e.g., see][]{saha-2018,KatariaandDas2018}. Alternatively, bars can be destroyed by a massive central mass concentration and/or inflow of the interstellar gas in the central region \citep[e.g., see][]{Pfenniger1990,ShenSellwood2004,Athanassoulaetal2005,Bournaudetal2005,HozumiHernqusit2005,Athanassoula2013}. Furthermore, recent study by \citet{Ghoshetal2021} revealed that a minor merger can lead to a substantial weakening of bars, and in extreme cases, even a complete destruction of bars in galaxies.

Since most of the galaxies in our sample showed extraplanar material in both \textit{UV} and H$\alpha$ emission, in the next sections we discuss some general results from the sample in order to understand the origin of the eDIG and \textit{UV}-haloes detected.

\subsection{The eDIG detection}\label{Sec:ddd}

\begin{figure*}
\centering 
\includegraphics[width=0.8\hsize]{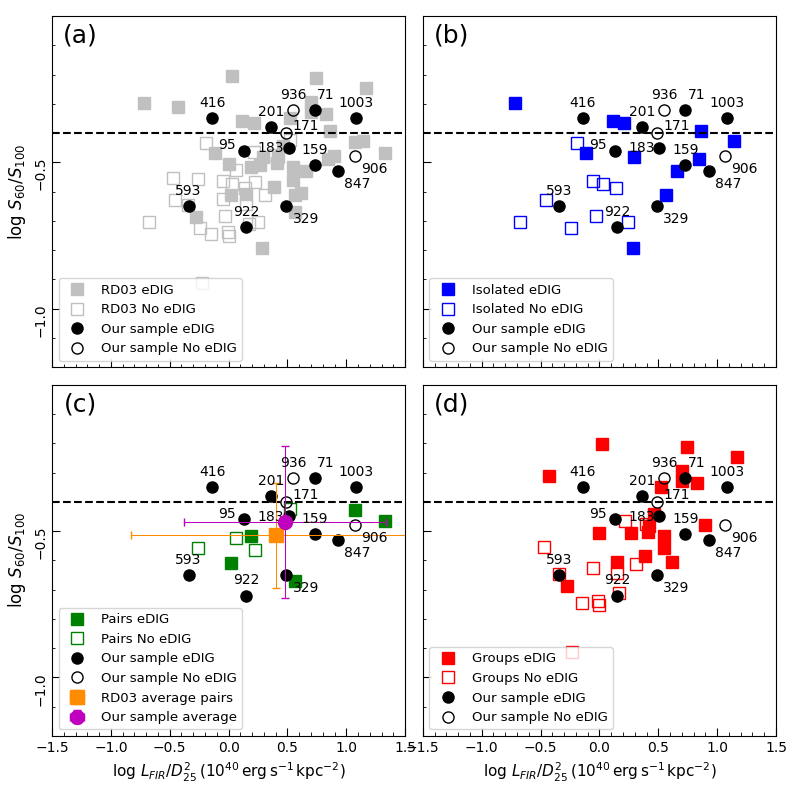}
\caption{
Diagnostic DIG diagram \citep{rossa-2003-i} showing the flux densities ratio at 60$\,\mu m$ and 100$\,\mu m$ ($S_{60}/S_{100}$)  versus the ratio of the \textit{FIR} luminosity ($L_{FIR}$) divided by the optical diameter of the $25^{th}\, \mathrm{mag\,\mathrm{arcsec^{-1}}}$ isophote squared ($D_{25}^{2}$) in units of $10^{40}\,\mathrm{\,ergs\,s^{-1}kpc^{-2}}$. The horizontal dashed line marks the threshold for warm galaxies at $S_{60}/S_{100}\geq 0.4$. 
We compare our sample (black circles with the CIG catalogue numbers indicated) with the whole sample (frame~a) of \citet[]{rossa-2003-i} (squares) labelled RD03. Then, we use the hierarchy published in databases to distinguish whether the galaxies in the \citet{rossa-2003-i} sample are isolated~(frame~b), in pairs~(frame~c) or in groups~(frame~d) (see Table\ref{tab:RD03}). 
Due to the low number of galactic pairs in the \citet{rossa-2003-i} sample, in panel~(c) we also compare the mean values and confidence intervals \citep[see][]{gehrels-1986} of this subsample (square) and that of our sample (circle). In all panels we distinguish  between galaxies that have (filled markers) or do not have eDIG (empty markers). 
}
\label{Fig:ddd}
\end{figure*}

In general, not all edge-on galaxies show detectable eDIG emission, i.e. a gaseous halo above the layer of \ion{H}{ii} regions associated with the disc.  
The presence of eDIG in late-type spirals is a direct consequence of SFR in the underlying galactic disc \citep[e.g][]{rossa-2000, rossa-2003-i}. 
It has been found that the prominence of gaseous haloes correlates with tracers of star formation in the disc, such as the surface density of \textit{FIR} emission, the SFR determined by H$\alpha$ luminosity, and dust temperatures \citep[e.g.][]{heald-2006, lu-changes-2023}. For instance,  \cite{rossa-2000} constructed a diagnostic DIG diagram (DDD) that shows the ratio of the flux densities at 60 and 100$\,\mu$m expressed as $S_{60}/S_{100}$ versus the ratio of the \textit{FIR} luminosity ($L_{FIR}$) divided by the optical diameter of the $25^{th}\,\mathrm{mag\,arcsec^{-2}}$ isophote squared  ($D_{25}^{2}$), with which they demonstrated that galaxies suggesting features associated with starburst activity as 
 $S_{60}/S_{100}\geq 0.4$ and $L_{FIR}/D_{25}^{2}\simeq 1\times 10^{40}\mathrm{erg\,s^{-1}\,kpc^{-2}}$ are candidates to display eDIG layers.

In order to locate our sample in the DDD, we computed the \textit{FIR}-luminosity using the expression:  
\begin{equation}
L_{FIR}=3.1\times 10 ^{39}\,d^{2}[2.58 S_{\nu}(60)+S_{\nu}(100)],
\label{ec:LFIR-rossa00Eq8}
\end{equation}
with $S_{\nu}(60)$ and $S_{\nu}(100)$ in Jy (obtained from \cite{lisenfeld-2007} and presented in Table~\ref{table:general}) and $d$ the distance to the galaxy in Mpc \citep{rossa-2003-i}. The term $D_{25}^{2}$ was expressed in $\mathrm{kpc^{2}}$.

From our maps, showed in Section~\ref{Sec:results} and in Appendix~\ref{Sec:individualMaps}, 
we detected galaxies showing one or more morphological eDIG features \citep[see e.g.][]{rossa-2003-i, heald-2006, Rosado-2013},  summarized in Table~\ref{table:edig}.  
The brightest \ion{H}{ii} regions in Local Group galaxies have a surface brightness detection between magnitude orders of  $10^{-14}$ and $10^{-15}\,\mathrm{erg\,s^{-1}\,cm^{2}\,arcsec^{-2}}$ \citep[e.g.][]{hodge-1990PASP..102...26H, azimlu-2011AJ....142..139A, cedres-2012A&A...545A..43C}, while the estimated mean sensitivity for the eDIG observed in other galaxies, such as NGC~891 or NGC~4565, is  on the order of $10^{-18}\,\mathrm{erg\,s^{-1}\,cm^{2}\,arcsec^{-2}}$ \citep[e.g.][]{rand-1996, rossa-2003-ii, ho-2016, jones-manga-2017}. For this work, the surface brightness detection limit of GHASP  is $2.5\times10^{-17}\,\mathrm{erg\,s^{-1}\,cm^{2}\,arcsec^{-2}}$ \citep{epinat-2008-ii, gomezl-2019, sardaneta-2022}, which is shallower than needed for detection of the faintest eDIG, but still within an acceptable range of sensitivity to offer a reliable perspective for our study.

By discriminating the H$\alpha$ emission in $z$-direction from the galaxies displaying  H$\alpha$ emitting gas extended radially, almost all the galaxies in our sample showed H$\alpha$ emitting gas extended radially outside from the stellar disc, CIG\,1003 being the only exception, whose H$\alpha$ emission is concentrated in the central regions drawing a cone shape and  ionized gas filaments. 
In the vertical direction, the H$\alpha$ monochromatic maps display eDIG configurations such as 
filamentary structures in CIG\,71, 329, 593 and 1003; 
layers of diffuse gas in CIG\,201, 329, 416; 
and patches or detached clouds of ionized gas emission in CIG\,95, 159, 593, 922, 847 and 1003. 
In some cases, the ionized gas falls in the outskirts of the stellar disc tracing a  
warped structure likely from a recent interaction as in CIG\,71 and 183.  
We have not detected any extraplanar H$\alpha$ emission in the vertical direction in CIG\, 171,  906 and 936.

As shown in the top left panel of Figure~\ref{Fig:ddd}, we compare the position in the DDD of our galaxies with those in the \cite{rossa-2003-i} sample which contains 62 spiral galaxies with inclination $i>76^{\circ}$, an optical diameter of $3\leq D_{25}\leq12$ and radial velocity $V_{rad}\leq6000\,\mathrm{km\,s^{-1}}$, all of them detected with IRAS in the \textit{FIR}. We must consider that when defining extraplanar emission, \cite{rossa-2003-i} did not distinguish between radially or vertically extended H$\alpha$ emission. In this case, all our galaxies would have eDIG. To ensure a more reliable comparison, in Figure~\ref{Fig:ddd}, we differentiate between galaxies in our sample with and without vertically extended H$\alpha$ emission.

Figure~\ref{Fig:ddd} shows that 6 out of 14 galaxies (42\%) of our sample lie above the threshold of warm galaxies at $S_{60}/S_{100}\geq 0.4$ (horizontal dashed line).  Half of these galaxies, namely CIG\,71, 936 and 1003, are located outside  the region of bona fide isolated galaxies of the AMIGA sample (see Figure~\ref{Fig:isolation}). In this context, we explore the possibility that  eDIG incidence is correlated with the environment. In fact, \citet{lisenfeld-2007} found that  the value of $S_{60}/S_{100}$ of the AMIGA sample is lower than the one of interacting samples from the literature indicating that the interaction can increase the dust temperature.  The  panel~(a) of Figure~\ref{Fig:ddd} also shows  that our sample and the \cite{rossa-2003-i} sample share a similar distribution: galaxies with eDIG from both samples populate the entire plane.

\cite{rossa-2003-i} rejected galaxies in compact groups but included widely spaced pairs of galaxies.  We associated an environmental classification 
to their sample using NED and Simbad\footnote{"The SIMBAD astronomical database", \citep{simbad-Wenger-2000}.} databases, by distinguishing groups, galaxy pairs and relatively isolated galaxies, whereby galaxies with siblings at a distance at $d<1^{\circ}$ were included in galaxies belonging to a galactic group and, galaxies with siblings at $d>1^{\circ}$ and galaxies with no record were considered isolated (see Table~\ref{tab:RD03}). 
These sub-samples are plotted in the (b), (c) and (d)~panels of Figure~\ref{Fig:ddd}. \cite{rossa-2003-i} detected eDIG in 5 out of 9 (56\%) paired galaxies and in 21 out of 34 (62\%) galaxies in groups: that is, in 26 out of 45 (58\%) interacting galaxies, while, for the sub-sample of isolated galaxies, the eDIG was detected in 10 out of 19 (53\%) targets. If we do not distinguish between radial and vertically extended emission, we detected H$\alpha$ emission extended outside the disc in all galaxies in our sample. However, due to the inaccuracies that may exist in the definition of the stellar disc (see Section~\ref{Sec:stellarDisc}), we consider having detected eDIG only in 11 out of 14 galaxies (79\%) where the diffuse ionized gas extends out of the disc vertically.

From the comparison with the entire sample of \cite{rossa-2003-i} we suggest that the environment is not connected with the presence of the eDIG. Confirming that   the incidence of eDIG in late-type spirals is a direct consequence of the SFR in the underlying galactic disc as found by \cite{rossa-2003-i}. The interaction could trigger the SFR for a short period of time. However, we also detect eDIG in a high percentage of galaxies that have been isolated for a long period of time.

\begin{table*}
\caption{Maximum radial ($r$) and vertical ($z$) distance reached by the extra-planar component in \textit{UV} and H$\alpha$ emission.}
\begin{center}
\begin{tabular}{c|c|cc|cc|cccc}
\hline
 &  \textit{MIR} & \multicolumn{2}{c}{\textit{NUV}}  &  \multicolumn{2}{c}{\textit{FUV}}  &  \multicolumn{2}{c}{H$\alpha$} & \multicolumn{1}{c}{H$\alpha$}& \multicolumn{1}{c}{Morphological} \\ 
\cmidrule(lr){3-4}  \cmidrule(lr){5-6} \cmidrule(lr){7-8}
CIG & M$_{*}$ & $z$ & $r$ & $z$ & $r$ & $z$ & $r$ & \multicolumn{1}{c}{eDIG morphology} & \multicolumn{1}{c}{features}\\ 
Name & (10$^{9}$~M$_{\odot}$) & (kpc) & (kpc) & (kpc) & (kpc) & (kpc) & (kpc) & \multicolumn{1}{l}{} & \multicolumn{1}{l}{}\\ 
(1) & (2) & (3) & (4) & (5) & (6) & (7) & (8) & \multicolumn{1}{c}{(9)}& \multicolumn{1}{c}{(10)} \\ 
\hline
71 & 14.1 & 4.2 & 8.8 & 2.0 & 3.9 & 1.2 & 4.7 & E(r) & A, Ph, UV,  \\ 
95 & 3.6 & 1.8 & 12.1 & 1.8 & 12.1 & 1.8 & 10.4 & E(r), \ion{H}{ii}-R & Ph, UV \\ 
159 & 3.8 & 1.4 & 10.7 & ... & 1.7 & 0.5 & 9.6 & E(r), \ion{H}{ii}-R & W, L(H$\alpha$), L(FUV), T \\ 
171 & 1.4 & 1.9 & 6.2 & ... & 5.6 & ... & 4.8 & E(r), \ion{H}{ii}-R & A, UV \\ 
183 & 8.1 & 4.5 & 8.0 & 2.6 & 8.0 & 2.8 & 8.0 & E(r)(z), \ion{H}{ii}-R, P & A, W, K, Ph \\ 
201 & 5.1 & 2.3 & 10.0 & 2.3 & 8.9 & 1.8 & 8.7 & E(r)(z), P, F & A, K \\ 
329 & 31.8 & 3.3 & 8.1 & 1.8 & 6.2 & 1.7 & 4.1 & E(r)(z), F & B, A, UV \\ 
416 & 2.6 & 1.2 & 7.1 & 1.1 & 6.2 & 1.0 & 2.9 & E(r)(z), \ion{H}{ii}-R, F & A, K, UV \\ 
593 & 12.1 & 2.5 & 13.0 & 2.1 & 11.0 & 1.6 & 9.3 & \ion{H}{ii}-R & A, W, UV, L(H$\alpha$)  \\ 
847 & 11.9 & 0.9 & 8.3 & 1.0 & 8.6 & 0.3 & 4.3 & E(r), \ion{H}{ii}-R & W, K, Ph \\ 
906 & 5.1 & 3.1 & 8.0 & 2.9 & 8.0 & ... & 1.4 & F & Ph, UV, L(H$\alpha$) \\ 
922 & 4.3 & 1.9 & 9.1 & 1.5 & 9.1 & 1.0 & 4.5 & E(r), \ion{H}{ii}-R & A, W, UV \\ 
936 & 1.0 & 1.7 & 18.4 & ... & 17.7 & ... & 1.4 & \ion{H}{ii}-R & UV, L(H$\alpha$), T \\ 
1003 & 6.0 & 3.2 & 4.4 & 1.3 & 1.2 & 1.2 & ... & E(z), F & A, W, K, UV \\ 
\hline
\end{tabular}
\end{center}
\label{table:edig}
Columns: (1)~CIG galaxy name; (2)~Stellar mass (M$_{*}$) computed with \textit{Wide-field Infrared Survey Explorer} (WISE) data at \textit{MIR} band (see equation~\ref{ec:1-cluver14}); (3)~and~(4),  (5)~and~(6) and (7)~and~(8) averaged maximum radial ($r$) and vertical ($z$) distance reached by the extra-planar component
at \textit{NUV}, \textit{NUV} and H$\alpha$ emission, respectively; (9)~morphological description of the extraplanar warm gas component or eDIG: E:~extended emission in radial~(r)- or vertical~(z)-direction mostly, P:~patches, F:~filaments, \ion{H}{ii}-R: extraplanar \ion{H}{ii}~region; (10)~morphological features in different wavelengths: 
B:~bar, A:~asymmetric disc, W:~warped disc, K:~multiple inner emission knots, Ph:~unmatching photometric maxima, UV:~  disc dominated by the UV halo, L($\lambda$): Low emission detected in the wavelength band $\lambda$.
 T: thin disc structure.
\end{table*}

\subsection{The \textit{UV}-halo}\label{Sec:uv-halo}

From the comparison between the H$\alpha$ and the \textit{UV} imagery, \cite{thilker-2007-i} discovered that 30\% of low inclined galaxies ($i<80^{\circ}$) in the local universe present extended \textit{UV} emission in the extreme outer disc where the number of \ion{H}{ii} regions expected is negligible  with respect to the \textit{UV} emission regions.   
In the case of high inclined galaxies ($i>80^{\circ}$), extended diffuse  \textit{UV} emission up to 15-20\,kpc above the disc midplane has been detected  \citep[e.g.][]{hodges-k-2014}, which always coincides with the presence of the extraplanar H$\alpha$ emission, although sometimes the ionized gas component is only detected at lower heights \citep{hodges-k-2016, jo-2018}.

From the individual maps displayed in Figure~\ref{Fig:maps-example} and Figures~\ref{Fig:maps-c95} to \ref{Fig:maps-c1003}, 
we found that all galaxies in our sample present radial or vertically extended \textit{NUV} emission, that all galaxies have radially extended \textit{FUV} emission, and that 11 out of 14 galaxies show vertically extended \textit{FUV} emission. These 11 galaxies displaying  \textit{UV} halo agree with the  galaxies in our sample showing diffuse ionized gas extended out of the disc vertically (eDIG) (see Table~\ref{table:edig}). We also observed that in most galaxies in our sample presenting the ionized gas extraplanar component, the H$\alpha$ and \textit{UV} emissions have very similar morphology and the \textit{UV} halo persists reaching larger heights  where the ionized gas is no longer detected, as observed previously by \cite{hoopes-2001} and \cite{hodges-k-2016}. And, finally, even in galaxies where we did not observe any extraplanar ($z-$axis direction) H$\alpha$ emission (CIG\,159, 171 and 936), we still have detected an \textit{UV} halo in the \textit{NUV} emission.

We found that 7 out of 14 (50\%) galaxies of our sample showing a radially extended \textit{UV} disc presented at least one feature of what could be perceived as a recent interaction such as disturbed ionized gas disc (CIG\,71, 95, 171, 183, 416, 593, 847) and/or mismatch of the gaseous and stellar maxima (CIG\,71, 95, 183, 593, 847) suggesting that the environment might be a potential cause of these events, even within isolated galaxies. 
From studies to low-inclined galaxies, it has been proposed that the extended \textit{UV} disc might be a consequence of young stars associated with low-mass stellar associations located at large galactocentric distances \citep{gil-de-paz-2007-ii} which were likely triggered by recent/ongoing interaction events or a high specific rate of gas accretion \citep{thilker-2007-i}. 
However, it is assumed that isolated galaxies have not experienced gravitational influences from its close neighbours over the past few billion years \cite[e.g.][]{Karachentseva-2009, verdesm-2005, rampazzo-2016}. Moreover, the lack of signals from a short-term event as a past major merger implies that the mechanism that supplies the gas needed to form the \textit{UV} extension has been working for a prolonged duration to achieve this level of richness \citep{thilker-2007-i}. Studying  \textit{UV} photometry of cluster galaxies, \cite{cortese-2012} interpreted the extended \textit{UV} emission as the result of the consumption of their \ion{H}{i} reservoir, inferring that this growth could be stopped and even reversed if the atomic hydrogen is removed via some kind of environmental eﬀect. Therefore, the observed \textit{UV} extension could potentially be the result of scattering dust particles stemming from ongoing regions of star formation \citep[see e.g.][]{shin-2015}, potentially introducing attenuation effects that shape the observed \textit{UV} emission.

We observed that the \textit{UV} halo (vertically extended \textit{UV} emission) in galaxies CIG\,201, 329 and 1003 does not trace a structured disc, these haloes are more extensive and have complex filamentary structures that match the H$\alpha$ and \textit{NIR} morphology.  Moreover, the \textit{NIR}, H$\alpha$ and \textit{UV} emission detected in CIG\,201 
and 1003 shows signs of a cone-shape suggesting the presence of a central outflow  (see individual galaxy notes in the Appendix\,\ref{Sec:individualResults}). 
From the comparison between \textit{UV} and \textit{NIR} imagery, it has been shown that \textit{UV} halo along the major axis relative to the optical radius $R_{25}$ is unrelated to the prominence of the bulge and, therefore, the radially concentrated \textit{UV} haloes are unlikely light from the bulge outskirts \citep{hodges-k-2016}. 
In the case of CIG\,201, 329 and 1003, the \textit{UV} diffuse emission might be expected as a result of scattered \textit{UV} light that escape from the disc dust (Diffuse extraplanar Dust, eDust) or emission from hot core-helium-burning stars \citep{hoopes-2005}. The simultaneity between the eDust and the  extraplanar H$\alpha$ emission suggests that a large fraction of the H$\alpha$ emission could be  originated from the galactic plane and is scattered by the eDust into the galactic halo \citep{hoopes-2005,  seon-2011, jo-2018}.

On the other hand, the existence of an UV halo does not always mean the existence of eDust \citep[e.g.][]{shin-2019}. We observed that in 10 out of 14 galaxies (71\%) of our sample presenting an UV halo (CIG~71, 95, 159, 171, 183, 416, 593, 847, 906, 922), the UV~light reveals an intense brightness and a vertical extension above the galactic centre, which gradually decreases with increasing distance from the galactic centre, rather indicating a disc extended structure. Other possible mechanisms producers of diffuse-and-global UV~haloes are galactic radiation, (magneto-)hydrodynamic phenomena or accretion from the CGM or ICM \citep{hodges-k-2014, shin-2015, hodges-k-2016, shin-2019}. Conversely, \cite{hodges-k-2016} observed that the UV haloes are brightest over the parts of the disc with bright H$\alpha$, X-ray, or radio emission, suggesting that they are connected to star formation in the disc, but not necessarily in a way that requires ionizing photons or winds to escape the disc.

Previous studies about the origin of the eDIG have found regions consistent with mixed OB and hot low-mass evolved stars (HOLMES), and OB–shock ionization as the primary driver of eDIG ionization \citep{FloresFajardo-2011, jones-manga-2017, levy-2019, rautio-2022}. The NUV light is dominated by stars in the turn-off point in the HR-diagram (turn-off stars), while, the FUV emission of old `normal’ stellar populations is dominated by post asymptotic giant branch (PAGB) stars \citep[e.g.][]{marino-2011-mnras}. HOLMES, which include PAGB stars and white dwarfs, are found abundantly in the thick discs and lower haloes of galaxies. Their considerable vertical distribution compared to OB stars and their significant contribution to the UV radiation of galaxies \citep[see e.g.][and references therein]{rautio-2022}, make them candidates to be a source of eDIG ionization and, consequently, of UV haloes. 

However, while late-type galaxies in nearby groups, observed from various inclination angles (ranging from 58° to 90°), display \textit{FUV} and \textit{NUV} images practically identical \citep[e.g.][]{marino-2010}, our  isolated galaxies, as previously noted in early-type galaxies, reveal that the \textit{FUV} emission appears more compact and luminous  compared to the \textit{NUV} emission
\citep[e.g.][]{gil-de-paz-2007, rampazzo-2007, jeong-2007, jeong-2009, marino-2010, marino-2011-s0, marino-2011-mnras}. Following \cite{cortese-2012}, if the extended \textit{UV} emission arises from the conversion of atomic hydrogen gas into new stars, it suggests that the stellar population is becoming younger or more  metal-poor. As a result, the turn-off stars shift to bluer colours and higher luminosity, increasing the \textit{NUV} emission. Simultaneously, the contribution to the \textit{FUV} from PAGB stars decrease because, despite their higher luminosity, the duration of the PAGB phase gets much shorter due to decreased available fuel \cite[][]{marino-2011-mnras}. 

In summary, the nature of \textit{UV} haloes in galaxies are subject to a complex interplay of factors, including SF, ionization processes, and possible CGM or ICM contributions. Further investigation of the specific mechanisms driving \textit{UV} haloes is needed to fully understand their origin and importance in different galactic environments.

\section{Summary and conclusions}\label{Sec:Conc}

Isolated galaxies constitute a reference model to study the environmental influence on galaxy evolution. 
In this work we studied the environmental impact on the incidence of eDIG in a sample of 14 nearby ($z\leq$0.02) isolated late-type edge-on ($i\geq80^{\circ}$)  galaxies from the Catalogue of Isolated Galaxies \citep[CIG,][]{Karachentseva-1973}.  We presented the H$\alpha$ emission maps of galaxies in our sample with data obtained from the scanning Fabry-Perot interferometer, GHASP, offering  a complete two-dimensional coverage of the emitting line region. We aimed to determine the galactic plane and to examine the distribution of the {\it warm} gaseous component with respect to the old and young stellar populations. Thus, we compared monochromatic H$\alpha$ images with the \textit{NIR} 2MASS \textit{K$_{s}$}-band and \textit{UV} from GALEX archive images.

We started by defining the stellar disc using the ellipse fitted to the 3$\sigma$ on the background level of the 2MASS \textit{K$_{s}$}-band image, which delimits the region occupied by both old and young stellar populations. However, we are aware of selecting the boundary of the stellar disc might be another parameter causing discordance in the published results regarding the detection of eDIG. Thus, we consider having detected eDIG only in the 11 out of 14 galaxies (79\%) where the diffuse ionized gas extends out of the disc vertically. If we do not distinguish between radial and vertically extended emission, we would have detected H$\alpha$ emission extended outside the disc in all galaxies in our sample. In this case, we demonstrated that our sample is comparable to the broader sample studied by \cite{rossa-2003-i} when studying the incidence of eDIG.

We found that all galaxies in our sample present extraplanar \textit{NUV} emission in both radial and vertical directions. While  11 out of 14 galaxies (79\%) have vertically extended \textit{FUV} emission, the same galaxies presenting eDIG. Because we showed that at least 79\%~($\sim$2/3) of isolated late-type, high inclined galaxies have both extended UV discs and UV haloes, our study reinforces the observations made by \cite{thilker-2007-i} who found that one-third of the low-inclined galaxies in the local Universe have extended UV discs. Hence, suggesting that the oldest star formation occurring between $\sim$10 and 100~Myrs extends well beyond the disc defined by the H$\alpha$ map which only traces the most recent star formation (younger than 10~Myrs).

We find that the PA obtained with \textit{NIR}, optical and \textit{NUV} images varies on average $\Delta$PA$\simeq\pm2.5^{\circ}$, which is expected for galaxies that have not experienced strong interactions in the last billions of years. However, when computing the inclination, some cases showed $\Delta i\simeq\pm10^{\circ}$ with respect to the optical inclination as a consequence of the presence of old stellar population centralized discs and \textit{UV} extended discs and haloes.

Our study suggests that the eDIG is revealed in a significant fraction of galaxies  
(79\% of the galaxies presented H$\alpha$ emission vertically extended) 
which have been isolated from a long period of time. When compared with the generic sample of highly inclined galaxies by \cite{rossa-2003-i} we did not find evidence that the environment plays a role in the presence of the eDIG incidence. This confirms that the presence of eDIG in late-type spirals is a direct consequence of  the SFR in the underlying galactic disc as suggested by \cite{rossa-2003-i}.

We foresee to discuss the H$\alpha$ kinematics of the galaxies in our sample in a forthcoming paper. Furthermore, since GALEX images are quite faint we are acquiring {\tt Astrosat-UVIT} \citep{tandon-2017}  pointed observations  of our galaxies in the \textit{FUV} band.

\section*{Acknowledgements}

Based on observations taken with the GHASP spectrograph at the Observatoire de Haute Provence (OHP, France), operated by the French CNRS. The authors acknowledge the technical assistance provided by the late Olivier Boissin \Cross\, from LAM and the OHP team before and during the observations, namely the night team: Jean Balcaen, St\'ephane Favard, Jean-Pierre Troncin, Didier Gravallon and the day team led by Fran\c{c}ois Moreau. 
This research has made use of the NASA/IPAC Extragalactic Database (NED), which is operated by the Jet Propulsion Laboratory, California Institute of Technology, under contract with the National Aeronautics and Space Administration. 
This research has made use of the SIMBAD database, operated at CDS, Strasbourg, France. 
M.M.S. thanks the ``Programa de Becas Posdoctorales en la UNAM'' of DGAPA-UNAM. 
M.R. acknowledges the project CONACyT~CF-86367. 
I.F.C. acknowledges the financial support of SIP-IPN grant no.\,20232054. 

\section*{Data Availability}

The data used in this work can be found online on the data servers of \textit{The STScI Digitized Sky Survey}\footnote{\textit{The STScI Digitized Sky Survey} \url{https://archive.stsci.edu/cgi-bin/dss_form}}, \textit{Two Micron All Sky Survey} (2MASS)\footnote{\textit{Two Micron All Sky Survey} (2MASS) \url{https://irsa.ipac.caltech.edu/Missions/2mass.html}} and the \textit{Galaxy Evolution Explorer} (GALEX) \footnote{\textit{Galaxy Evolution Explorer} (GALEX) \url{https://archive.stsci.edu/missions-and-data/galex}}. The FP data underlying this article will be shared on reasonable request to the corresponding author. 


\bibliographystyle{mnras}
\bibliography{biblio_2023} 


\appendix

\section{Spatial signal enhancement}\label{sec:VT}

\begin{figure}
\centering
\includegraphics[width=0.85\hsize]{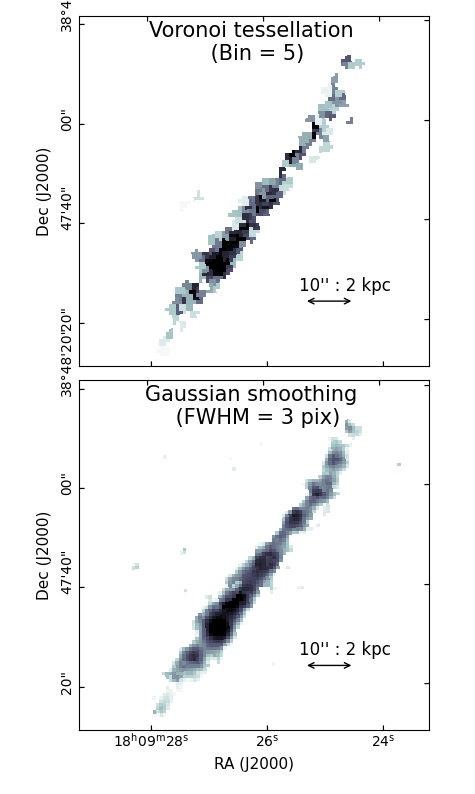}
\caption{H$\alpha$ monochromatic map of CIG\,847 (UGC\,11132) obtained after being applied a Voronoi tessellation (VT) with signal-to-noise ratio SNR\,$\simeq5$ (top panel) and a spatial Gaussian smoothing with $\mathrm{FWHM}\simeq 2"$ (3\,pix) (bottom panel). We use the VT binned map to mask the Gaussian smoothed map to ensure that all regions of this map have SNR\,$\geq 5$. 
}
\label{Fig:mono}
\end{figure}

To enhance the spatial signal-to-noise ratio (SNR), we employed two techniques on the wavelength data cubes. Firstly, we implemented adaptive spatial binning through Voronoi tessellation \citep{capellari-2003, daigle-2006-fp, daigle-2006-i}, aiming for a target SNR of approximately 5~or~7. Additionally, we applied a Gaussian smoothing with a $\mathrm{FWHM}\simeq 2"$\,(3\,pix). To maintain a SNR threshold above~5 in the Gaussian smoothed data cube, we performed a comparison between both H$\alpha$ monochromatic maps and we set a limit for the detected emission in the Gaussian smoothed maps. 
Since we aim to study the morphology of the eDIG, traced by the H$\alpha$ line emission, we opted to utilize the H$\alpha$ monochromatic map resulting from Gaussian smoothing. This map yields a more comprehensive representation of the morphology of the detected H$\alpha$ emitting gas, in comparison to the map obtained through spatial binning via Voronoi tessellation. To illustrate this technique, Figure~\ref{Fig:mono} presents the H$\alpha$ monochromatic map for CIG~847 (UGC~11132).

\section{Calibration of the total H$\alpha$ flux}\label{sec:flux}

As   indicated by \cite{epinat-2008-ii}, our major scientific goal is not to use the Fabry–Perot technique to make photometric studies but kinematic ones. Therefore, during the observations, we decided not to calibrate our data, saving observing time. On the other hand, there is no H$\alpha$ calibrated images of galaxies in our sample in the literature yet to compare our targets which would provide us with a more accurate flux range. However, to obtain the total H$\alpha$ flux of our galaxies, we can  follow previous total H$\alpha$ flux indirect calibrations made by \cite{epinat-2008-ii} and by \cite{gomezl-2019} for large galaxy samples observed with the same instrument, GHASP.

\cite{epinat-2008-ii} calibrated the total H$\alpha$ flux of more than 200 galaxies from the GHASP sample by making a linear regression of the comparison between monochromatic FP images of 69 galaxies and images with calibrated fluxes from narrow-band filters including H$\alpha$ and [\ion{N}{ii}] which had been already published. The coefficient of the resulting line is $0.48\pm0.06\times10^{-16}\,\mathrm{W\,m^{-2}\,ph^{-1}\,s}$, the fitting  was forced to pass through the origin for obvious physical reasons. Later, \cite{gomezl-2019} studied the kinematics of a sample of 152~late-type galaxies from the HRS catalogue\footnote{\textit{The Herschel Reference survey} \citep[HRS; ][]{boselli-2010}}. 
In this context, they calibrated indirectly the FP monochromatic H$\alpha$~maps, this time using the calibrated fluxes from \cite{boselli-2015}, obtaining a similar coefficient of the resulting line fitted and a GHASP surface brightness detection limits of $\sim2.5\pm0.2\times10^{-17}\,\mathrm{ergs\,s^{-1}\,cm^{-2}\,arcsec^{-2}}$ 
for 2~hours exposure time,  resulting in objects with an average lower threshold of SNR$\geq5$. This value of surface brightness detection limit was also obtained by \cite{sardaneta-2022} when, studying the kinematics of the eDIG in the ram-pressure stripped Virgo galaxy NGC~4330, compared the GHASP H$\alpha$~monochromatic image of this galaxy with the deep CFHT~H$\alpha$ image from VESTIGE survey\footnote{\textit{A Virgo Environmental Survey Tracing Ionised Gas Emission} \citep[VESTIGE; ][]{boselli-2018}}.

Thus, from the results obtained for galaxies observed with the FP~GHASP, we estimated the integrated flux of each galaxy in the sample studied in this work. We list these integrated fluxes in Table~\ref{tab:flux}.

\begin{table}
\caption{H$\alpha$ integrated fluxes deduced from previous extrapolations to large galaxy surveys observed with GHASP \citep[see e.g.][]{epinat-2008-ii, gomezl-2019}}
\begin{center}
\begin{tabular}{cccc}
\hline 
CIG & \multicolumn{ 3}{c}{Flux} \\ 
\cmidrule(lr){2-4} 
Name & (ph s$^{-1}$) & (10$^{-16}$ W m$^{-2}$) & (10$^{-15}$ erg s$^{-1}$ cm$^{-2}$) \\ 
\hline 
71 & 0.61 & 0.29 & 2.93 \\ 
95 & 0.30 & 0.14 & 1.42 \\ 
159 & 0.42 & 0.20 & 2.03 \\ 
171 & 0.34 & 0.16 & 1.64 \\ 
183 & 0.44 & 0.21 & 2.14 \\ 
201 & 1.80 & 0.86 & 8.62 \\ 
329 & 0.83 & 0.40 & 3.97 \\ 
416 & 0.36 & 0.17 & 1.74 \\ 
593 & 0.06 & 0.03 & 0.28 \\ 
847 & 0.19 & 0.09 & 0.93 \\ 
906 & 0.20 & 0.10 & 0.96 \\ 
922 & 0.36 & 0.17 & 1.74 \\ 
936 & 0.19 & 0.09 & 0.93 \\ 
1003 & 0.26 & 0.12 & 1.23 \\ 
\hline 
\end{tabular}
\end{center}
\label{tab:flux}
\end{table}

\section{Stellar Mass from MIR}

\textit{Wide-field Infrared Survey Explorer} (WISE) is a NASA-funded Medium-Class Explorer mission that consists of a 40\,cm space infrared telescope, whose science instrumentation includes $1024\times 1024$\,pixel$^2$ arrays of  Si:As and HgCdTe. WISE mapped the entire sky at 3.4\,$\mu$m ($W_{1}$), 4.6\,$\mu$m ($W_{2}$), 12\,$\mu$m ($W_{3}$), and 22\,$\mu$m ($W_{4}$). Each band covers a field-of-view of $47\times 47\,\,\mathrm{arcmin^{2}}$ with an angular resolution of 6.0\,arcsec in the short band-pass and  12.0\,arcsec in the longest one and a 1.375\,arcsec\,pixel scale \citep{jarrett-2012}. 
We infer the stellar mass using the mass-to-light ($M/L$) ratio of galaxies  \citep[e.g.][]{korsaga-2019}.  For low$-z$ galaxies, the stellar $M/L$ has a linear trend with WISE $W_{1}\,(3.4\,\mu\mathrm{m})-W_{2}\,(4.6\,\mu\mathrm{m})$ colour \citep{cluver_2014}:
\begin{equation}
\log(M_{*}/L_{W_{1}}) = -2.54(W_{1} - W_{2})-0.17,
\label{ec:1-cluver14}
\end{equation}
with $L_{W1}(L_{\odot})=10^{- 0.4(M-M_{\odot})}$, where $M$ is  the  absolute  magnitude  of  the  source  in  $W_{1}$, $M_{\odot} = 3.24$\,mag, and $W_{1}-W_{2}$ is the rest-frame colour of the source. Mid-infrared magnitudes and stellar masses are  listed in Table~\ref{table:mir}.

\begin{table}
\caption{Stellar mass and luminosity computed with the mid- and far-infrared magnitudes respectively. }
\begin{center}
\begin{tabular}{cccc}
\hline
CIG & $W_{1}$ ($3.4\,\mu$m) & $W_{2}$ ($4.6\,\mu$m) & $M_{*}$  \\ 
Name & (mag) & (mag) & ($10^{10}\,\mathrm{M_{\odot}}$) \\ 
(1) & (2) & (3) & (4)  \\ 
\hline
71 & 11.42 & 11.32 & 1.41  \\ 
95 & 12.59 & 12.53 & 0.36  \\ 
159 & 11.06 & 10.93 & 0.38 \\ 
171 & 11.58 & 11.18 & 0.14 \\ 
183 & 11.99 & 12.01 & 0.81 \\ 
201 & 11.10 & 11.01 & 0.51 \\ 
329 & 10.46 & 10.53 & 3.18 \\ 
416 & 12.40 & 12.28 & 0.26 \\ 
593 & 11.44 & 11.48 & 1.21 \\ 
847 & 11.28 & 11.18 & 1.19 \\ 
906 & 10.83 & 10.58 & 0.51 \\ 
922 & 12.03 & 11.95 & 0.43 \\ 
936 & 11.57 & 11.36 & 0.10 \\ 
1003 & 11.00 & 10.75 & 0.60 \\ 
\hline
\end{tabular}
\end{center}
Columns: (1)\,CIG galaxy name; (2)\,and\,(3)\,\textit{Wide-field Infrared Survey Explorer} (WISE) 3.4 and 4.6\,$\mu$m colour, respectively; (4)\,stellar mass computed using the relation \ref{ec:1-cluver14}.
\label{table:mir}
\end{table}

\section{Individual galaxies}\label{Sec:individualResults}

We provide the individual notes as discussed in Section~\ref{Sec:results}.

\begin{landscape}
\begin{figure}
\centering
\includegraphics[width=.95\hsize]{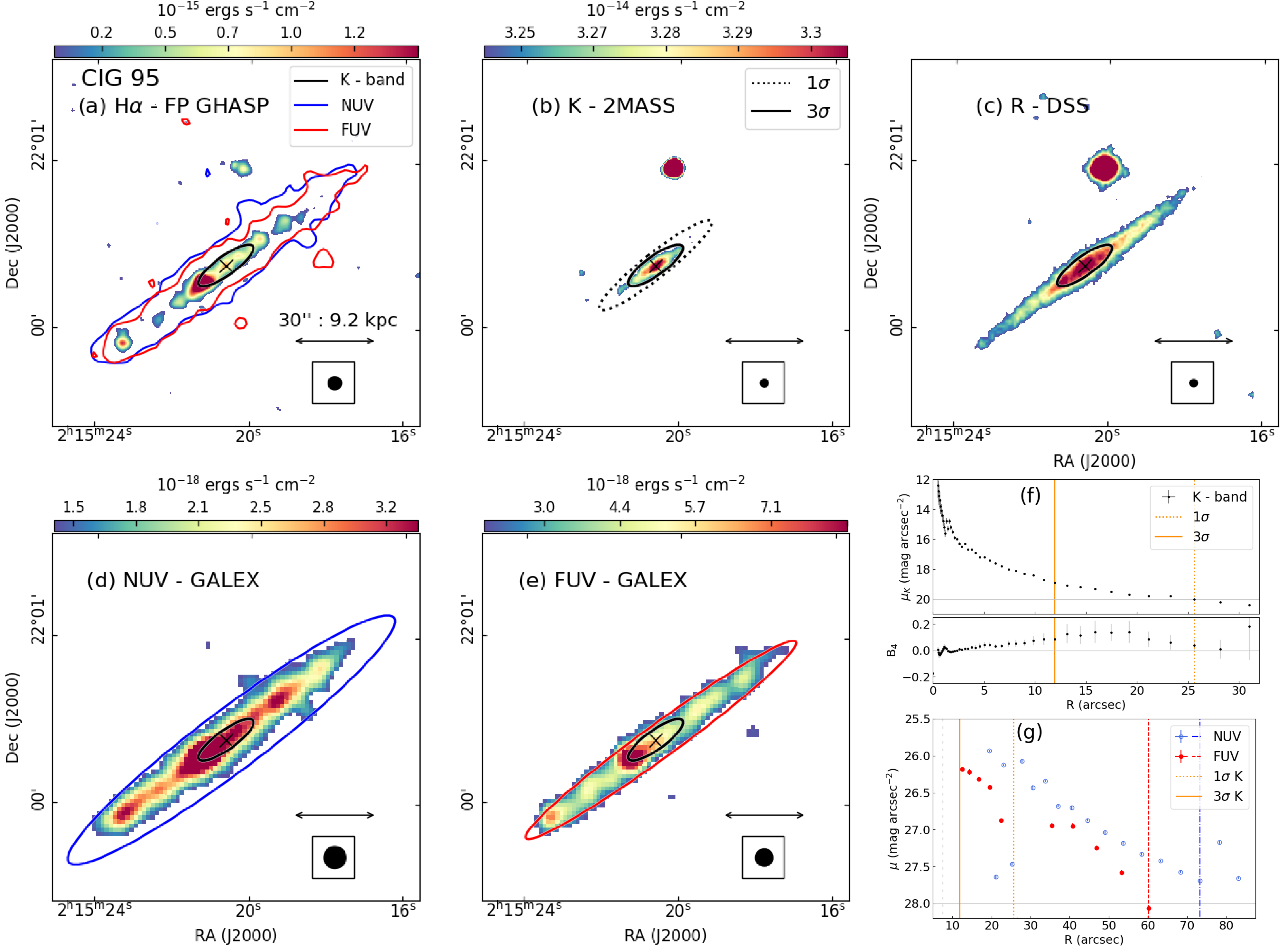}
\caption{CIG\,95 (UGC\,1733). Same as Figure~\ref{Fig:maps-example}.}
\label{Fig:maps-c95}
\end{figure}
\end{landscape}

\paragraph*{\textbf{CIG\,95 (UGC\,1733).}} Figure~\ref{Fig:maps-c95}: 
The H$\alpha$ map shows a detached cloud from the disc that exceeds radially the old stellar population disc observed in the \textit{NIR}-band emission. The brightest  H$\alpha$ emission region is much wider than the old stellar disc and exceeds it in the z-direction.  
The 2.22$\mu$m image reveals a central bulge surrounded by an asymmetric warped old stellar population disc that extends Southward. 
The \textit{FUV} emission displays an extended and warped disc. 
The \textit{NUV} emission traces a well-bound large disc with a brighter Southeastern tip and a sharpened, fainter Northwestern tip. 
The H$\alpha$, \textit{FUV} and \textit{NUV} emissions match in maxima emission but it is located to the Southeastern edge of the old stellar population disc. 
From the SDSS optical images, \cite{buta-2019} defined CIG\,95 as a Sc warped (see Table~\ref{table:general}). Although the galaxy appears to be slightly pulled to the Southeast, indicating a possible interaction, CIG\,95 meets all the isolation criteria of the AMIGA sample (see Section~\ref{sec:isolation}).

\begin{landscape}
\begin{figure}
\centering
\includegraphics[width=0.95\hsize]{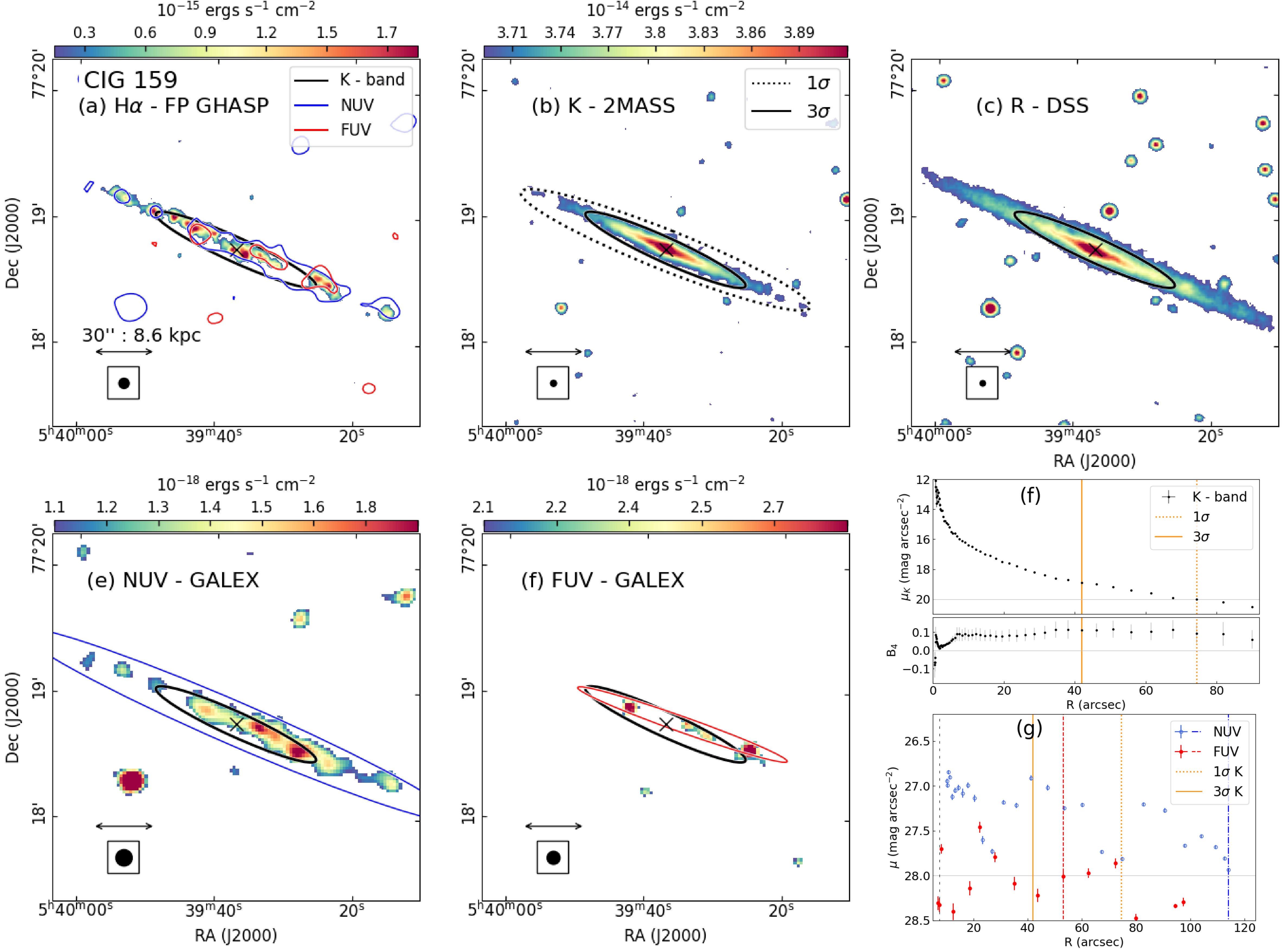}
\caption{CIG\,159 (UGC\,3326). Same as Figure~\ref{Fig:maps-example}. }
\label{Fig:maps-c159}
\end{figure}
\end{landscape}

\paragraph*{\textbf{CIG\,159 (UGC\,3326).}} Figure~\ref{Fig:maps-c159}: 
The H$\alpha$ emission 
displays a discontinuous disc composed of bright clouds uniformly distributed along the major axis with a size similar to that of the \textit{K$_{s}$}-band 1$\sigma$ isophote. 
The \textit{K$_{s}$}-band image of the galaxy shows a large thin warped disc with a central bulge. 
The GALEX \textit{FUV} image shows a few faint clouds mostly at the  
photometric centre. 
The \textit{NUV} emission draws a warped disc brighter on the East side than on the West side. The GALEX \textit{NUV} and \textit{FUV} images have low SNR which may yield spurious surface  brightness profiles. 
The location of the H$\alpha$ emission is consistent with the \textit{NUV} emission. 
The emission maxima of the four emissions presented in Figure~\ref{Fig:maps-c159} coincide within $\sim$5~arcsec.  
Further spectral data and deeper, higher resolution \textit{UV} images are required to draw a conclusion about the morphology of this apparently quiescent galaxy. However, \cite{Jones2018} found that in the \ion{H}{i} emission CIG\,159 is a not an isolated galaxy

\begin{landscape}
\begin{figure}
\centering
\includegraphics[width=0.95\hsize]{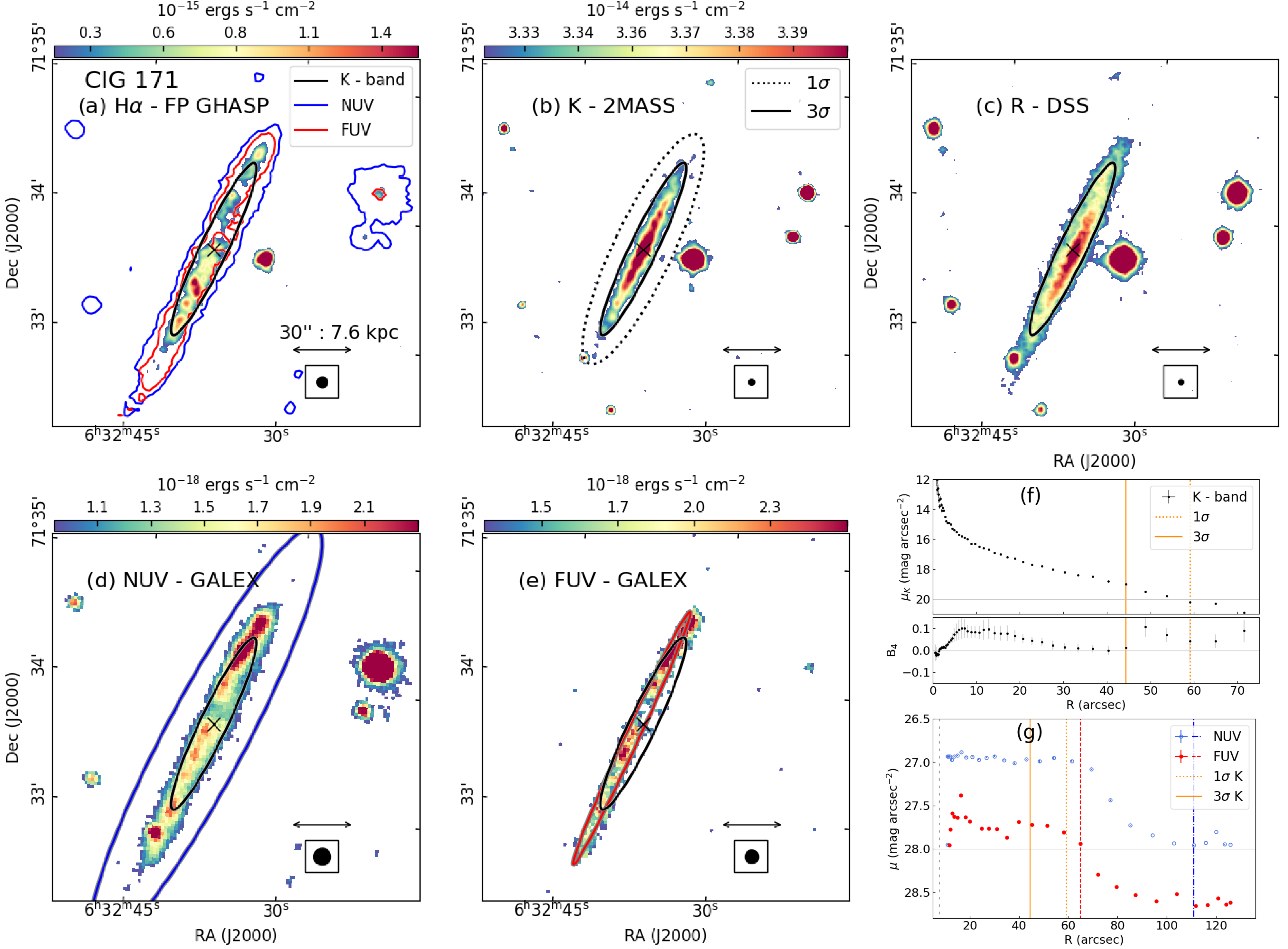}
\caption{CIG\,171 (UGC\,3474). Same as Figure~\ref{Fig:maps-example}.}
\label{Fig:maps-c171}
\end{figure}
\end{landscape}

\paragraph*{\textbf{CIG\,171 (UGC\,3474).}} Figure~\ref{Fig:maps-c171}: 
The H$\alpha$ emission is brighter in the South than the North likely shadowed by dust and stars emitting in UV. In the South the ionized gas has a dense distribution while in the North it traces detached clouds. 
The old stellar population disc is mostly symmetric with feathering along the edges. 
The \textit{NUV} and \textit{FUV} emissions trace a slightly warped  disc brighter in the North than in the South of the \textit{NIR} photometric centre. The \textit{UV}  surface brightness profiles are flat in the inner $\sim$60\,arcsec and fall exponentially outward. 
There is no H$\alpha$ emission over the stellar disc in the z-direction, but the H$\alpha$ and \textit{UV} discs are more extended radially than the old stellar population disc. The maximum emissions of \textit{NIR}, H$\alpha$, \textit{NUV}, and \textit{FUV} agree within a $\sim$5\,arcsec at the centre of the galaxy. \cite{Jones2018} classified CIG\,171 as not  isolated using \ion{H}{i} emission data.

\begin{landscape}
\begin{figure}
\centering
\includegraphics[width=0.95\hsize]{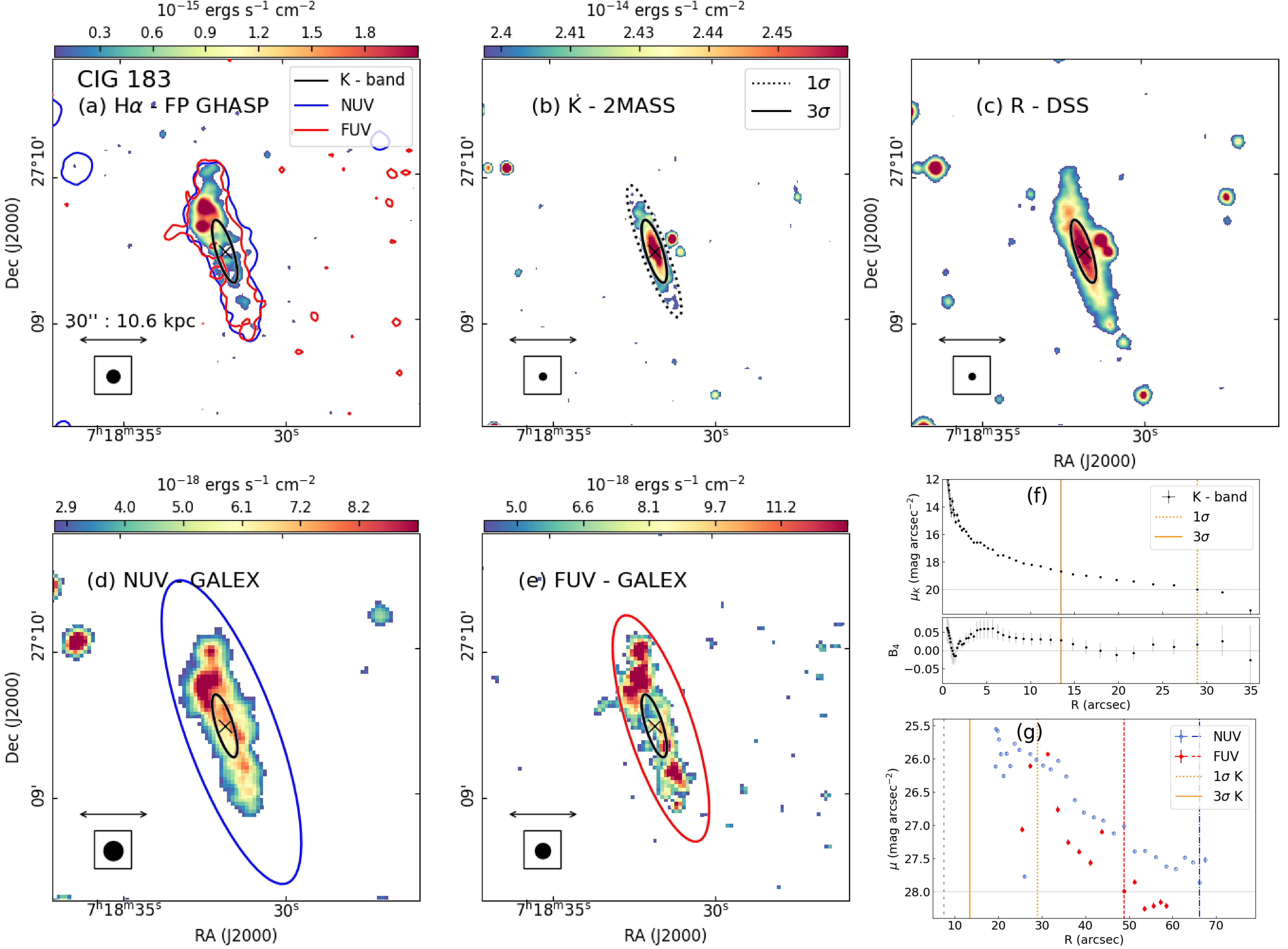}
\caption{CIG\,183 (UGC\,3791). Same as Figure~\ref{Fig:maps-example}. }
\label{Fig:maps-c183}
\end{figure}
\end{landscape}

\paragraph*{\textbf{CIG\,183 (UGC\,3791).}} Figure~\ref{Fig:maps-c183}: 
The H$\alpha$ emission displays an asymmetric disc displaced to the Northeast outwards of the old stellar population disc. It also shows two big bright regions located at the North of the photometric centre and several remnant clouds that draw two tails in both sides of the H$\alpha$ disc. 
The 2.22$\mu$m emission shows a mostly symmetrical disc with a bulge and two plumes at the ends of the galactic major axis. 
The \textit{NUV} and \textit{FUV} emissions agree with the structure of the H$\alpha$ monochromatic map. 
The maxima of the \textit{NUV}, \textit{FUV} and 2.22$\mu$m emissions match within a circle of $\sim$5~arcsec radius in the galactic centre. 
No additional spectral or photometric data are available yet in the literature to allow a conclusion about the morphology of this galaxy.

\begin{landscape}
\begin{figure}
\centering
\includegraphics[width=0.95\hsize]{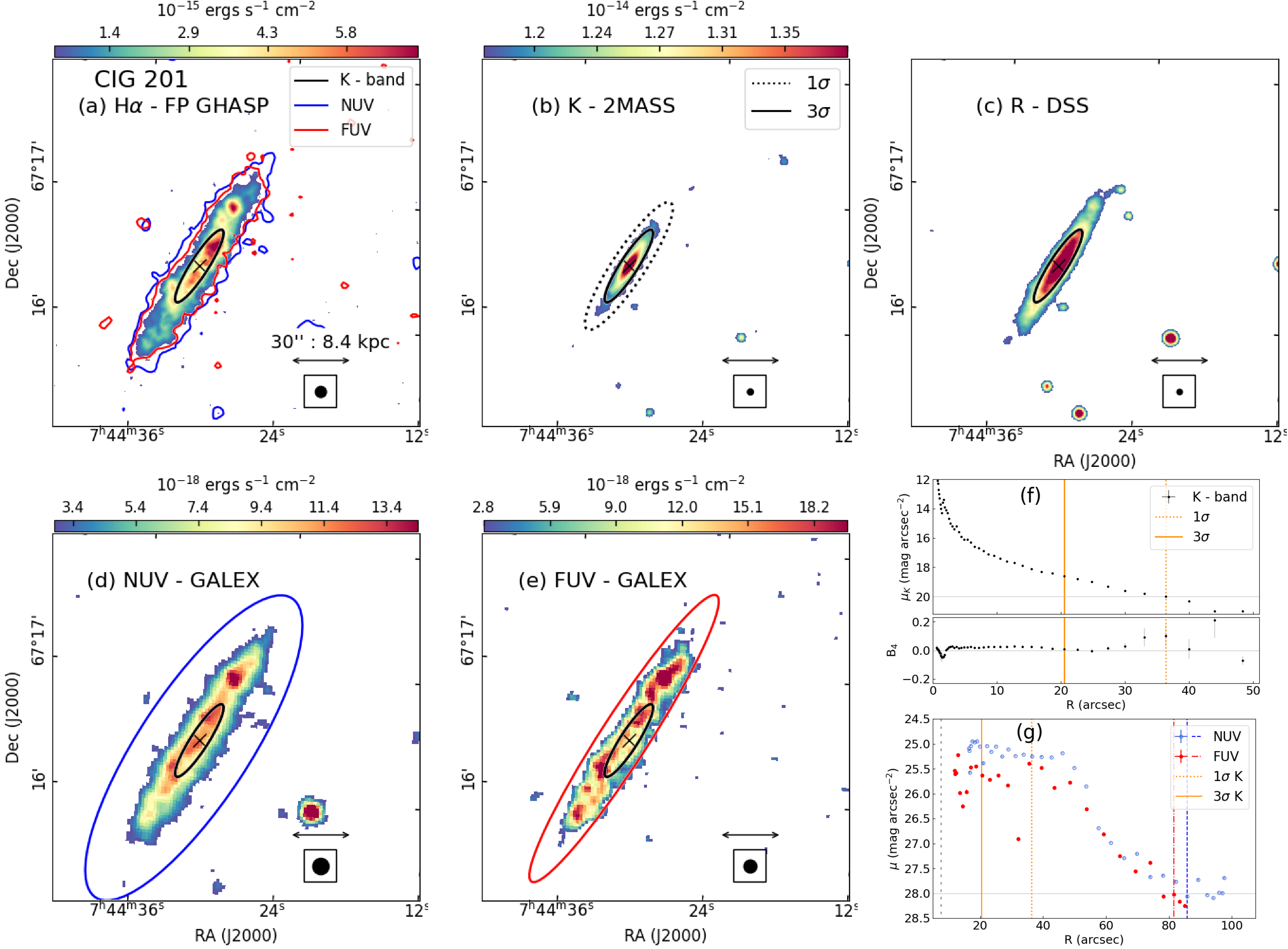}
\caption{CIG\,201 (UGC\,3979). Same as Figure~\ref{Fig:maps-example}.}
\label{Fig:maps-c201}
\end{figure}
\end{landscape}

\paragraph*{\textbf{CIG\,201 (UGC\,3979).}} Figure~\ref{Fig:maps-c201}: 
The H$\alpha$ emission map reveals a concentrated distribution of ionized gas, forming bright knots predominantly located in the central region of the galaxy. 
A filamentary structure displayed by the ionized gas is consistently observed across all wavelengths. 
The \textit{K$_{s}$}-band image shows a small disc that is symmetric but outwardly warped, tracing a similar shape to the H$\alpha$ and \textit{UV} emission. 
The \textit{NUV}, \textit{FUV}, and H$\alpha$ maps display extended discs that exceed more than twice the area occupied by the \textit{NIR} emission. 
The location of the H$\alpha$ bright knots match  with the brightest knots seen in  \textit{NUV} and \textit{FUV} emissions. 
The plateau shape of the \textit{NUV} and \textit{FUV} surface  brightness profiles in the inner $\sim$40\,arcsec may be due to the emission of these bright regions uniformly distributed within a specified area. 
All the maps of CIG~201, presented in Figure~\ref{Fig:maps-c201}, outline the same filamentary structure and display a cone shape at the galactic centre pointing Eastwards reaching different heights above the stellar disc. 
From SDSS data, \cite{buta-2019} classified CIG~201 as a SAc galaxy and identified an inner pseudoring (see Table~\ref{table:general}). In addition, from \ion{H}{i} emission data, \cite{Jones2018} determined that this galaxy is not interacting. 
The multi-wavelength morphology of CIG~201 suggests that the galactic disc may host a galactic wind.

\begin{landscape}
\begin{figure}
\centering
\includegraphics[width=0.95\hsize]{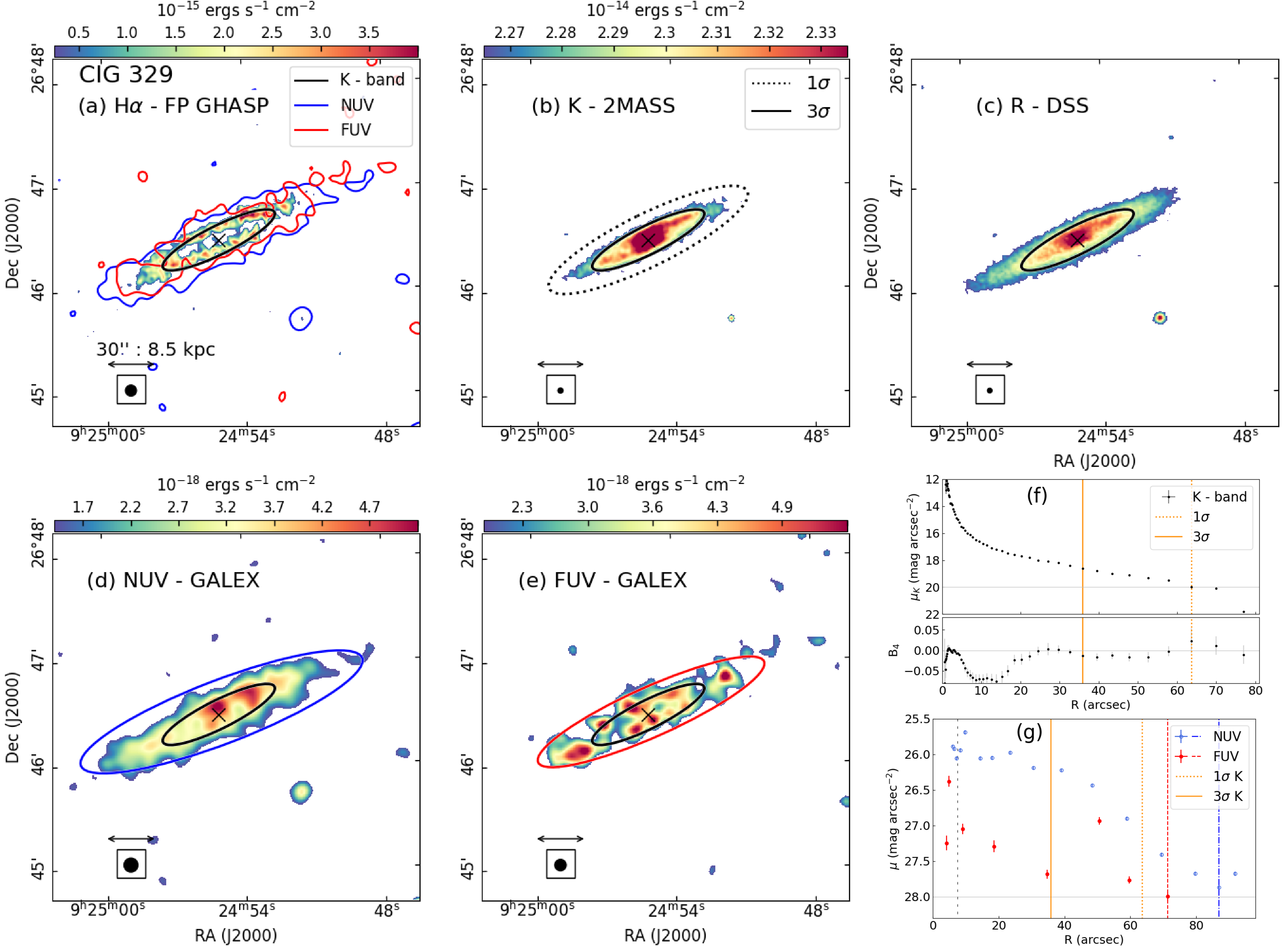}
\caption{CIG\,329 (UGC\,5010). Same as Figure~\ref{Fig:maps-example}.}
\label{Fig:maps-c329}
\end{figure}
\end{landscape}

\paragraph*{\textbf{CIG\,329 (UGC\,5010).}} Figure~\ref{Fig:maps-c329}: 
The H$\alpha$ emission is distributed in a nearly symmetric ring structure. Indeed, \cite{buta-2019} classified it as a SAb galaxy identifying an inner pseudoring (see Table~\ref{table:general}). 
The 2.22$\mu$m emission displays a long and symmetric disc with a central bulge and dust lane, with filaments along the edge. 
Following the photometric analysis of the \textit{K$_{s}$}-band image, this galaxy has a `boxy' bulge ($B_4<0$). 
The brightest \textit{FUV} emission regions agree with the H$\alpha$ emission, while the faintest ones draw a filamentary warped disc and few detached clouds in the North. 
The \textit{NUV} emission reveals a central dense distribution of the young stellar population, the  \textit{NUV} brightest clouds are consistent with the 2.22$\mu$m emission and the faintest draw filaments along the disc and the Northern cloud traced by the \textit{FUV} emission.  
Based on the SDSS spectroscopy, \cite{sabater-2012} classified CIG~329 as a LINER. CIG~329 satisfies all the AMIGA sample isolation criteria (see Section~\ref{sec:isolation}).

\begin{landscape}
\begin{figure}
\centering
\includegraphics[width=0.95\hsize]{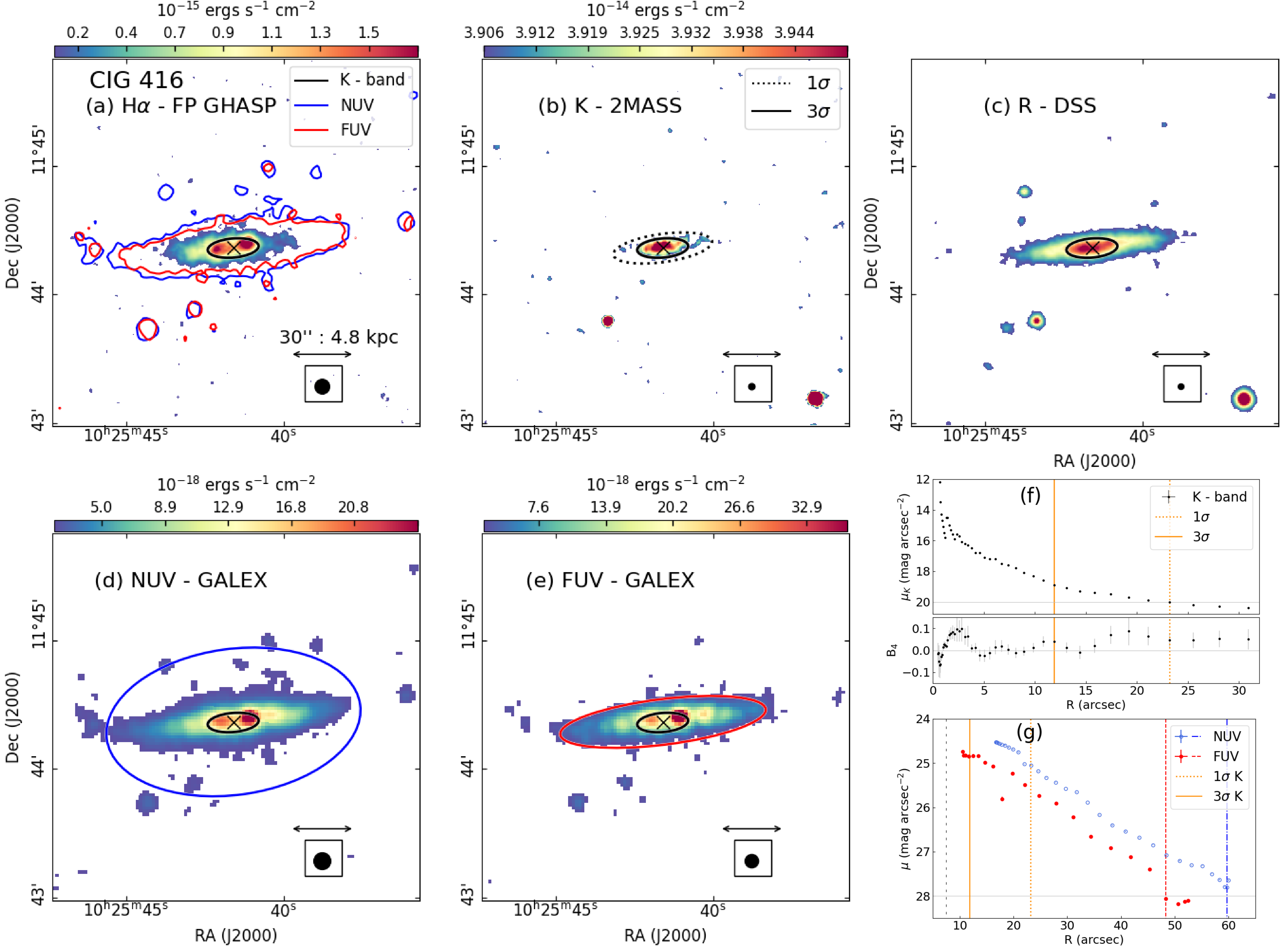}
\caption{CIG\,416 (UGC\,5642). Same as Figure~\ref{Fig:maps-example}.}
\label{Fig:maps-c416}
\end{figure}
\end{landscape}

\paragraph*{\textbf{CIG\,416 (UGC\,5642).}} Figure~\ref{Fig:maps-c416}: 
The H$\alpha$ emission map displays an asymmetric filamentary wide and radially extended disc with a tail-shape at the Western tip and a few detached ionized gas clouds. 
The \textit{K$_{s}$}-band image of this galaxy shows a small disc dominated by a bright central bulge. Beyond the \textit{K$_{s}$}-band isophotal perimeter of the 3$\sigma$ level, the old stellar population disc  is asymmetric with a filamentary structure and a tail shape pointing Northwards at the West extreme. The phenomenon triggering the Western tail
could be the reason for the non-linear relationship the $B_4$ parameter and the radius,  however, neither the graphics nor the \textit{K$_{s}$}-band image show the presence of a bar. 
The \textit{NUV} and \textit{FUV} images show even more extended asymmetric discs and larger detached clouds than the H$\alpha$ emission image. The images of these three bands present bright knots inside the disc.  
The photometric centre of the H$\alpha$, \textit{NIR}, \textit{NUV} and \textit{FUV} emissions match quite well. 
The asymmetric disc is also observed at the optical wavelength since \cite{buta-2019} classified CIG\,416 as Sdm peculiar using SDSS data. CIG\,416 meets the isolation criteria of AMIGA sample in the optical band, nonetheless, it does not accomplish these criteria using radio band data \citep{Jones2018}. Therefore, the asymmetries, removed clouds, filaments and tails may be a signature of a past interaction.

\begin{landscape}
\begin{figure}
\centering
\includegraphics[width=0.95\hsize]{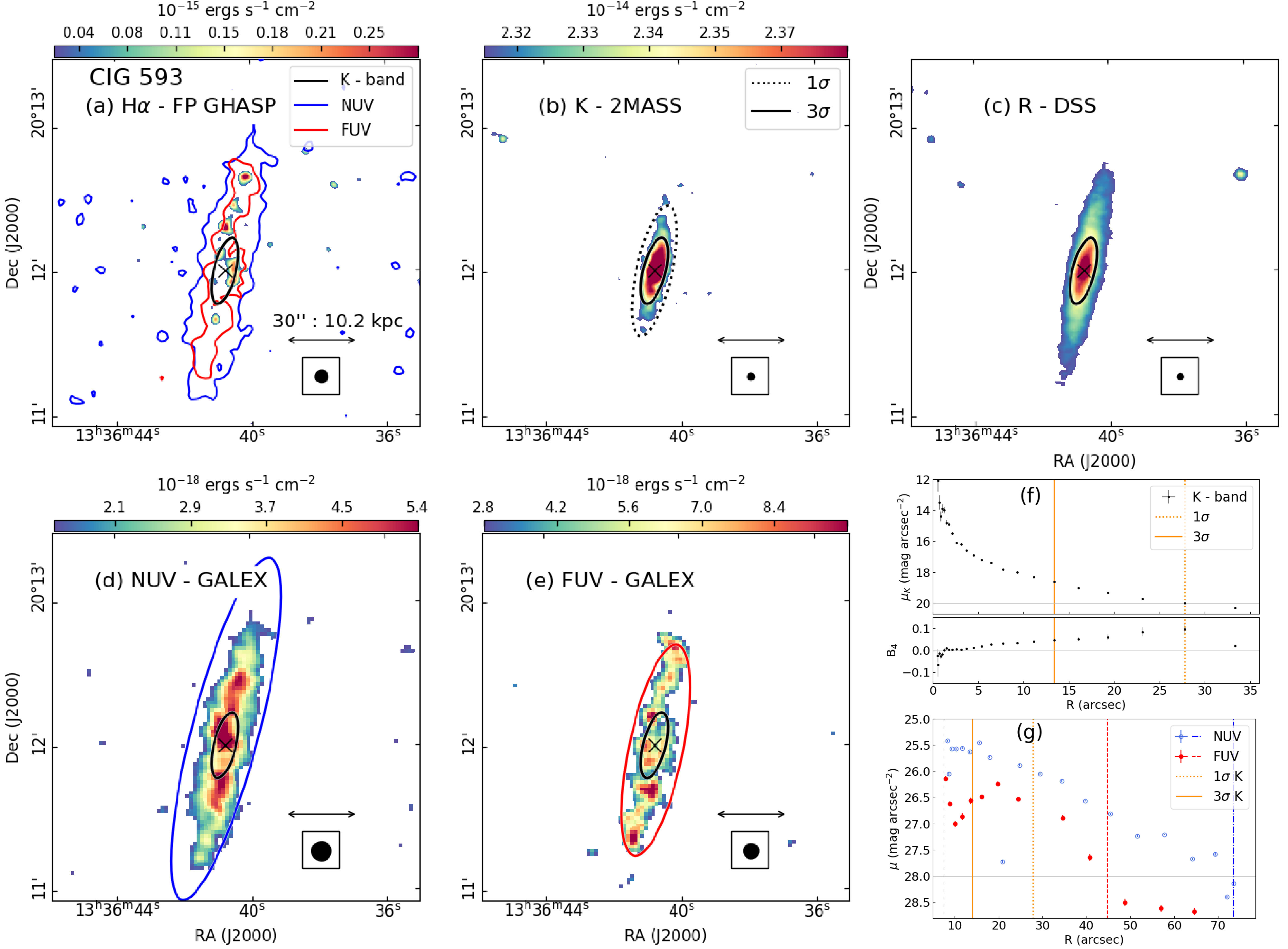}
\caption{CIG\,593 (UGC\,8598). Same as Figure~\ref{Fig:maps-example}.}
\label{Fig:maps-c593}
\end{figure}
\end{landscape}

\paragraph*{\textbf{CIG\,593 (UGC\,8598).}} Figure~\ref{Fig:maps-c593}: 
Despite of a long exposure of 223 minutes ($\sim$4\,hours), we only detected a few faint H$\alpha$ clouds tracing an asymmetric and truncated ionized gas distribution only visible on the North side, slightly  displaced with respect to the major axis of the old stellar population. It looks an `integral sign arm' starting from a small bar. Indeed, \cite{buta-2019} classified this galaxy as a strongly barred (SB, see Table~\ref{table:general}).
The stellar disc shows a central bright bulge. The most diffuse \textit{NIR} emission displays an asymmetric and truncated disc. 
The extended \textit{NUV} emission disc covers the entire galaxy and draws a warped disc as the one traced by the \textit{NIR} emission. The \textit{FUV} emission brightest knots at the North are located in the same position as the H$\alpha$ clouds. The maxima of the \textit{NUV}, \textit{FUV} and \textit{NIR} emissions are displaced by $\sim$10\,arcsec in the galactic centre. 
CIG~593 meets  all the AMIGA sample isolation criteria (see Section~\ref{sec:isolation}).

\begin{landscape}
\begin{figure}
\centering
\includegraphics[width=0.95\hsize]{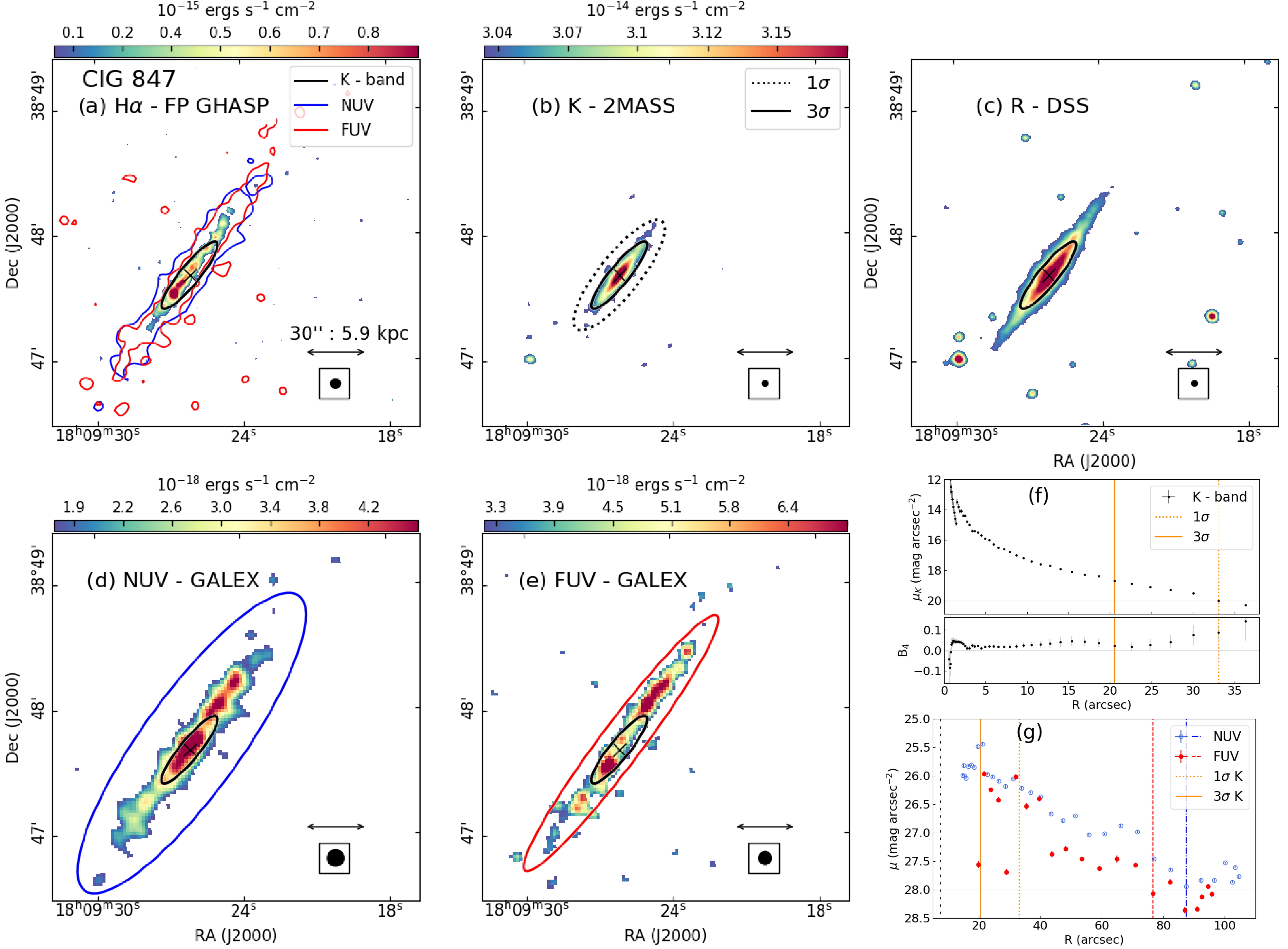}
\caption{CIG\,847 (UGC\,11132). Same as Figure~\ref{Fig:maps-example}. }
\label{Fig:maps-c847}
\end{figure}
\end{landscape}

\paragraph*{\textbf{CIG\,847 (UGC\,11132).}} Figure~\ref{Fig:maps-c847}: 
The H$\alpha$ emission depicts a radially extended asymmetric and warped disc more prominent and brighter towards the South than to the North along the major axis with the maximum located towards the Southeast from the \textit{NIR} photometric centre. The galaxy has a ionized gas tail-shape at the Southern tip and a detached cloud at the Northern edge.  
The 2.22$\mu$m emission reveals an almost symmetric elongated disc with small warp. 
The \textit{FUV} emission only shows large disengaged clouds along the major axis instead of a dense disc giving as a result an inaccurate surface brightness profile. 
The \textit{NUV} emission traces a long and warped disc with feathering along the borders of the disc, populating mainly the Northwest and with the maximum immediately to the North of the \textit{NIR} photometric centre. 
Both H$\alpha$ and \textit{NUV} emissions trace a tail shape at the Southeastern tip. The maxima of the \textit{NUV} and \textit{NIR} emissions match within $\sim$5\,arcsec in the galactic centre, while the \textit{FUV} and H$\alpha$ emission maxima are located $\sim$10\,arcsec and $\sim$15\,arcsec Southwards from them, respectively. Using the \ion{H}{i} image of CIG~847, \cite{Jones2018} found that this galaxy fulfils the most strict isolation criteria of AMIGA sample (see Section~\ref{Sec:Sample}).

\begin{landscape}
\begin{figure}
\centering
\includegraphics[width=0.95\hsize]{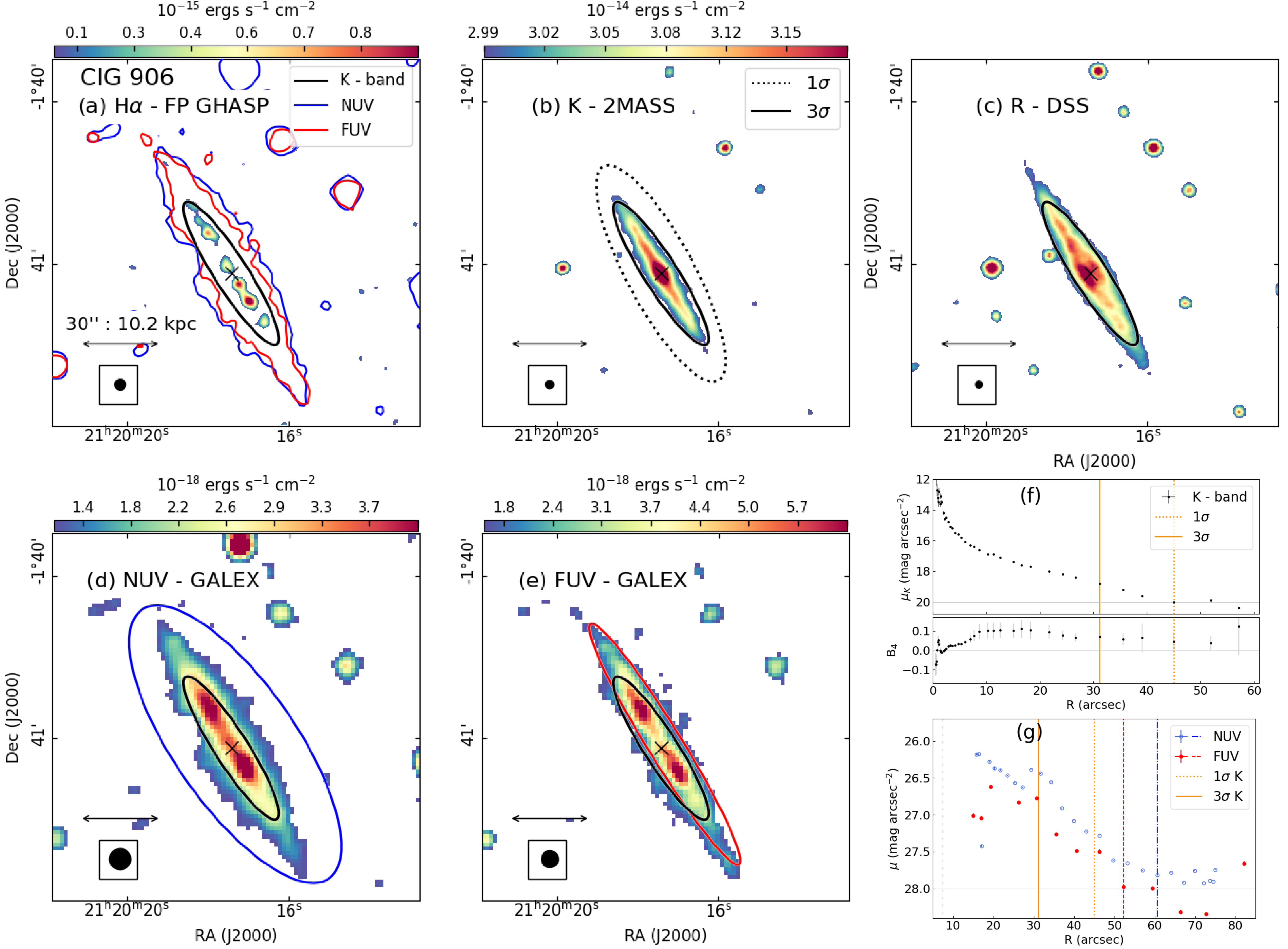}
\caption{CIG\,906 (UGC\,11723). Same as Figure~\ref{Fig:maps-example}. }
\label{Fig:maps-c906}
\end{figure}
\end{landscape}

\paragraph*{\textbf{CIG\,906 (UGC\,11723).}} Figure~\ref{Fig:maps-c906}: 
After an exposure of 186 minutes ($\sim$3~hours) with FP GHASP, only faint H$\alpha$ clouds as knots were detected along the galactic major axis. A faint ionized gas cloud is located over the Northern edge of the stellar disc. 
The \textit{K$_{s}$}-band image shows an old stellar population disc elongated and symmetrical with some filaments along the disc and a plume at the Southwest edge and a well-defined dust lane. 
The \textit{NUV} and \textit{FUV} emission display wider and brighter discs than the \textit{NIR} one, both show filaments all along the edges of the disc and even some detached clouds. The surface brightness profiles of \textit{NUV} and \textit{FUV} emission show a discontinuity between the ellipses fitted to the 1$\sigma$ and 3$\sigma$ brightness levels of the \textit{K$_{s}$}-band image, this is from $\sim$30 and 45\,arcsec. 
The maxima of the H$\alpha$, \textit{NUV} and \textit{FUV} emissions are displaced $\sim$10\arcsec Southwards from the \textit{NIR} one. 
Despite \cite{buta-2019} classified CIG~906 as Sbc, our analysis to the surface brightness profile of the 2.22~$\mu$m image showed a ‘disc-like' shape. According with \cite{Jones2018}, from \ion{H}{i} data, CIG~906 is a well isolated galaxy.

\begin{landscape}
\begin{figure}
\centering
\includegraphics[width=0.95\hsize]{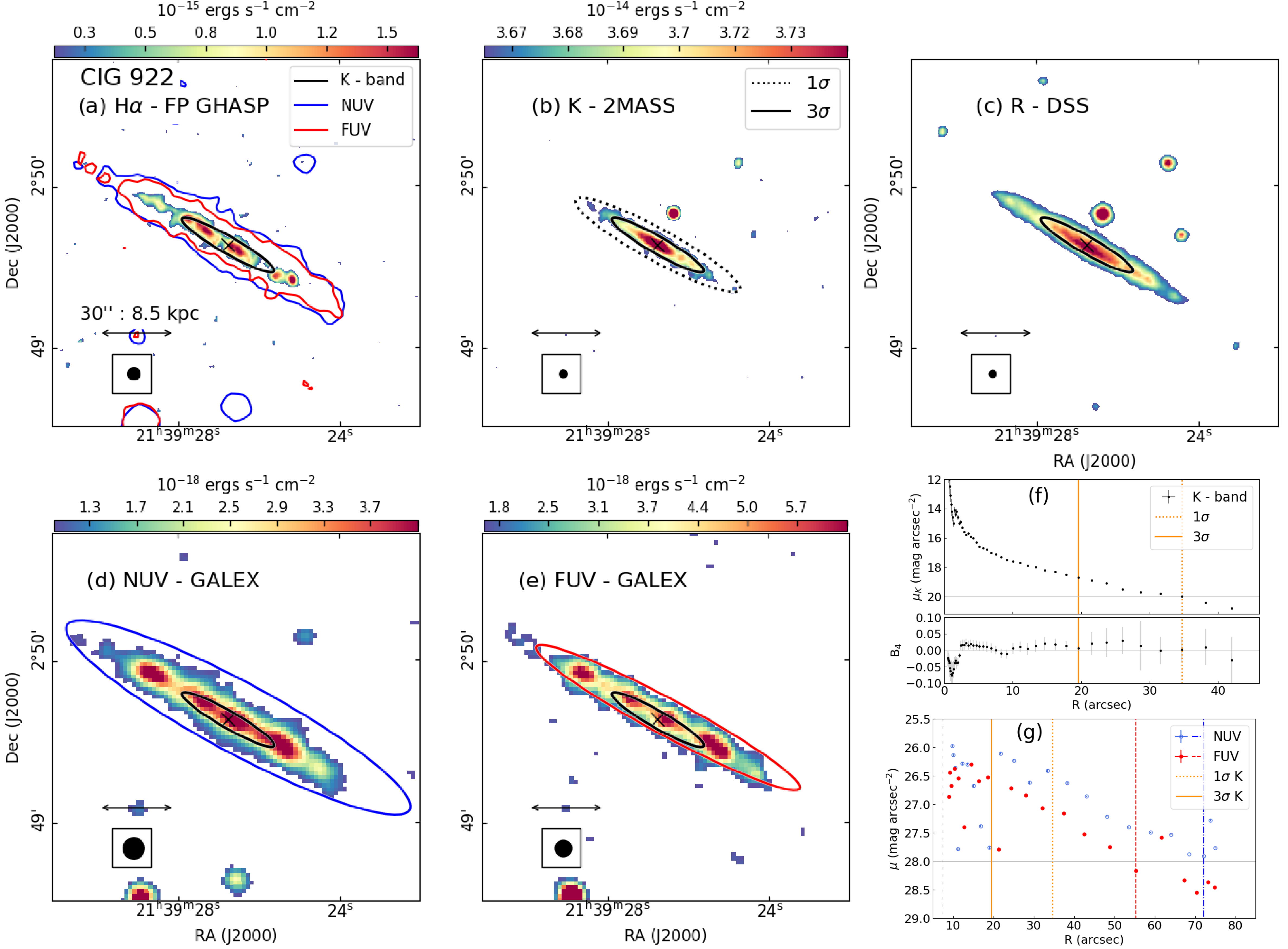}
\caption{CIG\,922 (UGC\,11785). Same as Figure~\ref{Fig:maps-example}.}
\label{Fig:maps-c922}
\end{figure}
\end{landscape}

\paragraph*{\textbf{CIG\,922 (UGC\,11785).}} Figure~\ref{Fig:maps-c922}:  
The H$\alpha$ emission disc is asymmetric and warped, it dominates the Northern galactic region and it is composed by large detached clouds. The ionized gas disc  is wider and longer than the old stellar disc. The H$\alpha$ cloud furthest North is faint and exceeds radially the old stellar disc tracing an arrow-shape, while to the South there are two H$\alpha$ detached bright knots just outside of the \textit{NIR} disc edge.   
The old stellar population disc is mostly symmetric and elongated, it has a central bulge.  The faintest \textit{NIR} emission shows a warp at the Northeastern and a plume outside the Southwestern tip of the stellar disc, both peculiarities are pointing to the Southeast. 
The \textit{FUV} emission traces a wider and denser disc than the H$\alpha$ emission disc but both match in shape: the rounded Southern tip and  the Northern arrow-shaped. 
The \textit{NUV} emission displays the most extended disc, with large plumes at one side of the disc pointing to the Northwest, and a detached cloud at the Northern tip. The maxima of the four emissions matches within  the central $\sim$5\,arcsec. 
\cite{buta-2019} classified CIG\,922 as Scd warped galaxy (see Table~\ref{table:general}) and, from \ion{H}{i} data, \cite{Jones2018} determined that this galaxy is not strictly isolated: the morphology of CIG~922 might be a result of some interaction.

\begin{landscape}\begin{figure}
\centering
\includegraphics[width=0.95\hsize]{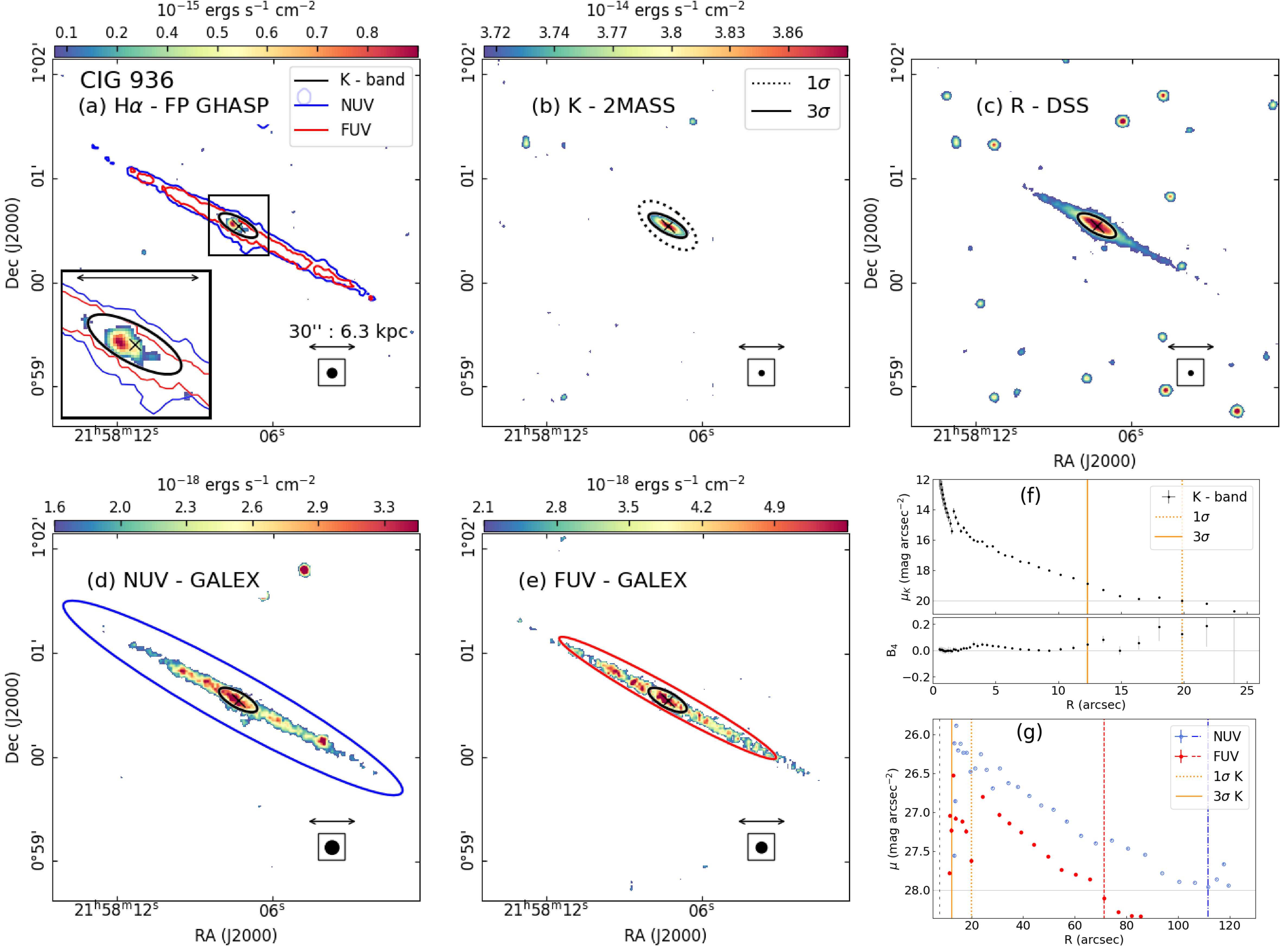}
\caption{CIG\,936 (UGC\,11859). Same as Figure~\ref{Fig:maps-example}.}
\label{Fig:maps-c936}
\end{figure}
\end{landscape}

\paragraph*{\textbf{CIG\,936 (UGC\,11859).}} Figure~\ref{Fig:maps-c936}: 
The \textit{K$_{s}$}-band and most of H$\alpha$ emissions were detected in the central 30\,arcsec of a galaxy with an optical diameter of $D_{25}(B)=185.4$\,arcsec (see Table \ref{table:general} and Figure~\ref{Fig:maps-c936}\textcolor{blue}{a}). 
The H$\alpha$ monochromatic image shows most of the warm gas compressed at the centre within the old stellar population disc, drawing a cone-shape. 
The \textit{K$_{s}$}-band image shows an asymmetric disc with a central bulge. 
The \textit{FUV} emission draws an extended thin disc (isophotal radius $r_{\mathrm{\textit{FUV}}}=47.6$\,arcsec, see Table\,\ref{table:ellipseParams}) whose tips are warped pointing to the Northwest. 
The \textit{NUV} disc emission is wider  than the \textit{FUV} one, with filaments mainly along the Southeastern side of the disc with the largest and broadest filament at the galactic centre. 
At $\sim$70\,arcsec Southwards from the galactic centre, there is a \textit{NUV} bright knot on the disc matching with the DSS R-band image and a faint H$\alpha$ cloud, which is probably a SF region because it is not detected at the \textit{NIR} emission. 
The maxima of the four emissions matches in the galactic centre. 
\cite{Jones2018} classified this galaxy as \ion{H}{i}-not-isolated (see Section~\ref{sec:isolation}).

\begin{landscape}
\begin{figure}
\centering
\includegraphics[width=0.95\hsize]{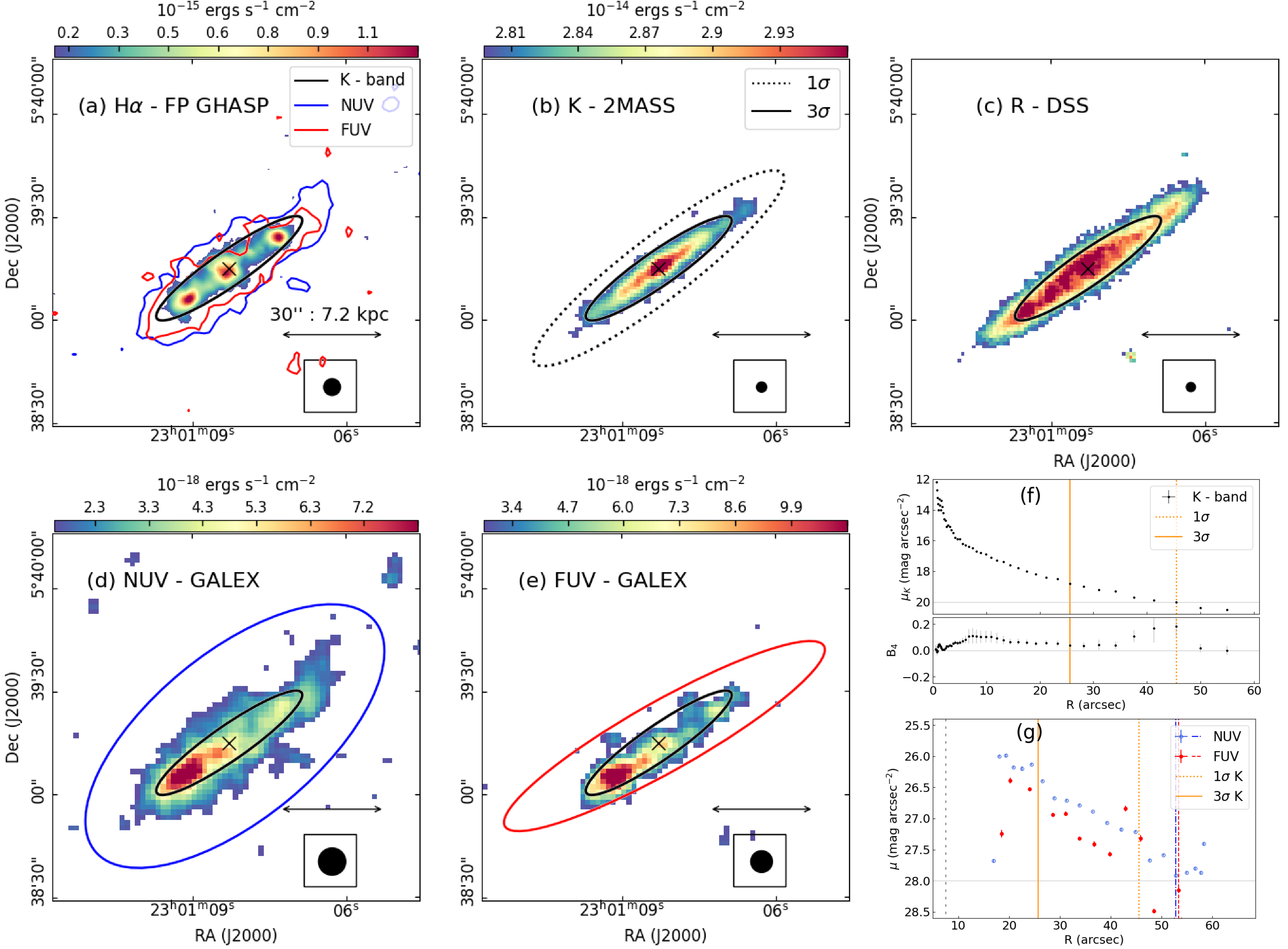}
\caption{CIG\,1003 (UGC\,12304). Same as Figure~\ref{Fig:maps-example}.}
\label{Fig:maps-c1003}
\end{figure}
\end{landscape}

\paragraph*{\textbf{CIG\,1003 (UGC\,12304).}} Figure~\ref{Fig:maps-c1003}: 
The ionized gas is compressed in the central region of the disc which presents three bright knots, one at the centre and the other two symmetrically placed $\sim$15\,arcsec away. 
The H$\alpha$ emission map displays filaments of ionized gas in all the edges, but the most prominent of them are around the central knot tracing a cone-shape pointing to the  Southwest.    
The \textit{K$_{s}$}-band image shows an elongated and symmetric stellar disc with a central bulge. 
The diffuse \textit{NIR} emission displays a cone-shape at the galactic centre pointing Southwest and a hook-shaped extension at the Northern tip pointing Northeast. 
The \textit{NUV} emission image shows a prominent filament of $\sim$15\,arcsec length pointing to the Southwest; in addition, the cone-shape in the central region is more noticeable in this emission. 
The \textit{FUV} and \textit{NUV} emission maps reveal very extended asymmetric and warped discs.  
Also, the 
brightest knots in these bands, match the Southern H$\alpha$ knot. 
All the photometric centres defined by the ellipse fitting lay on the central H$\alpha$ knot agreeing within $\sim$5\,arcsec. 
CIG\,1003 displays thinner discs Northwards and wider Southwards in the \textit{NIR}, \textit{FUV}, and \textit{NUV} emissions. 
CIG\,1003's multiwavelength morphology, position on the region of bona fide isolated galaxies (see Figure~\ref{Fig:isolation}) and \ion{H}{i} classification as not isolated \citep{Jones2018}, suggest that it has a galactic wind, possibly due to a recent interaction.

\section{Individual maps}\label{Sec:individualMaps}

\clearpage

\section{Rossa \& Dettmar (2003a) sample}

\begin{figure}
\centering
\includegraphics[width=0.95\hsize]{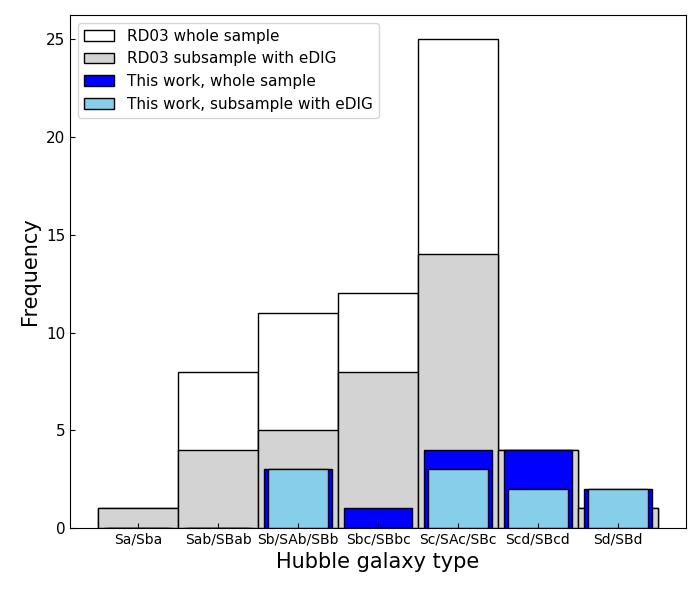}
\caption{Histogram of the Hubble galaxy type distribution of the whole galaxy sample of  \citet{rossa-2003-i} (RD03) (see Table~\ref{tab:RD03}) and their subsample of galaxies presenting eDIG compared with the Hubble galaxy type of our galaxy sample and our subsample of galaxies presenting eDIG.}
\label{Fig:histogram}
\end{figure}

\clearpage
\onecolumn
\footnotesize{
\begin{longtable}{ccccccccccc}
\caption{The sample of \citet{rossa-2003-i}.} 
\label{tab:RD03} \\
\hline
Object Name & RA & Dec  & Hubble &  V$_{\mathrm{sys}}$ & Log$_{10}$ M$_{*}$ & L$_{FIR}/D_{25}^{2}$ & S$_{60}$/S$_{100}$ & eDIG & Interact & Ref. \\ 
     &   J2000.0    &   J2000.0  & type  &  $\mathrm{km\,s^{-1}}$  & ($M_{\odot}$) & ($10^{40}\,\mathrm{\,ergs\,s^{-1}kpc^{-2}}$) &  &  &     \\ 
(1) & (2) & (3) & (4) & (5) & (6) & (7) & (8) & (9)  & (10) & (11) \\ 
\hline
\endhead
NGC 24 & 00h09m56.54s & -24d57m47.3s & Sc & 554 & 8.0 & 0.77 & 0.3396 &  & p & (i) \\ 
NGC 100& 00h24m02.84s & +16d29m11.0s & Sc & 844 & 8.3 & 0.21 & 0.1971 &  & i & (i) \\ 
UGC 260& 00h27m02.92s & +11d35m01.7s & Sc & 2134 & 9.8 & 3.30 & 0.3761 & * & p & (i) \\ 
ESO 540–16& 00h42m14.69s & -18d09m42.5s & SBcd & 1559 & 10.0 & 0.64 & 0.3692 & * & i & (ii) \\ 
NGC 360& 01h02m51.45s & -65d36m35.9s & Sc & 2306 & 8.9 & 1.94 & 0.1609 & * & i & (iii) \\ 
NGC 669& 01h47m16.15s & +35d33m47.9s & Sab & 4674 & 8.0 & 1.04 & 0.2455 &  & g & (i) \\ 
NGC 891& 02h22m33.41s & +42d20m56.9s & Sb & 528 & 5.9 & 3.19 & 0.2089 & * & g & (i) \\ 
UGC 2082& 02h36m16.15s & +25d25m25.8s & Sc & 708 & 5.5 & 0.35 & 0.2356 &  & i & (i) \\ 
IC 1862& 02h51m58.80s & -33d20m24.6s & Sbc & 6395 & 7.0 & 2.48 & 0.3357 &  & g & (i) \\ 
NGC 1247& 03h12m14.32s & -10d28m52.0s & Sab & 3947 & 8.8 & 2.92 & 0.3604 & * & g & (i) \\ 
ESO 117–19& 04h02m32.13s & -62d18m56.2s & SBbc & 5335 & 6.2 & 3.51 & 0.2759 & * & g & (ii) \\ 
IC 2058& 04h17m54.35s & -55d55m58.4s & Sc & 1379 & 5.1 & 1.15 & 0.2982 &  & p & (i) \\ 
ESO 362–11& 05h16m38.80s & -37d06m09.1s & Sbc & 1345 & 7.6 & 2.63 & 0.3295 & * & g & (ii) \\ 
IC 2135& 05h33m12.90s & -36d23m55.8s & Sc* & 1312 & 7.3 & 7.28 & 0.4051 & * & i & (ii) \\ 
ESO 121–6& 06h07m29.85s & -61d48m27.3s & Sc & 1203 & 10.2 & 7.05 & 0.3256 & * & i & (ii) \\ 
NGC 2188& 06h10m09.53s & -34d06m22.3s & SBcd & 747 & 8.4 & 1.63 & 0.4301 & * & i & (iii) \\ 
ESO 209–9& 07h58m15.01s & -49d51m05.1s & SBc & 1126 & 9.1 & 4.56 & 0.2948 & * & i & (ii) \\ 
UGC 4559& 08h44m07.61s & +30d07m08.9s & Sb & 2085 & 7.8 & 0.89 & 0.2721 &  & i & (iii) \\ 
NGC 2654& 08h49m11.87s & +60d13m16.0s & SBab & 1339 & 7.3 & 0.57 & 0.1882 &  & i & (i) \\ 
NGC 2683& 08h52m41.33s & +33d25m18.3s & Sb & 411 & 3.7 & 1.76 & 0.1976 &  & i & (i) \\ 
NGC 3003& 09h48m36.05s & +33d25m17.4s & SBc & 1480 & 6.7 & 1.65 & 0.3434 &  & g & (i) \\ 
NGC 3044& 09h53m40.88s & +01d34m46.7s & Sbc* & 1289 & 9.9 & 6.72 & 0.4633 & * & g & (i) \\ 
NGC 3221& 10h22m19.98s & +21d34m10.5s & SBcd & 4111 & 10.5 & 13.80 & 0.3738 & * & i & (iii) \\ 
NGC 3600& 11h15m52.01s & +41d35m27.7s & Sab & 703 & 9.3 & 1.30 & 0.4362 & * & i & (i) \\ 
NGC 3628& 11h20m16.97s & +13d35m22.9s & Sb & 846 & 8.9 & 5.02 & 0.4694 & * & g & (i) \\ 
NGC 3877& 11h46m07.70s & +47d29m39.6s & Sc & 895 & 6.8 & 4.15 & 0.2486 & * & g & (i) \\ 
NGC 3936& 11h52m20.59s & -26d54m21.2s & SBbc & 2013 & 9.4 & 2.03 & 0.2447 &  & g & (i) \\ 
ESO 379–6& 11h53m03.27s & -36d38m19.8s & Sbc & 2944 & 8.8 & 3.67 & 0.2452 & * & i & (ii) \\ 
NGC 4206& 12h15m16.81s & +13d01m26.3s & Sc & 703 & 10.7 & 0.53 & 0.2054 & * & g & (i) \\ 
NGC 4216& 12h15m54.44s & +13d08m57.8s & SBab & 131 & 10.6 & 0.59 & 0.1221 &  & g & (i) \\ 
NGC 4235& 12h17m09.88s & +07d11m29.7s & Sa & 2263 & 8.7 & 0.37 & 0.4896 & * & g & (i) \\ 
NGC 4256& 12h18m43.09s & +65d53m53.7s & Sb & 2489 & 4.6 & 0.71 & 0.1794 &  & g & (i) \\ 
NGC 4302& 12h21m42.48s & +14d35m53.9s & Sc* & 1098 & 8.5 & 3.69 & 0.2136 & * & p & (i) \\ 
NGC 4388& 12h25m46.75s & +12d39m43.5s & Sb & 2524 & 10.5 & 5.53 & 0.6127 & * & g & (i) \\ 
NGC 4402& 12h26m07.65s & +13d06m48.0s & Sb* & 237 & 7.0 & 7.82 & 0.3320 & * & g & (i) \\ 
NGC 4634& 12h42m40.96s & +14d17m45.0s & Sbc* & 139 & 5.0 & 11.84 & 0.3720 & * & p & (i) \\ 
NGC 4700& 12h49m08.15s & -11d24m35.5s & SBc & 1409 & 8.3 & 4.99 & 0.5074 & * & g & (i) \\ 
NGC 4945& 13h05m27.48s & -49d28m05.6s & SBc & 563 & 4.8 & 14.79 & 0.5661 & * & g & (i) \\ 
NGC 5170& 13h29m48.79s & -17d57m59.1s & Sc* & 1501 & 8.8 & 0.34 & 0.2796 &  & g & (i) \\ 
NGC 5290& 13h45m19.18s & +41d42m45.3s & Sbc & 2549 & 7.8 & 2.57 & 0.3139 & * & g & (i) \\
NGC 5297& 13h46m23.67s & +43d52m20.4s & SBbc & 2409 & 7.2 & 1.69 & 0.2714 &  & p & (i) \\ 
IC 4351 & 13h57m54.26s & -29d18m56.6s & Sb* & 2674 & 7.9 & 1.48 & 0.1947 &  & g & (i) \\ 
NGC 5775& 14h53m57.60s & +03d32m40.0s & SBc & 1676 & 6.4 & 21.49 & 0.3415 & * & p & (i) \\ 
ESO 274–1& 15h14m13.43s & -46d48m33.1s & Sd & 524 & 8.6 & 0.19 & 0.5055 & * & i & (ii) \\ 
NGC 5965& 15h34m02.46s & +56d41m08.2s & Sb & 3377 & 6.3 & 0.55 & 0.2769 &  & p & (i) \\ 
UGC 10288& 16h14m24.80s & -00d12m27.1s & SABc & 2044 & 8.5 & 0.89 & 0.2370 &  & g & (i) \\ 
IC 4837 & 19h15m14.64s & -54d39m41.1s & Sab & 2668 & 8.3 & 2.42 & 0.2605 & * & g & (i) \\ 
NGC 6875& 20h13m12.47s & -46d09m41.9s & SBc & 3121 & 9.0 & 1.56 & 0.3048 & * & p & (i) \\ 
MCG-01-53-012 & 20h49m52.23s & -07d01m18.5s & Sc & 6024 & 8.6 & 0.94 & 0.2067 &  & i & (ii) \\ 
IC 5052 & 20h52m05.57s & -69d12m05.9s & SBcd & 584 & 9.5 & 1.00 & 0.3112 & * & g & (i) \\ 
IC 5071 & 21h01m19.74s & -72d38m33.8s & SABb & 3123 & 8.0 & 1.41 & 0.2472 & * & g & (i) \\ 
IC 5096 & 21h18m21.54s & -63d45m38.4s & Sbc & 3144 & 7.4 & 1.42 & 0.2280 &  & g & (i) \\ 
NGC 7064& 21h29m02.98s & -52d46m03.4s & SBc & 767 & 5.1 & 1.06 & 0.6250 & * & g & (i) \\ 
NGC 7090& 21h36m28.86s & -54d33m26.4s & SBc & 847 & 4.6 & 1.95 & 0.3287 & * & g & (i) \\ 
NGC 7184& 22h02m39.82s & -20d48m46.2s & SBc & 2623 & 9.6 & 1.38 & 0.2588 &  & i & (iii) \\ 
IC 5171 & 22h10m56.70s & -46d04m53.3s & SBb & 2847 & 7.8 & 1.86 & 0.3101 & * & g & (ii) \\ 
IC 5176 & 22h14m55.93s & -66d50m57.9s & SBbc & 1748 & 5.9 & 3.51 & 0.3044 & * & g & (ii) \\ 
NGC 7361& 22h42m17.91s & -30d03m27.6s & Sc* & 1249 & 7.8 & 1.07 & 0.2669 &  & i & (i) \\ 
UGC 12281& 22h59m12.80s & +13d36m24.0s & Sc & 2562 & 5.3 & 0.98 & 0.1830 &  & g & (i) \\ 
NGC 7462& 23h02m46.49s & -40d50m06.9s & SBbc & 1064 & 10.0 & 3.34 & 0.4492 & * & g & (i) \\ 
NGC 7640& 23h22m06.58s & +40d50m43.5s & SBc & 369 & 8.1 & 0.45 & 0.2259 &  & g & (i) \\ 
ESO 240–11& 23h37m49.42s & -47d43m38.4s & Sb & 2836 & 5.9 & 1.00 & 0.1776 &  & g & (ii) \\ 
\hline \\
\multicolumn{10}{l}{
Columns: (1)~object name; (2)~and~(3)~wcs~coordinates (J2000); 
}\\
\multicolumn{10}{l}{
(4)~Hubble type from \citet{rossa-2003-ii}, when unavailable, marked by (*) from HyperLeda \citep{hyperleda};
}\\
\multicolumn{10}{l}{
(5)~heliocentric velocity ($V_{\mathrm{sys}}$); 
(6)~logarithm of the stellar mass computed with \textit{MIR}-band photometry (see equation \ref{ec:1-cluver14}); 
}\\
\multicolumn{10}{l}{
(7)~and~(8)~DDD parameters from \citet{rossa-2003-i}: 
\textit{FIR} luminosity ($L_{FIR}$) divided by the optical diameter of the $25^{th}\, \mathrm{mag/\mathrm{arcsec}}$ 
}\\
\multicolumn{10}{l}{
isophote squared ($D_{25}^{2}$) and flux densities ratio at 60$\,\mu m$ and 100$\,\mu m$ ($S_{60}/S_{100}$), respectively; 
} \\
\multicolumn{10}{l}{
(9)~cases where the eDIG was detected by \citet{rossa-2003-i}; 
} \\
\multicolumn{10}{l}{
(10)~hierarchy reported so far in the literature: i: isolated or no records, p: pair member, g: group member; 
}\\
\multicolumn{10}{l}{
(11) hierarchy reference: (\textit{i})~NED,  (\textit{ii})~SIMBAD, (\textit{iii})~There is no record.
}
\centering
\end{longtable}
}

\clearpage
\twocolumn



\bsp	
\label{lastpage}
\end{document}